\documentclass[12pt]{article}

\usepackage[scaled]{helvet}
\usepackage[T1]{fontenc}
\usepackage{setspace}
\usepackage{graphics}
\usepackage{latexsym}
\usepackage{rotating}
\usepackage{amsmath}
\usepackage{color}
\usepackage{mathpazo}
\usepackage{fullpage}
\usepackage{psfrag}
\usepackage{pifont}
\usepackage{amssymb}
\usepackage{indentfirst}
\usepackage{slidesec}
\usepackage{graphicx,subfigure}
\usepackage{epsfig}
\usepackage{fancybox}
\usepackage{color}
\usepackage{amssymb,amsmath}
\usepackage{epsfig, graphicx}
\usepackage{epsfig}
\usepackage{mathrsfs}
\usepackage{verbatim}
\usepackage{xcolor}
\usepackage{xspace}
\usepackage{hyperref}

\usepackage{bbm}
\usepackage{natbib}
\bibpunct[, ]{(}{)}{,}{a}{}{;}
\usepackage{ctable}
\usepackage{footnote}
\usepackage{enumerate}

\makesavenoteenv{tabular}

\newtheorem{theorem}{Theorem}

\newtheorem{lem}{Lemma}

\newtheorem{remark}{Remark}
\newtheorem{definition}{Definition}
\newcommand{\defn}[1]{\textbf{\emph{#1}}}

\def \btheta{\boldsymbol{\theta}}
\def \bthetahat {\widehat{\btheta}}
\def \bthetahatn {\widehat{\btheta}_n}
\def \bthetaast {\btheta^{\ast}}
\def \bthetaastn {\btheta_n^{\ast}}

\def \Hhat{\widehat{H}}
\def \Uhat{\widehat U}

\def \bS{\textbf{S}}
\def \bV{\textbf{V}}
\def \bR{\textbf{R}}

\def \bShatn{\widehat{\bS}_n}
\def \bShatnjk{\widehat{\bS}_{n,jk}}
\def \bVhatn{\widehat{\bV}_n}
\def \bVhatnjk{\widehat{\bV}_{n,jk}}
\def \bSast{\bS^\ast}
\def \bVast{\bV^\ast}
\def \bA{{\textbf A}}
\def \Acal{\mathcal A}
\def \sumin{\sum_{i=1}^n}
\def \bP {\textbf{P}}
\def \bM {\textbf{M}}
\def \Cbbm{\mathbbm C}
\def \cbbm{\mathbbm c}
\def \szero{{\mbox{\tiny 0}}}
\def \Ezero{\mathbbm{E}^\szero}

\begin{document}

\title{Information matrix equivalence in the presence of censoring: A goodness-of-fit test for semiparametric copula models with multivariate survival data}
\author{Qian M. Zhou\\
\small Department of Mathematics and Statistics, Mississippi State University,\\
\small MS, USA, qz70@msstate.edu}
\date{}
\maketitle

\abstract{Various goodness-of-fit tests are designed based on the so-called \textit{information matrix equivalence}: if the assumed model is correctly specified, two information matrices that are derived from the likelihood function are equivalent. In the literature, this principle has been established for the likelihood function with fully observed data, but it has not been verified under the likelihood for censored data. In this manuscript, we prove the information matrix equivalence in the framework of semiparametric copula models for multivariate censored survival data. Based on this equivalence, we propose an information ratio (IR) test for the specification of the copula function. The IR statistic is constructed via comparing consistent estimates of the two information matrices. We derive the asymptotic distribution of the IR statistic and propose a parametric bootstrap procedure for the finite-sample $P$-value calculation. The performance of the IR test is investigated via a simulation study and a real data example.}\\


\section{Introduction}

As a graduate student, one learned an important derivation about the likelihood method: assume a random variable $X$ has a distribution function $f(x;\btheta)$ (probability density function or probability mass function) with a $p$-dimensional parameter $\btheta$. Under certain regularity conditions \citep{white1982Maximum} assumed on $f(x;\btheta)$, we have the following equation:
\begin{equation}\label{equ:IME}
-\int \frac{\partial^2 \log f(x;\btheta)}{\partial \btheta\partial \btheta'}f(x;\btheta)dx =\int \left[\frac{\partial \log f(x;\btheta)}{\partial \btheta}\right]\left[\frac{\partial \log f(x;\btheta)}{\partial \btheta}\right]'f(x;\btheta)dx. 
\end{equation}

When $f(x;\btheta)$ is the true data generating mechanism of $X$, the left-side of Equation (\ref{equ:IME}) can be expressed as a $p\times p$ matrix 
\begin{equation*}\label{equ:sensitivity}
\Ezero\left[-\frac{\partial^2 \ell(\btheta;X)}{\partial \btheta \partial \btheta'}\right]  \triangleq \bSast(\btheta),
\end{equation*}
where $\ell(\btheta;x)=\log f(x;\btheta)$ is the log-likelihood function, and $\Ezero$ denotes the expectation with respect to (w.r.t.) the true distribution of $X$. This matrix is referred to as the Fisher information matrix, or the \textit{sensitivity matrix}. The right-side of Equation (\ref{equ:IME}) can be expressed as another $p\times p$ matrix 
\begin{equation*}\label{equ:variability}
\Ezero\left\{\left[\frac{\partial \ell(\btheta;X)}{\partial \btheta}\right]\left[\frac{\partial \ell(\btheta;X)}{\partial \btheta}\right]'\right\}  \triangleq \bVast(\btheta),
\end{equation*}
called the \textit{variability matrix} \citep{varin2011overview}. Equation (\ref{equ:IME}) becomes $\bSast(\btheta)=\bVast(\btheta)$, which is referred to as the second Bartlett identity \citep{bartlett1953Approximatea,bartlett1953Approximateb} or information matrix equivalence \citep{white1982Maximum}. 

Several goodness-of-fit (GoF) tests for detecting  model misspecification were designed through comparing these two information matrices. \cite{white1982Maximum} proposed an information matrix (IM) test based on the elements of $\bVast(\btheta)-\bSast(\btheta)$. \cite{zhou2012Information} proposed an information ratio test by comparing $\bSast(\btheta)^{-1}\bVast(\btheta)$ with a $p$-dimensional identity matrix. \cite{golden2013New} and \cite{golden2016Generalized} extended these two comparisons to a general framework, called generalized IM test, which covers a range of comparison forms. Later, these tests were applied to copula models for multivariate random variables \citep{huang2013Goodnessoffit, zhang2016Goodnessoffit, prokhorov2019Generalized}. 

Copulas have been a popular tool for modeling the dependence structure of multivariate data, such as multivariate time series \citep{chen2006Estimationa, chen2006Estimationb} and multivariate survival times \citep{clayton1978model,hougaard1986class,oakes1989Bivariate,shih1995Inferences}. In this manuscript, we are interested in a class of semiparametric survival copula models for multivariate survival times, denoted by $(T_{1},T_{2},\cdots,T_{d})$. The multivariate survival times can be times to different types of events collected on each subject, such as time to relapse or second cancer and time to cardiovascular disease among breast cancer survivors \citep{li2020Multiple}. Or they are times to the same type of event from different individuals within a cluster, such as the survival times of acute lymphoblastic leukemia patients from 104 institutions \citep{othus2010Gaussian}. 

A survival copula specifies the joint survival function $H(t_1,\cdots,t_d)=Pr(T_{1}>t_1,\cdots,T_{d}>t_d)$ as  
\begin{equation}\label{equ:copula}
H(t_1,\cdots,t_d)= \Cbbm\left(H_1(t_1),\cdots,H_d(t_d);\btheta\right), 
\end{equation}
where $H_r(t)=Pr(T_r>t)$, $r=1,\cdots,d$, are the marginal survival functions of individual survival times, and $\Cbbm(u_1,u_2,\cdots,u_d;\btheta): [0,1]^d \rightarrow [0,1]$ is a copula function with a $p$-dimensional parameter $\btheta$. Copulas were originally proposed for modeling the joint cumulative distribution function (CDF) of multivariate random variables, and its properties have been extensively studied \citep{mikosch2006copulas, nelsen2007introduction}. The way that the survival copula relates the joint survival function to marginal survival functions is completely analogous to the way that the original copula connects the joint CDF to marginal CDFs. Thus,  the survival copula satisfies the properties of the original copula \citep{georges2001multivariate, nelsen2007introduction}. 

Copulas enjoy the flexibility in coupling different marginal distributions with a wide variety of copula families that exhibit different dependence structures. A class of {\it semiparametric} copula models assumes a parametric form for the copula function but leaves the marginal distributions unspecified.  Thus, a crucial element in such a model is the specification of the copula function. Archimedean copula families, such as Clayton, Frank, and Joe, are the most popular choices \citep{nelsen2006Archimedean}. \cite{li2008Semiparametric} and \cite{othus2010Gaussian} considered the Gaussian copula, which belongs to the elliptical families (including Gaussian and $t$ copulas). Different copulas families display different features. For example, in terms of the tail dependence, Clayton has a lower-tail dependence; Joe has an upper-tail dependence; both Gaussian and Frank have no dependence for either lower-tail or upper-tail. Misspecification of the copula function can lead to incorrect estimation of the joint distribution as well as its derivatives, such as conditional distributions. 

The above-mentioned IM-based GoF tests \citep{huang2013Goodnessoffit, zhang2016Goodnessoffit, prokhorov2019Generalized} were proposed for detecting misspecification of the copula function under a semiparametric copula model. They can be regarded as the \textit{blanket} tests introduced in \cite{genest2009Goodnessoffit}: they can be applied to any copula families and do not require selection of smoothing parameters, weight functions, or kernel functions. However, they were designed based on the presumption that the data are fully observed, and may not be applicable for data with missing values. For example, survival times can be missing due to censoring, such as the termination of the follow-up or participants being lost to follow-up. 
  
For censored survival times, several copula specification tests were proposed, but most are limited to Archimedean families by using their unique properties. For example, \cite{shih1998Goodnessoffit} and \cite{emura2010Goodnessoffit} designed their test statistics using the cross-ratio function expressed as a function of the joint survival. The test statistics in \cite{wang2000Model}, \cite{wang2010Goodnessoffit}, and \cite{lakhal-chaieb2010Copula} used the Kendall distribution, expressed in terms of the generator function.   By contrast, \cite{yilmaz2011Likelihood} and \cite{lin2020Diagnostic} proposed tests for {\it any} copula families while imposing assumptions on the form of copulas under the alternative hypothesis. For example, in \cite{yilmaz2011Likelihood}, the null and alternative models are nested, i.e., the null is embedded in the alternative. \cite{lin2020Diagnostic} assumes a particular form for the alternative copulas. In addition, several tests, such as \cite{shih1998Goodnessoffit},  \cite{emura2010Goodnessoffit}, and \cite{andersen2005Class}, require the choice of a weight function or bandwidth, or the partition of the data. According to \cite{genest2009Goodnessoffit}, they are not blanket tests. 

Our goal is to propose a blanket test for multivariate censored survival times, and we adopt the information ratio (IR) test originally proposed in \cite{zhou2012Information}. First, the IR test can be applied to all parametric copula families. Second, it is likelihood-based and depends solely on the parametric form of the null copula. Thus, it does not impose any assumptions on the alternative copulas. Third, it does not require any smoothing parameters, weight functions, kernel functions, or partition of the data. However, the first problem we encountered was whether the information matrix equivalence, the foundation of the IR test, still holds under the likelihood for censored data. No existing work has verified it. Thus,  our first contribution is to prove this equivalence in the presence of censoring. 

The IR test was first proposed under the quasi-likelihood  for cross-sectional or longitudinal data \citep{zhou2012Information}. Later, this test was extended to various models for univariate and multivariate time series \citep{zhang2012Information, zhang2016Goodnessoffit, zhang2019GoodnessofFit, zhang2021Goodnessoffit}. The asymptotic properties of the IR statistic have been investigated for the above settings where data are fully observed, but not for censored data yet. Another contribution of our manuscript is to derive its asymptotic properties  when the marginal distributions and the copula parameters are estimated in the presence of censoring. 

In this paper, we will show that if the copula function is correctly specified, the IR statistic is asymptotically distributed as a normal random variable. However, the expression of its asymptotic variance is complicated, so it is difficult to use an analytic variance estimate to calculate $P$-values. Thus, we propose a bootstrap procedure to approximate the statistic's null distribution via generating replications of multivariate censored data from the null copula. 


\cite{zhang2016Goodnessoffit} established the asymptotic equivalence between the IR statistic and an in-and-out-of-sample pseudo (PIOS) likelihood ratio test statistic. The PIOS statistic is based on the comparison between two types of pseudo likelihood: the \textit{in-sample} likelihood, which is the full likelihood, and the \textit{out-of-sample} likelihood, which is a ``leave-one-out" cross-validated likelihood. In this manuscript, we will prove the asymptotic equivalence between these two test statistics with censored data. We created an R package called {\tt IRtests} that implements both tests for copula specification with bivariate censored data, and it is available at \url{https://github.com/michellezhou2009/IRtests}.

The remainder of the manuscript is organized as follows. In Section 2, we prove the information matrix equivalence under a semiparametric copula model for censored survival times.  We define the IR statistic in Section 3 and discuss its asymptotic properties.  In Section 4, we describe how to calculate $P$-values via bootstrap resampling  and how to use the $P$-values to select the best copula family. Section 5 presents the simulation studies for investigating the finite-sample performance of the proposed IR test and comparing it with the other two forms of generalized IM tests. In Section 6, we apply the IR test to a data example. Concluding remarks are given in Section 7.

\section{Information Matrix Equivalence in the Presence of Censoring} \label{sec:IMT}

For ease of illustration, we present the proposed methods in the context of bivariate event times $(T_{1},T_{2})$. We denote the true marginal survival functions by $H_r^\szero(t)$ with the probability density function $f_r^\szero(t)=- d H_r^\szero(t)/dt$. We assume a copula model $\Cbbm(u_1,u_2;\btheta)$ in Equation (\ref{equ:copula}) for the joint survival function of $(T_1,T_2)$ with $\btheta \in \Theta \subset R^p$. To differentiate from this {\it assumed} copula, we denote the {\it true} copula by $\Cbbm^\szero(u_1,u_2)$. According to \cite{sklar1959Fonctionsa}, for a continuous random vector $(T_1,T_2)$, there exists a unique copula function $\Cbbm^\szero(u_1,u_2)$ such that $Pr(T_1>t_1,T_2>t_2)=\Cbbm^\szero\left(H_1^\szero(t_1),H_2^\szero(t_2)\right)$ for all $(t_1,t_2)$.

\begin{definition}\label{defn:correct-spec}
The assumed copula $\Cbbm(u_1,u_2;\btheta)$ is said to be {\bf correctly specified}, denoted as $ \Cbbm^\szero (u_1,u_2)\in \mathcal C_{\theta} = \{\Cbbm(u_1,u_2;\btheta), \btheta \in \Theta\}$, if there exists $\btheta_0\in \Theta$ such that $\Cbbm\left(u_1,u_2;\btheta_0\right)=\Cbbm^\szero\left(u_1,u_2\right)$ for all $(u_1,u_2)\in (0,1)^2$. The value $\btheta_0$ is called the true value of the copula parameter. On the other hand, if for any $\btheta \in \Theta$, there exists some $(u_1,u_2)\in (0,1)^2$ such that $\Cbbm\left(u_1,u_2;\btheta\right)\neq \Cbbm^\szero\left(u_1,u_2\right)$, we say that the assumed copula $\Cbbm(u_1,u_2;\btheta)$ is {\bf misspecified}. 
\end{definition}

In the remainder of the manuscript, we let $g_{\btheta}$ and $g_{\btheta\btheta}$ denote the first-order and second-order partial derivatives of a function $g$ w.r.t. $\btheta$.

\subsection{Likelihood Function}\label{sec:likelihood}

As mentioned earlier, survival times $(T_1,T_2)$ are often subject to censoring. Let $(C_1,C_2)$ denote the bivariate censoring times. We assume independent censoring, i.e., $(C_{1},C_{2})$ are independent of $(T_1,T_2)$. The observed variables include \begin{equation}\label{equ:survival-outcome}
X_{r}=\min\{T_{r},C_{r}\},\, \text{and}\, \delta_{r}=I(T_{r}\leq C_{r}),\, r=1,2, 
\end{equation}
where $I(\cdot)$ is the identity function. Note that in some situations both event times are subject to a common censoring time, i.e., $C_1=C_2$. 

Under a semiparametric copula model, the parameters consist of the unspecified marginal survival functions and the copula parameter. Since our focus is the specification of the copula function, we regard $\btheta$ as the parameter of interest and marginal survival functions as nuisance parameters.  {\it For now, let us assume that the true marginal survival functions $H_r^\szero$, $r=1,2$, are known.} Thus, given $(X_1,X_2,\delta_1,\delta_2)$, the log-likelihood under the assumed copula is a function of the copula parameter $\btheta$. It can be written as the sum of two components: $\ell(\btheta)= \mathfrak C(\btheta, U_{1}^\szero, U_{2}^\szero, \delta_{1},\delta_{2}) + \mathfrak F(X_{1}, X_{2}, \delta_{1},\delta_{2})$, where $\mathfrak C$ is a function of the assumed copula on $(U_1^\szero,U_2^\szero)$ with $U_r^\szero=H_r^\szero(X_r)$, $r=1,2$. The second term $\mathfrak F=\delta_1 \log f_1^\szero(X_1) + \delta_2 \log f_2^\szero(X_2)$ is a function of marginal densities only, which can be regarded as a constant. Thus, the log-likelihood can be defined as 
\begin{align}
\nonumber & \ell(\btheta;U_1^\szero,U_2^\szero,\delta_1,\delta_2) =\mathfrak C(\btheta, U_{1}^\szero, U_{2}^\szero, \delta_{1},\delta_{2})  \\
\nonumber & =   \delta_{1}\delta_{2}\log\cbbm(U_1^\szero,U_2^\szero;\btheta) + \delta_{1}(1-\delta_{2})\log \cbbm_1\left(U_1^\szero,U_2^\szero;\btheta\right) \\
\label{equ:log-likelihood} &  \quad \quad +(1-\delta_{1})\delta_{2}\log \cbbm_2\left(U_1^\szero,U_2^\szero;\btheta\right) + (1-\delta_{1})(1-\delta_{2})\log \Cbbm\left(U_1^\szero,U_2^\szero;\btheta\right)
\end{align}
with $\cbbm_r(u_1,u_2;\btheta) = \frac{\partial \Cbbm(u_1,u_2;\btheta)}{\partial u_r}$ for $r=1,2$ and $\cbbm(u_1,u_2;\btheta)=\frac{\partial^2 \Cbbm(u_1,u_2;\btheta)}{\partial u_1\partial u_2}$. 
\smallskip

\begin{remark}\label{remark:log-likelihood}
The survival copulas $\Cbbm$ and $\Cbbm^\szero$ can be regarded as the {\it assumed} and {\it true} joint CDF for $(Y_1,Y_2)$ with $Y_r=H_r^\szero(T_r)$, $r=1,2$,  which are uniformly distributed on $(0,1)$. The above log-likelihood function is also the log-likelihood function for data $(U_1^\szero, U_2^\szero, \delta_1, \delta_2)$, where $U_r^\szero$ and $\delta_r$ can be expressed as $U_r^\szero = \max\{Y_r,H_r^\szero(C_r)\}$ and $\delta_r = I\left(Y_r\geq H_r^\szero (C_r)\right)$.
\end{remark}

\begin{remark}\label{remark:full-observed}
If the bivariate event times $(T_1,T_2)$ are fully observed, i.e., $\delta_1=\delta_2\equiv 1$, 
\begin{equation}\label{equ:log-likelihood-observed}
\mathfrak  \ell(\btheta;U_1^\szero,U_2^\szero,\delta_1,\delta_2) =\log \cbbm(U_1^\szero,U_2^\szero;\btheta).
\end{equation}
\end{remark}
\smallskip

\begin{definition}\label{def:pseudo-true}
Given the log-likelihood function in Equation (\ref{equ:log-likelihood}), 
we define
\begin{equation}\label{equ:pseudo-true}
\btheta^* = \arg\max_{\btheta \in \Theta} \Ezero\left[\ell(\btheta;U_1^\szero,U_2^\szero,\delta_1,\delta_2)\right], 
\end{equation}
as  the pseudo-true value of the parameter $\btheta$, where $\Ezero$ takes the expectation w.r.t. the true distributions of $(T_1,T_2)$ and $(C_1,C_2)$.  
\end{definition}

If the assumed copula is correctly specified, the pseudo-true value $\btheta^*=\btheta_0$ \citep{shih1995Inferences,chen2010Estimation}; if the assumed copula is misspecified, $\btheta^*$ might not be equal to $\btheta_0$.

\subsection{Information Matrix Equivalence}\label{sec:IME}

Given the above log-likelihood function in Equation (\ref{equ:log-likelihood}), the sensitivity and variability information matrices, both $p\times p$ dimensional, are defined as 
\begin{equation}\label{equ:IM-censor}
 \bSast(\btheta) =\Ezero\left[-\ell_{\btheta\btheta}\right]\, \text{and}\, \bVast(\btheta) =\Ezero\left[\ell_{\btheta}\ell_{\btheta}'\right].
\end{equation}
In the supplementary material, we provide expressions of $\ell_{\btheta}$ and $\ell_{\btheta\btheta}$ for four copula families: Clayton, Frank, Joe, and Gaussian.

The proof of the information matrix equivalence in Theorem \ref{thm:equivalence} requires the following regularity conditions R1 - R6. First, we introduce all the required notation. Let $\|x\|$ denote the usual Euclidean metric of any $p$-dimensional vector $x=(x_1,\cdots,x_p)$, i.e., $\|x\|=\sqrt{x_1^2+\cdots+x_p^2}$. For a $p\times p$ matrix $A$, define $\|A\| = \sqrt{\sum_{j,k=1}^p a_{jk}^2}$, where $a_{jk}$ is the $(j,k)$-th element of $A$. For simplicity, in the remaining of the manuscript, we suppress $\delta_1$ and $\delta_2$ from the log-likelihood function $\ell(\btheta,u_1,u_2,\delta_1,\delta_2)$ as well as its partial derivatives defined as follows. For $j,k=1,\cdots,p$, let $\ell_{\theta_j}=\partial \ell/ \partial \theta_j$ denote the $j$-th element of the $p\times 1$ vector $\ell_{\btheta}$, and let $\ell_{\theta_j\theta_k}(\btheta, u_1, u_2)=\partial^2 \ell / \partial \theta_j\partial\theta_k $ denote the $(j,k)$-th element of the $p\times p$ matrix $\ell_{\btheta\btheta}(\btheta, u_1, u_2)$. Define $\ell_{\theta_j,\btheta}=\partial \ell_{\theta_j}/\partial \btheta$ and $\ell_{\theta_j\theta_k,\btheta}=\partial \ell_{\theta_j\theta_k} / \partial \btheta$, both $p\times 1$ vectors. For $r=1,2$, let $\ell_{\theta_j,u_r}=\partial  \ell_{\theta_j}/ \partial u_r$ and $\ell_{\theta_j\theta_k,u_r}=\partial \ell_{\theta_j\theta_k} / \partial u_r$. Let $\ell_{\btheta,u_r}$ denote a $p\times 1$ vector with the $j$-th element $\ell_{\theta_j,u_r}$.  Let $\ell_{\btheta\btheta, u_r}$ denote a $p\times p$ matrix with the $(j,k)$-th element $\ell_{\theta_j\theta_k,u_r}$. 
\medskip

Our regularity conditions are:
\begin{itemize}
\item [{\bf R1}] 
\begin{itemize}
\item [(i)] $\{(T_{i1},T_{i2}),i=1,\cdots,n\}$ is an independent and identically distributed (i.i.d.) sample from an unknown joint survival function $\Cbbm^\szero(H_1^\szero(t_1),H_2^\szero(t_2))$ with continuous marginal survival functions $H_r^\szero(\cdot)$, $r=1,2$;
\item [(ii)] $\{(C_{i1},C_{i2}),i=1,\cdots,n\}$ is an i.i.d. sample with joint survival function $\mathcal G^\szero(t_1,t_2)=Pr(C_{i1}>t_1,C_{i2}>t_2)$ and marginal survival functions $G_r^{\szero}(t)=Pr(C_{ir}>t)$, $r=1,2$;
\item [(iii)] The censoring variables $(C_{i1},C_{i2})$ are independent of $(T_{i1},T_{i2})$ and there is no mass concentration at 0 in the sense that $G_r^\szero(\eta)\rightarrow 1$ as $\eta\rightarrow 0$.
\end{itemize}
\item [{\bf R2}] Let $\Theta$ be a compact space of $\mathcal R^p$. For every $\epsilon>0$,  
$$
\displaystyle \liminf_{\btheta \in \Theta: \|\btheta - \btheta^*\|\geq \epsilon} \Ezero[\ell(\btheta^*,U_1^\szero,U_2^\szero)] - \Ezero[\ell(\btheta,U_1^\szero,U_2^\szero)] > 0.
$$
\item [{\bf R3}] The true (unknown) copula function $\Cbbm^\szero(u_1,u_2)$ has continuous partial derivatives.

\item [{\bf R4}] For any $(u_1,u_2)$, $\ell(\btheta,u_1,u_2)$ is a continuous function of $\btheta \in \Theta$.

\item [{\bf R5}] Functions $\ell_{\theta_j}(\btheta,u_1,u_2)$, $\ell_{\theta_j,\theta_k}(\btheta,u_1,u_2)$, and $\ell_{\theta_j,u_r}(\btheta,u_1,u_2)$, $j,k=1,\cdots,p$, $r=1,2$, are well defined and continuous in $(u_1,u_2,\btheta) \in [0,1]^2\times \Theta$. 

\item [{\bf R6}] 
\begin{itemize}
\item [(i)] $\|\ell_{\btheta}(\btheta,u_1,u_2)\|\leq q\{u_1(1-u_1)\}^{-a_{1}}\{u_2(1-u_2)\}^{-a_{2}}$ for some $q>0$ and $a_1,a_2\geq 0$ such that $\Ezero[\{U^\szero_1(1-U^\szero_1)\}^{-2a_1}\{U_2^\szero(1-U_2^\szero)\}^{-2a_2}]<\infty$;
\item [(ii)] $\|\ell_{\btheta,u_r}(\btheta,u_1,u_2)\|\leq q\{u_s(1-u_s)\}^{-a_{s}}\{u_r(1-u_r)\}^{-a_{r}}$ for some $q$, $a_s$, $a_r$, and $s\neq r$ such that $\Ezero[\{U^\szero_s(1-U^\szero_s)\}^{\epsilon_s - a_s}\{U_r^\szero(1-U_r^\szero)\}^{-a_r}]<\infty$ for some $\epsilon_s \in (0,1/2)$;
\item [(iii)] $\|\ell_{\btheta\btheta}(\btheta,u_1,u_2)\|\leq q\{u_1(1-u_1)\}^{-a_1}\{u_2(1-u_2)\}^{-a_2}$ for some $q>0$ and $a_1,a_2\geq 0$ such that $\Ezero[\{U^\szero_1(1-U^\szero_1)\}^{-2a_1}\{U_2^\szero(1-U_2^\szero)\}^{-2a_2}]<\infty$.
\end{itemize}
\end{itemize}

These regularity conditions are similar to those listed in \cite{shih1995Inferences} and  \cite{chen2010Estimation}, which also focused on semiparametric survival copula models for censored survival data. \cite{shih1995Inferences} referred them to as the standard regularity conditions for maximum likelihood estimation \citep{white1982Maximum} in the copula context. It is worth pointing out that our assumptions of homogenous censoring distribution (i.e., $(C_{i1}, C_{i2})$ follows the same joint distribution across subjects) was also imposed in \cite{shih1995Inferences}. However, \cite{chen2010Estimation} allowed different censoring distributions across subjects. We defer more discussions on this assumption to Section \ref{sec:concluding}.  In addition, these regularity conditions except for those related to censoring were used in \cite{huang2013Goodnessoffit} to prove the information matrix equivalence under the log-likelihood function in Equation (\ref{equ:log-likelihood-observed}) for fully observed bivariate event times.

\smallskip

\begin{theorem}[Information Matrix Equivalence]\label{thm:equivalence} 
Assume that conditions R1 - R6  hold. If the assumed copula is correctly specified, under the log-likelihood function in Equation (\ref{equ:log-likelihood}), $\bSast(\bthetaast)=\bVast(\bthetaast)
$, where $\bSast(\btheta)$ and $\bVast(\btheta)$ are the sensitivity and variability matrices defined in Equation (\ref{equ:IM-censor}), and $\bthetaast$ is the pseudo-true value of the parameter $\btheta$ defined in Equation (\ref{equ:pseudo-true}).
\end{theorem}

We prove this theorem in  Appendix \ref{app:Thm1}, where we will show that $\bSast(\btheta) = \bVast(\btheta) - \bA(\btheta)$ for any assumed copula, where $\bA(\btheta)=\Ezero_{(C_1,C_2)}\left\{\Ezero_{(T_1,T_2)}\left[\Delta|C_1,C_2\right]\right\}$ with
\begin{align*}
\Ezero_{(T_1,T_2)}\left[\Delta|C_1,C_2\right]   &  = \sum_{\delta_1,\delta_2=0,1}\int\int_{ \Omega_{\delta_1,\delta_2}} \cbbm_{\btheta\btheta}(u_1,u_2;\btheta) w_{\delta_1,\delta_2}(u_1,u_2;\btheta)du_1du_2.
\end{align*}
Here, for each censoring status $(\delta_1,\delta_2)$,  $\Omega_{\delta_1,\delta_2}$ is the corresponding region for $\left(Y_1,Y_2\right)$ (Remark \ref{remark:log-likelihood}) given $(C_1,C_2)$ with $\displaystyle \bigcup_{\delta_1,\delta_2=0,1} \Omega_{\delta_1,\delta_2} = [0,1]^2$. In addition, $w_{\delta_1,\delta_2}$ is a ratio of the {\it true} copula function versus the {\it assumed} copula or a ratio of their partial derivatives.

If the assumed copula is correctly specified, $w_{\delta_1,\delta_2}(u_1,u_2;\btheta^*)\equiv 1$ for all $(u_1,u_2)$ and $(\delta_1,\delta_2)$.  It leads to 
\begin{equation}\label{equ:interchange}
\Ezero_{(T_1,T_2)}\left[\Delta|C_1,C_2\right]   =  \int_0^1 \int_0^1 \cbbm_{\btheta\btheta}(u_1,u_2;\btheta^*) du_1du_2 = 0,
\end{equation}
because of the interchangeability between integrals and derivatives ensured by the regularity conditions. Equation (\ref{equ:interchange}) makes $\bA(\btheta^*)=0$, resulting in the information matrix equivalence stated in Theorem \ref{thm:equivalence}. On the other hand, if the assumed copula is misspecified, $w_{\delta_1,\delta_2}(u_1,u_2;\btheta^*)\neq 1$ for some $(u_1,u_2)$ and $(\delta_1,\delta_2)$, and thus, Equation (\ref{equ:interchange}) will not hold, indicating $\bSast(\bthetaast)\neq \bVast(\bthetaast)$.
\medskip 

We define the following \defn{information matrix ratio}: $\bR^*(\btheta^*)=\bSast(\btheta^*)^{-1}\bVast(\btheta^*)$. If the assumed copula is correctly specified, $\bR^*(\btheta^*)=I_p$, a $p$-dimensional identity matrix; otherwise, $\bR^*(\btheta^*)\neq I_p$. The discrepancy between $\bR^*(\btheta^*)$ and $I_p$ implies copula misspecification, and it can be quantified by a scalar metric: $tr[\bR^*(\btheta^*)]-p$, where $tr(\cdot)$ denotes the trace of a matrix. This is the motivation for the IR test we will propose for detecting copula misspecification.

\section{Information Ratio Statistic under Semiparametric Copula Models}

In this section, we propose an IR statistic, which is a consistent estimator of $tr[\bR^*(\btheta^*)]$ under the semiparametric copula model.

\subsection{IR Statistic}

As described earlier, a semiparametric copula model assumes that the {\it true} marginal survival functions $H_1^\szero(t)$ and $H_2^\szero(t)$ are unspecified. Thus, obtaining a consistent estimator  of $tr[\bR^*(\btheta^*)]$ requires the consistent estimation for the marginal survival functions, the copula parameter, and two information matrices. Let $\{(X_{i1},X_{i2}, \allowbreak \, \delta_{i1},\delta_{i2}),i=1,\cdots,n\}$ be $n$ independent realizations of $(X_1,X_2,\delta_1,\delta_2)$. 
\smallskip

\noindent{\bf Consistent estimation of marginal survival functions and copula parameter.} \cite{shih1995Inferences} proposed the following two-step procedure: at the first step, each marginal survival function is estimated by a nonparametric estimator $\Hhat_r(t)$ with data $\{(X_{ir},\delta_{ir}),i=1,\cdots,n\}$, $r=1,2$. Under the assumption that the censoring times are independent of the event times, we consider the Kaplan-Meier estimator, which is a consistent estimator for the marginal survival function \citep{kaplan1958Nonparametric}. Thus, $\Uhat_{ir}=\Hhat_r(X_{ir})$ is a consistent estimate of $U_{ir}^\szero=H_r^\szero(X_{ir})$, $r=1,2$, $i=1,\cdots,n$.

At the second step, the copula parameter $\btheta$ is estimated by a pseudo maximum likelihood estimator (PMLE), which maximizes the {\it psuedo log-likelihood function} given as $\ell_n(\btheta) = \sum_{i=1}^n  \ell(\btheta,\Uhat_{i1}, \Uhat_{i2})$. Specifically, the PMLE of $\btheta$ is given as 
\begin{equation}\label{equ:PMLE-theta}
\bthetahatn = \arg\max_{\btheta}\ell_n(\btheta). 
\end{equation}
\cite{chen2010Estimation} established the asymptotic properties of $\bthetahatn$. One of them is that, under certain conditions, the PMLE $\bthetahatn$  converges in probability to the pseudo-true value $\bthetaast$ defined in Equation (\ref{equ:pseudo-true}) as $n\rightarrow \infty$, {\it regardless of whether the assumed copula function is correctly specified or not}. Thus, $\bthetahatn$ is a consistent estimate of $\btheta^*$. \\
\medskip

\noindent{\bf Consistent estimation of information matrices.} By Equation (\ref{equ:IM-censor}), for a given value $\btheta$, $\bSast(\btheta)$ and $\bVast(\btheta)$ are the distributional means. If the true marginal survival functions are known, these two information matrices can be consistently estimated by the empirical means: 
$$
- n^{-1}\sum_{i=1}^n \ell_{\btheta\btheta}(\btheta; U_{i1}^\szero, U_{i2}^\szero)\, \text{and}\, n^{-1}\sum_{i=1}^n \ell_{\btheta}(\btheta; U_{i1}^\szero, U_{i2}^\szero)\ell_{\btheta}(\btheta; U_{i1}^\szero, U_{i2}^\szero)'.
$$ 
\cite{chen2010Estimation} provided the following consistent estimator for $\bSast(\bthetaast)$: 
$$
\widehat {\bS}_n(\bthetahatn) = -\frac{1}{n} \sum_{i=1}^n \ell_{\btheta\btheta}(\bthetahatn; \Uhat_{i1},\Uhat_{i2}),
$$ 
where $U_{ir}^\szero$ is estimated by $\Uhat_{ir}$, and $\bthetaast$ is estimated by the PMLE $\bthetahat_n$. Following the same idea, we propose the following consistent estimator for $\bVast(\bthetaast)$:
$$
\widehat {\bV}_n(\bthetahatn) = \frac{1}{n} \sum_{i=1}^n \ell_{\btheta}(\bthetahatn; \Uhat_{i1},\Uhat_{i2})\ell_{\btheta}(\bthetahatn; \Uhat_{i1},\Uhat_{i2})'.
$$
With the above estimators, the IR statistic is defined as
$$
R_n = tr\left[\widehat {\bS}_n(\bthetahatn)^{-1}\widehat {\bV}_n(\bthetahatn)\right].
$$
Next, we will present two key asymptotic properties of this IR statistic. First, Theorem \ref{thm:convergence} establishes the consistency of $R_n$, i.e., it converges in probability to $tr[\bR^*(\btheta^*)]$ for {\it any} assumed copula. Second, Theorem \ref{thm:normal} states the asymptotic normality of $R_n$ if the assumed copula is correctly specified. This result will be used for designing the IR test and copula selection in Section \ref{sec:test-selection}.

\subsection{Asymptotic Properties of IR statistic}

The consistency and asymptotic normality of the IR statistic $R_n$ requires the regularity conditions R1 - R6 listed in Section \ref{sec:IME} and the following additional conditions: 
\begin{itemize}
\item[{\bf C1}]  
\begin{itemize}
\item [(i)] Let $L=\sup_{\btheta \in \Theta}|\ell(\btheta,U_1^\szero,U_2^\szero)|$ and $L_{\btheta}=\sup_{\btheta \in \Theta}|\ell_{\btheta}(\btheta,U_1^\szero,U_2^\szero)|$. Then 
$$
\lim_{K\rightarrow \infty} \Ezero\left[L\ I(L\geq K) + L_{\btheta}I(L_{\btheta}\geq K)\right] = 0;
$$
\item [(ii)] For any $\eta>0$ and any $\epsilon>0$, there is $K>0$ such that $|\ell(\btheta,u_1,u_2)|\leq K|\ell(\btheta,u_1',u_2')|$ for all $\btheta \in \Theta$ and all $u_r\in [\eta,1)$ such that $1-u_r\geq \epsilon(1-u_r')$, $r=1,2$.
\end{itemize}
\item [{\bf C2}] For $r=1,2$, if $T_r$ is subject to non-trivial censoring (i.e., $C_r\neq \infty$), then the Kaplan-Meier estimator $\Hhat_r$ is truncated at the tail in the sense that for some $\tau_r$, $\Hhat_r(t) = \Hhat_r(\tau_r)$ for all $t\geq \tau_r$ and $G_r^\szero(\tau_r)H_r^\szero(\tau_r)>0$.
\item [{\bf C3}]
\begin{itemize}
\item [(i)] Regularity condition R2 holds with $\btheta^* \in int(\Theta^*)$, where $\Theta^*$ is a compact subset of $\Theta$;
\item [(ii)] $\Ezero\left[-\ell_{\btheta\btheta}(\btheta^*,U_1^\szero, U_2^\szero)\right]$ has all its eigenvalues bounded below and above by some finite positive constants;
\item [(iii)] $Var^\szero\left[\ell_{\btheta}(\btheta^*,U_1^\szero,U_2^\szero) + W_1(\btheta^*, X_1,\delta_1) + W_2(\btheta^*,X_2,\delta_2)\right]$ has all its eigenvalues bounded below
and above by some finite positive constants, where for $r=1,2$, $W_r(\btheta^*, X_r,\delta_r) $ is defined in Equation (\ref{equ:Wr}) of Appendix \ref{app:Thm3};
\item [(iv)] $\ell_{\btheta}(\btheta^*,U_1^\szero,U_2^\szero) + W_1(\btheta^*, X_1,\delta_1) + W_2(\btheta^*,X_2,\delta_2)$ satisfies Lindeberg condition. 
\end{itemize}

\item [{\bf C4}] 
\begin{itemize}
\item [(i)] Let $L_{\btheta,u_r}=\sup_{\btheta\in\Theta}\|\ell_{\btheta,u_r}(\btheta,U_1^\szero,U_2^\szero)\|$ and $L_{\btheta\btheta}=\sup_{\btheta\in\Theta}\|\ell_{\btheta\btheta}(\btheta,U_1^\szero,U_2^\szero)\|$. Then 
$$
\lim_{K\rightarrow \infty} \Ezero\left[L_{\btheta,u_r}I(L_{\btheta,u_r}\geq K)+L_{\btheta\btheta}I(L_{\btheta\btheta}\geq K)\right] = 0;
$$
\item [(ii)] Let $Q(\btheta,u_1,u_2)=\|\ell_{\btheta}(\btheta,u_1,u_2)\|+\|\ell_{\btheta\btheta}(\btheta,u_1,u_2)\|$. For any $\eta>0$ and any $\epsilon>0$, there is $K>0$ such that $Q(\btheta,u_1,u_2)\leq KQ(\btheta,u_1',u_2')$ for all $\btheta \in \Theta$ and all $u_r\in [\eta,1)$ such that $1-u_r\geq \epsilon(1-u_r')$, $r=1,2$.
\end{itemize}

\item [{\bf C5}] 
\begin{itemize}
\item [(i)] For $j,k=1,\cdots,p$, $r=1,2$, let \begin{align*}
LL_{\theta_j\theta_k,u_r} & =\sup_{\btheta \in \Theta}|\ell_{\theta_j,u_r}(\btheta,U_1,U_2)\ell_{\theta_k}(\btheta,U_1,U_2) + \ell_{\theta_k,u_r}(\btheta,U_1,U_2)  \ell_{\theta_j}(\btheta,U_1,U_2)|,\\
LL_{\theta_j\theta_k,\btheta} & =\sup_{\btheta \in \Theta}\|\ell_{\theta_j,\btheta}(\btheta,U_1,U_2)\ell_{\theta_k}(\btheta,U_1,U_2)+\ell_{\theta_k,\btheta}(\btheta,U_1,U_2)\, \allowbreak\ell_{\theta_j}(\btheta,U_1,U_2)\|. 
\end{align*}
Then, 
$$
\lim_{K\rightarrow \infty} \sup_{(j,k)}\Ezero [LL_{\theta_j\theta_k,u_r}I(LL_{\theta_j\theta_k,u_r}\geq K)+ LL_{\theta_j\theta_k,\btheta} I(LL_{\theta_j\theta_k,\btheta} \geq K)]= 0;
$$ 
\item [(ii)] For $j,k=1,\cdots,p$, let 
\begin{align*}
Q_{1,\theta_j\theta_k}(\btheta,u_1,u_2) & =  |\ell_{\theta_j}(\btheta,u_1,u_2)\ell_{\theta_k}(\btheta,u_1,u_2)|  \\
& +  \| \ell_{\theta_j,\btheta}(\btheta,u_1,u_2)  
\ell_{\theta_k}(\btheta,u_1,u_2)+\ell_{\theta_k,\btheta}(\btheta,u_1,u_2)\ell_{\theta_j}(\btheta,u_1,u_2)\|. 
\end{align*}
For any $\eta>0$ and any $\epsilon>0$, there is $K>0$, such that  $ Q_{1,\theta_j\theta_k}(\btheta,u_1,u_2)\leq K Q_{1,\theta_j\theta_k}(\btheta,u'_1,u'_2)$ for all $\btheta \in \Theta$ and all $u_r\in [\eta,1)$ such that $1-u_r\geq \epsilon(1-u'_r)$, $r=1,2$.
\end{itemize}

\item [{\bf C6}] 
\begin{itemize}
\item[(i)] Functions  $\ell_{\theta_j\theta_k, \btheta}(\btheta,u_1,u_2)$ and $\ell_{\theta_j\theta_k, u_r}(\btheta,u_1,u_2)$, $j,k=1,\cdots,p$, $r=1,2$, are well-defined and continuous in $(\btheta,u_1,u_2)\in  \Theta\times (0,1)^2$;
\item [(ii)] For $j,k=1,\cdots,p$, $r=1,2$, let $L_{\theta_j\theta_k,u_r} =\sup_{\btheta \in \Theta}|\ell_{\theta_j\theta_k,u_r}(\btheta,U_1^\szero,U_2^\szero)|$ and $L_{\theta_j\theta_k,\btheta} =\sup_{\btheta \in \Theta}\|\ell_{\theta_j\theta_k,\btheta}(\btheta,U_1^\szero,\, \allowbreak U_2^\szero)\|$. Then, 
$$
\lim_{K\rightarrow \infty} \sup_{(j,k)}\Ezero\left[L_{\theta_j\theta_k,u_r}I(L_{\theta_j\theta_k,u_r}\geq K)+L_{\theta_j\theta_k,\btheta}I(L_{\theta_j\theta_k,\btheta}\geq K)\right] = 0;
$$ 
\item [(iii)] For $j,k=1,\cdots,p$, let $Q_{2,\theta_j\theta_k}(\btheta,u_1,u_2) = |\ell_{\theta_j\theta_k}(\btheta,u_1,u_2)|+\| \ell_{\theta_j\theta_k,\btheta}(\btheta,u_1,u_2)\|$. For any $\eta>0$ and any $\epsilon>0$, there is $K>0$, such that 
$Q_{2,\theta_j\theta_k}(\btheta,u_1,u_2) \leq K Q_{2,\theta_j\theta_k}(\btheta,u'_1,u'_2) $ for all $\btheta \in \Theta$ and all $u_r\in [\eta,1)$ such that $1-u_r\geq \epsilon(1-u'_r)$, $r=1,2$.
\end{itemize}
\item [{\bf C7}] 
\begin{itemize}
\item [(i)] $\|\ell_{\btheta\btheta,u_r}(\btheta^\ast,u_1,u_2)\|\leq q  \{u_s(1-u_s)\}^{-a_s}\{u_r(1-u_r)\}^{-a_r}$ for some $q$, $a_s$, $a_r$, and $s\neq r$ such that $\Ezero[\{U_s^\szero(1-U_{s}^\szero)\}^{\epsilon_s- a_s}\{U_{r}^\szero(1-U_{r}^\szero)\}^{-a_r}]<\infty$ for some $\epsilon_s \in (0,1/2)$;
\item [(ii)] $\|\ell_{\btheta}(\btheta^{\ast},u_1,u_2)\ell_{\btheta}(\btheta^{\ast},u_1,u_2)'\| \leq q \{u_1(1-u_1)\}^{-a_1} \{u_2(1-u_2)\}^{-a_2}$ for some $q>0$ and $a_1, a_2 \geq 0$ such that $\Ezero[\{U_1^\szero(1-U_1^\szero)\}^{-2a_1}\{U_2^\szero(1-U_2^\szero)\}^{-2a_2}]$ $<\infty$;
\item [(iii)] $\|\ell_{\btheta,u_r}(\btheta^\ast,u_1,u_2)\ell_{\btheta}(\btheta^\ast,u_1,u_2)'\|\leq q  \{u_s(1-u_s)\}^{-a_s}\{u_r(1-u_r)\}^{-a_r}$ for some $q$, $a_s$, $a_r$, and $s\neq r$ such that $\Ezero[\{U_s^\szero(1-U_{s}^\szero)\}^{\epsilon_s- a_s}\{U_{r}^\szero(1-U_{r}^\szero)\}^{-a_r}]<\infty$ for some $\epsilon_s \in (0,1/2)$.
\end{itemize}
\end{itemize}

Our regularity conditions R1 - R6 combined with the above conditions C1  - C4 are the conditions C1 - C5 and A1 - A4 of \cite{chen2010Estimation} for the existence, consistency, and asymptotic normality of the PMLE $\bthetahat_n$. 
\begin{theorem}\label{thm:convergence}
Under conditions R1 - R6 and C1 - C5, we have $R_n\rightarrow tr\left[\bR^*(\btheta^*)\right]$ in probability as $n\rightarrow \infty$.
\end{theorem}

The proof of this theorem (Appendix \ref{app:Thm2}) requires the consistency of  $\bShatn(\bthetahatn)$ and $\bVhatn(\bthetahatn)$. \cite{chen2010Estimation} has proved the consistency of $\bShatn(\bthetahatn)$, which requires their condition A4.  We follow their arguments to prove the consistency of $\bVhatn(\bthetahatn)$, where our conditions C5 is analogous to Chen et al.'s condition A4. 

\begin{theorem}\label{thm:normal}
Assume conditions R1 - R6 and C1 - C7 hold. Define the null hypothesis $H_0$: $\Cbbm^\szero (u_1,u_2)\in \mathcal C_{\theta} = \{\Cbbm(u_1,u_2;\btheta), \btheta \in \Theta\}$, i.e., the assumed copula is correctly specified. If the null hypothesis $H_0$ is true, $R_n$ converges to $p$ in probability, and $\sqrt{n}(R_n-p)$ converges in distribution to a normal random variable with mean 0 and variance $\sigma_R^2=Var[h_R(X_{i1},X_{i2},\delta_{i1},\delta_{i2},\btheta)]$, where $h_R(X_{i1},X_{i2},\delta_{i1},\delta_{i2},\btheta)$ is given by Equation (\ref{equ:hR}) in Appendix \ref{app:Thm3}.
\end{theorem}

The proof of this theorem (Appendix \ref{app:Thm3}) utilizes the Taylor expansion of $\bShatn(\bthetahatn)$ and $\bVhatn(\bthetahatn)$. One step requires the consistency for the first-order derivative of $\bShatn(\bthetahatn)$ and of $\bVhatn(\bthetahatn)$ w.r.t. $\btheta$. Again, we follow the arguments of \cite{chen2010Estimation} for proving the consistency of  $\bShatn(\bthetahatn)$, where our condition C6 (i) is analogous to Chen et al.'s Condition A2, and our conditions C6 (ii) \& (iii) together are analogous to Chen et al.'s condition A4. Another component in the proof involves the expansion of the estimated pseudo-observations $\Uhat_{ir} - U_{ir}^\szero$ using the asymptotic properties of the Kaplan-Meier estimator for the marginal survival functions. Our condition R6 (iii) is analogous to Chen et al.'s condition A3 (i), and so is our condition C7 (ii). Our condition C7 (i) is analogous to Chen et al.'s condition A3 (ii), and so is our condition C7 (iii). We want to point out that the expression of $h_R(X_{i1},X_{i2},\delta_{i1},\delta_{i2},\btheta)$ is different from the expansion with fully observed data derived in \cite{zhang2016Goodnessoffit, zhang2021Goodnessoffit}. In their settings, the marginal distributions are estimated by the empirical distribution functions whose expansions are different from those of Kaplan-Meier estimators.

\subsection{Asymptotic Equivalence to the In-and-Out-of-Sample Pseudo Likelihood Ratio Statistic}

For semiparametric copula models with fully observed data, \cite{zhang2016Goodnessoffit} showed that the IR statistic $R_n$ is asymptotically equivalent to a class of in-and-out-of-sample pseudo  (PIOS) likelihood ratio test statistic. Theorem \ref{thm:IR-PIOS} below states this asymptotic equivalence still holds in the presence of censoring. The PIOS statistic is defined as a difference between two types of pseudo log-likelihood  functions: in-sample and out-of-sample. Under our log-likelihood function in Equation (\ref{equ:log-likelihood}), the in-sample pseudo log-likelihood is defined as $\ell_{n}^{in} = \sum_{i=1}^n \ell(\bthetahatn,\Uhat_{i1},\Uhat_{i2})$, where $\bthetahatn$ is obtained from Equation (\ref{equ:PMLE-theta}) using all the observations. The out-of-sample pseudo log-likelihood employs the leave-one-out technique and is defined as $\ell_{n}^{out} = \sum_{i=1}^n \ell(\bthetahat_{(-i)},\, \allowbreak \Uhat_{i1}, \Uhat_{i2})$, where $\bthetahat_{(-i)}=\arg\max_{\btheta}\sum_{s=1,s\neq i}^n \ell(\btheta,\Uhat_{s1},\Uhat_{s2})$ is the PMLE using the data with the $i$-th observation deleted. The PIOS test statistic is defined as $T_n = \ell_n^{in} - \ell_n^{out}$. A large value of $T_n$ suggests that the assumed copula model is a poor fit to the data since it is sensitive to the deletion of individual observations. 

\begin{theorem}\label{thm:IR-PIOS}
Under condition R1 - R6 and C1 - C4, $|R_n - T_n| = o_p(1)$.
\end{theorem}
The proof is provided in Appendix \ref{app:Thm4}. Because of this asymptotic equivalence, if the null hypothesis $H_0$ is true, the PIOS statistic $T_n$ also converges to $p$ in probability, and $\sqrt{n}(T_n-p)$ also converges in distribution to a normal random variable with mean 0 and the same variance $\sigma_R^2$. 

\section{Information Ratio Test and Copula Selection}\label{sec:test-selection}

In practice, it is challenging to calculate $P$-values using an analytical estimate of the asymptotic variance $\sigma_R^2$ because its expression is complicated. To address this issue, we suggest a parametric bootstrap resampling procedure for the $P$-value calculation. This approach is commonly employed in GoF tests, including those based on information matrix equivalence \citep{horowitz1994bootstrap, dhaene2004information, golden2013New, huang2013Goodnessoffit, golden2016Generalized, prokhorov2019Generalized}. \cite{genest2008validity} provided the validation of this procedure in the general setting of semi-parametric models. 

\subsection{$P$-value Calculation via Bootstrap Resampling}\label{sec:bootstrap}

The key idea is to approximate null distribution of $R_n$ by the test statistics values calculated from a large number of data replicates generated under the null copula (the copula family tested as the null hypothesis). These data replicates are referred to as the bootstrapped data, denoted by $\mathfrak D^{(b)}$; in contrast, we denote the original data by $\mathfrak D$.  The bootstraped data $\mathfrak D^{(b)}$ is obtained by generating bootstrapped resamples of the bivariate event times $(T_{i1}^{(b)},T_{i2} ^{(b)})$ and bivariate censoring times $(C_{i1}^{(b)},C_{i2}^{(b)})$. 

\paragraph{Generation of $(T_{i1}^{(b)},T_{i2}^{(b)})$ under the null copula.} For example, we test Clayton copula as the null hypothesis, i.e., 
$$
H_0: \Cbbm^0(u_1, u_2) = \Cbbm(u_1,u_2;\btheta) = (u_1^{-\theta}+u_2^{-\theta}-1)^{-1/\theta},\, \text{for some $\theta > 0$.}
$$ 
Let $\widehat\theta_n$ be the PMLE of $\theta$ from the log-likelihood function using the original data $\mathfrak D$ and the above parametric form of Clayton copula function. First, we generate a bivariate variable $\left(U_{i1}^{(b)}, U_{i2}^{(b)}\right)$ from the Clayton copula with parameter value $\widehat\theta_n$. This step can be implemented using R function \texttt{rCopula} of \texttt{copula} package. Second, we obtain $T_{ir}^{(b)}=\widehat H_r^{-1}(U_{ir} ^{(b)})$, $r=1,2$, where $\widehat H_r^{-1}$ is the inverse function of the Kaplan-Meier estimator of the marginal survival functions. 

\paragraph{Generation of $(C_{i1}^{(b)},C_{i2}^{(b)})$.}  The censoring times $C_{i1}$ and $C_{i2}$ might be correlated, but our method does not rely on their joint distribution. Thus, we can simulate them separately from their own marginal distributions. Under the assumption of independent censoring, the survival function $G_r(t)$ of $C_{ir}$ can be consistently estimated by a Kaplan-Meier estimator $\widehat G_{r}(t)$ using the data $\{(X_{ir}, 1- \delta_{ir}), i = 1, \cdots, n\}$. For each $r=1, 2$, we first generate a random number $v_{ir}^{(b)}$ from a uniform distribution between 0 and 1, and then obtain $C_{ir}^{(b)}=\widehat G_r^{-1}\left(v_{ir}^{(b)}\right)$. In some cases, both event times are subject to the same censoring time, i.e., $C_{i1}=C_{i2}=C_i$, its sole survival function $G(t)$ can be estimated using the data $\{(\max\{X_{i1},X_{i2}\},1-\delta_{i1}\delta_{i2}),i=1,\cdots,n\}$.

\paragraph{Bootstrap resampling.} The resampling procedure includes the following steps:
\begin{description}
\item [Step 1:] Generate a bootstrapped resample of $\{(T_{i1}^{(b)},T_{i2}^{(b)}, C_{i1}^{(b)},C_{i2}^{(b)}), i=1,\cdots,n\}$ with the same sample size of the original data following the above description. This forms a bootstrapped data $\mathfrak D^{(b)}=\{(X_{i1}^{(b)},\, \allowbreak X_{i2}^{(b)},\delta_{i1}^{(b)},\delta_{i2}^{(b)}),i=1,\cdots,n\}$, where $X_{ir}^{(b)}=\min\{T_{ir}^{(b)}, C_{ir}^{(b)}\}$ and $\delta_{ir}^{(b)}=I(T_{ir}^{(b)} \leq C_{ir}^{(b)})$, $r=1,2$.\item [Step 2:] Based on the bootstrapped data $\mathfrak D^{(b)}$, calculate the test statistic, denoted as $R_n^{(b)}$, referred to as a bootstrap resample of $R_n$. 
\item [Step 3:] Repeat Steps 1 and 2 $B$ times, producing  $B$ bootstrap resamples $\{R_n^{(b)},b=1,\cdots,B\}$.
\end{description}
The bootstrap resamples $\{\sqrt{n}(R_n^{(b)}-p),b=1,\cdots,B\}$ approximate the null distribution of $\sqrt{n}(R_n-p)$, and their sample variance approximates the asymptotic variance $\sigma_R^2$. Thus, we calculate 
$$
\sigma^{\mathbbm b} = \sqrt{\frac{1}{B-1}\sum_{b=1}^B \left[R_n^{(b)} - \overline R_n^{\mathbbm b}\right]^2} 
$$
where $\overline R_n^{\mathbbm b}$ is the average of $\{R_n^{(b)},b=1,\cdots,B\}$. The $P$-value of the IR test is 
$$
p\text{-value} = 2 \times \left[1-\Phi\left(\frac{|R_n-p|}{\sigma^{\mathbbm b}}\right)\right],
$$
where $\Phi(\cdot)$ is the CDF of the standard normal distribution. 

If the calculated $p$-value is smaller than a significance level $\alpha$, we reject the null hypothesis and conclude significant evidence suggesting copula misspecification. Alternatively, we can use critical values to make conclusions. Let $z_{\alpha/2}$ denote the upper $100*(\alpha/2)$\% quantile of the standard normal distribution. If $\frac{|R_n-p|}{\sigma^{\mathbbm b}} > z_{\alpha/2}$, we reject the null hypothesis.

\subsection{Selection of the Best Copula Family}\label{sec:selection}

For some data, a GoF test would fail to reject several copula families. It might be because the sample size is small or the censoring rate is high or both. As a result, the data do not contain sufficient information to reject the null hypothesis. In addition,  if the level of dependence is not strong, several families appear similar, and consequently, it is more difficult for a test to tell them apart. In some situations, the underlying true dependence structure might be complicated, and any parametric copula family is merely an approximation. For these cases, we are more concerned with selecting the {\it best} copula family from several candidates in the sense that the data exhibit the weakest evidence against it, i.e., showing the highest agreement between the assumed copula and the data. Here, we propose using the $P$-value of the IR test as the selection criteria: the best copula family is the one with the largest $P$-value. 

\section{Simulation}\label{sec:simulation}

In this section, we investigate the finite-sample performance of the proposed IR test through two simulation studies, where we consider different sample sizes, copula families with various dependence strength, and censoring rate (proportions of censored event times). The first study focuses on the null distribution of the IR statistic, i.e., the distribution when the null copula is the true copula. We compare it with the normal distribution and PIOS's null distribution. The second study examines the type I error rate and power of IR test as well as the performance of using IR's $p$-value for copula selection. As mentioned earlier, IR is a specific form of generalized IM tests. Thus, we compare our IR with two other forms: White test (difference between two IMs), and log IM test (difference between logarithms of two IMs).

\subsection{Simulation Setting}

We consider four copula families: Clayton, Frank, Joe, and Gaussian, each with a scalar copula parameter $\theta$. The value of $\theta$ is determined by Kendall's $\tau$ coefficient, which reflects the dependence strength \citep{kendall1938new}. The relationship between Kendall's $\tau$  and $\theta$ for each of the above copula families is described in the supplementary material.

Given a copula family $\Cbbm$ with a parameter value $\theta$, we generate $(T_{i1},T_{i2})$ whose marginal distributions are both exponential distribution with mean 1 and joint survival function follows the given copula family $\Cbbm$. For example, the copula is Clayton with parameter $\theta = 2$, corresponding to Kendall's $\tau = 0.5$.  Following a similar procedure described in Section \ref{sec:bootstrap}, we first generate $(U_{i1}, U_{i2})$ from Clayton copula with $\theta = 2$ using R function {\tt rCopula}. Second, calculate $T_{ir} = - \log(U_{ir})$, $r=1,2$. Note that $-\log(x)$ is the inverse function of the survival function for exponential distribution with mean 1.

In this simulation, both event times are subject to a common censoring time $C_i$, generated from an exponential distribution with mean $4$, $3/2$ or $3/7$,  that correspond to a censoring rate of 20\%, 40\%, or 70\% for individual event times. In addition, we include a no-censoring setting, i.e., $T_{i1}$ and $T_{i2}$ are fully observed, to investigate the effect of censoring on the performance of the IR test. Thus, there are four censoring scenarios, denoted as ``no-censoring", ``20\%-censored", ``40\%-censored",  and ``70\%-censored". Figures 1 - 4 in the supplementary material plot the estimated pseudo-observations $(\Uhat_{i1},\Uhat_{i2})$ obtained from one replication of the simulated bivariate censored survival times of sample size $n=100$ or $600$ generated from each of the four copula families with Kendall's $\tau=0.3$ or $0.7$.

\subsection{Study I: Null Distributions of IR and PIOS statistics}\label{sec:simu_null}

In this study, we generate data from a copula family and test for the same copula family, i.e., the null copula is the true copula. Figure 5 - 8 in the supplementary material plot the normal quantile-quantile (QQ) plots of 500 replications of the IR and PIOS statistics under Clayton, Frank, Joe, or Gaussian with Kendall's $\tau=0.5$ at sample size $n=100,300,600$.  These plots allow us to examine (1) whether IR's null distribution is close to normal, and (2) whether the null distributions of IR and PIOS statistics are similar to each other. 

First, we focus on comparing the IR's null distribution with normality. For a given sample size, the distribution gets more skewed to the right as the censoring rate increases. For each censoring scenario, as the sample size increases, it is getting closer to the normal distribution, which confirms the asymptotic normality of the IR statistic (Theorem \ref{thm:normal}). 

Second, we compare the distributions of IR and PIOS. The QQ plots clearly show that their distributions are close, and they get more similar as the sample size  increases. It confirms the asymptotic equivalence between IR and PIOS (Theorem \ref{thm:IR-PIOS}). However, their computational times are substantially different. The PIOS statistic requires repeated ($n$ times) estimation of the copula parameter, $\bthetahat_{(-i)}$, when obtaining the out-of-sample peudo log-likelihood. Thus, its computational burden is more intensive than IR. In addition, as the sample size increases, IR is more computationally efficient. Specifically, using a Dell desktop computer with 3.20 GHz Intel(R) Core(TM) i7-8700 CPU, the average computational time with sample size $n=100$ is 0.0072 seconds for calculating the IR statistic and 0.11 seconds for PIOS (about 15 times of IR's time). When the sample size increases to 600, the average computational time is 0.02 seconds for IR and 1.1 seconds for PIOS (about 55 times of IR's time). Note that the computation of the PIOS statistic has been optimized via parallel computation using R packages "parallel", "foreach", and "doSNOW". If without the parallel computation, it would take even longer.

\subsection{Study II: Test Size and Power of IR Test}

\begin{figure}
\centering
\caption{Simulation results: Proportions of rejecting {\bf Clayton} when the true copula is 
Clayton, Frank, Joe, or Gaussian and the sample size is 600. The dashed lines represent the significance level 0.05.}\label{fig:rej_clayton}
\vskip 0.3cm
\includegraphics[width=1\textwidth]{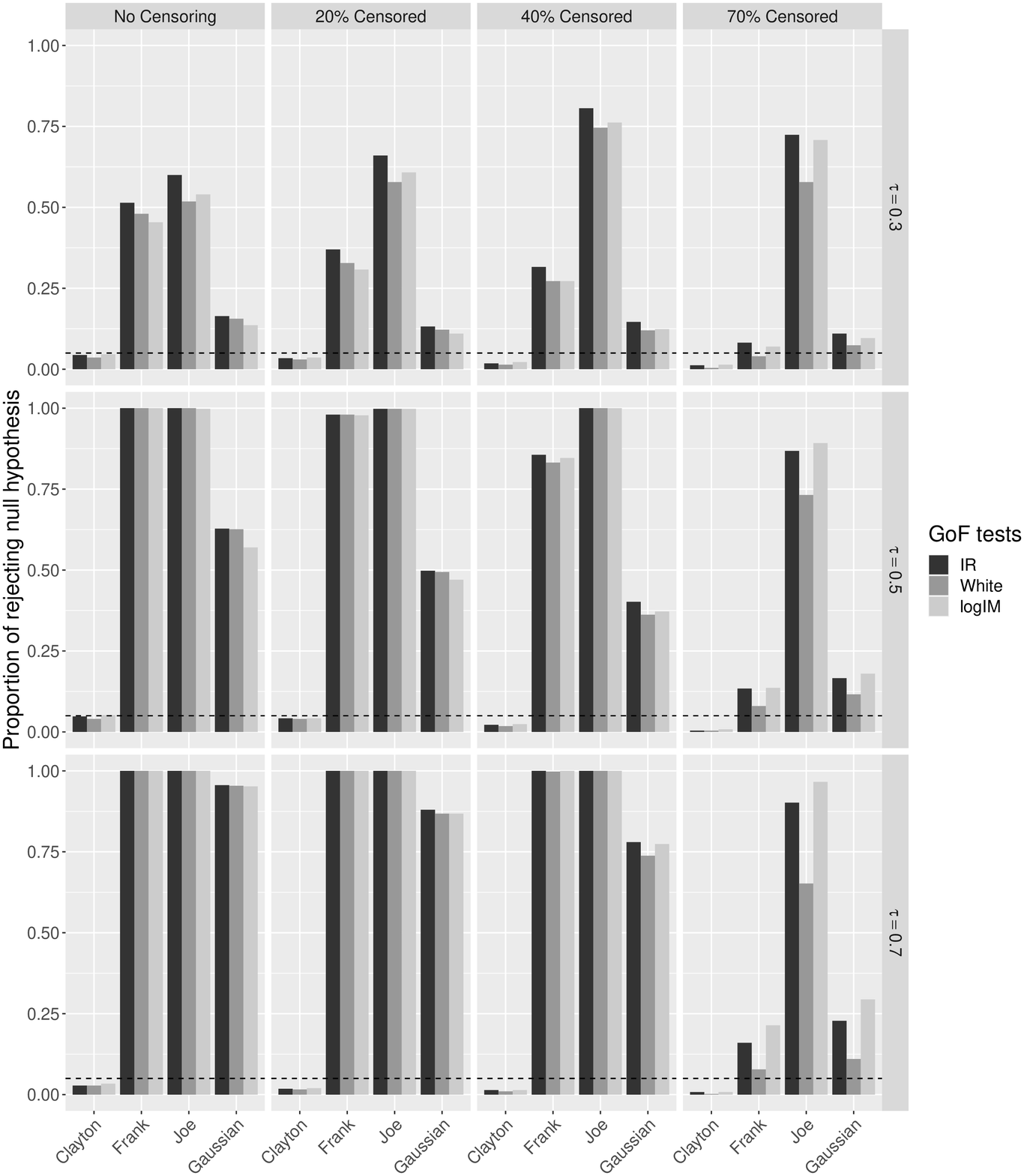}
\end{figure}

\begin{figure}
\centering
\caption{Simulation results: Proportions of rejecting {\bf Frank} when the true copula is 
Clayton, Frank, Joe, or Gaussian and the sample size is 600. The dashed lines represent the significance level 0.05.}\label{fig:rej_frank}
\vskip 0.3cm
\includegraphics[width=1\textwidth]{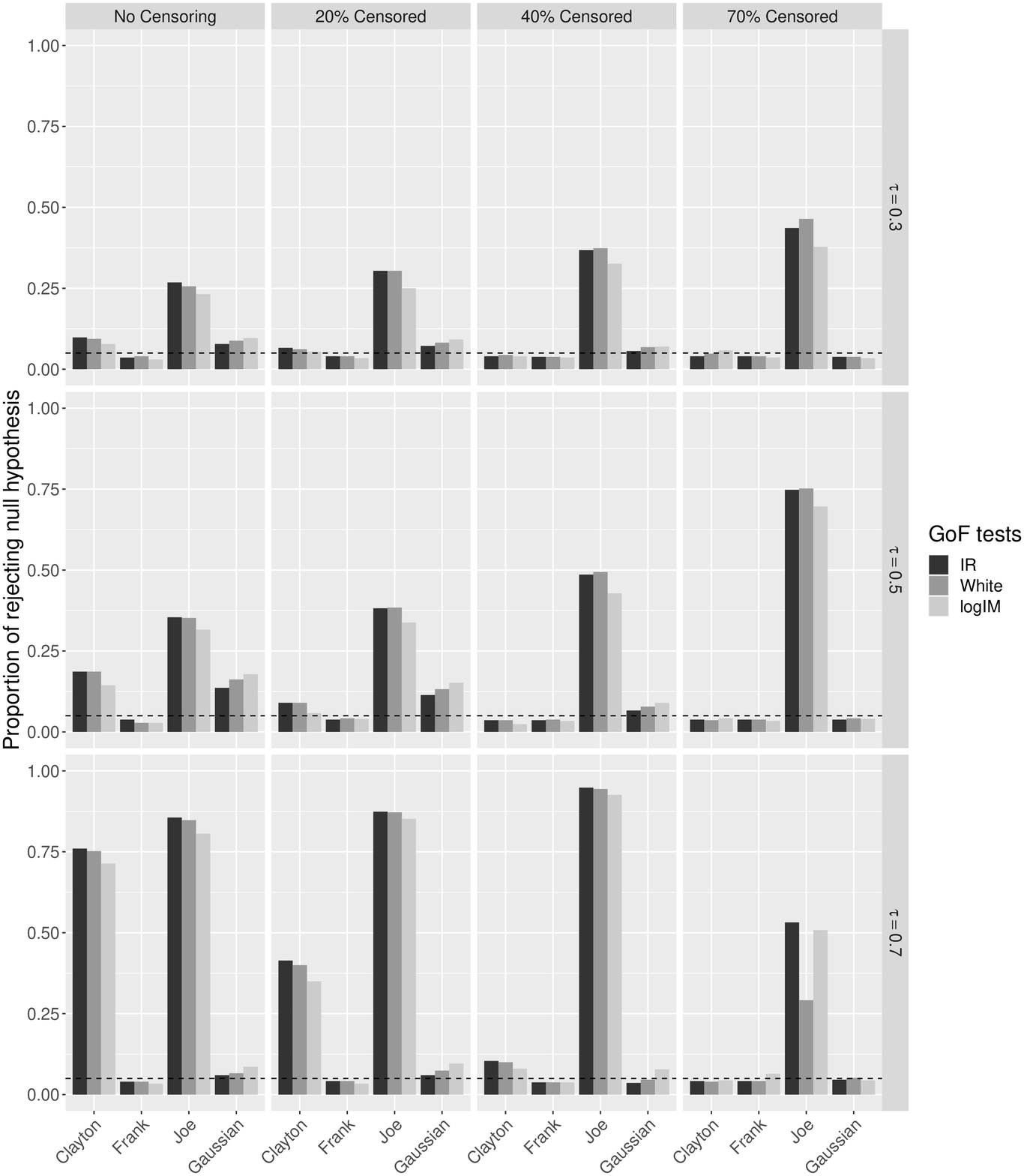}
\end{figure}

\begin{figure}
\centering
\caption{Simulation results: Proportions of rejecting {\bf Joe} when the true copula is 
Clayton, Frank, Joe, or Gaussian and the sample size is 600. The dashed lines represent the significance level 0.05.}\label{fig:rej_joe}
\vskip 0.3cm
\includegraphics[width=1\textwidth]{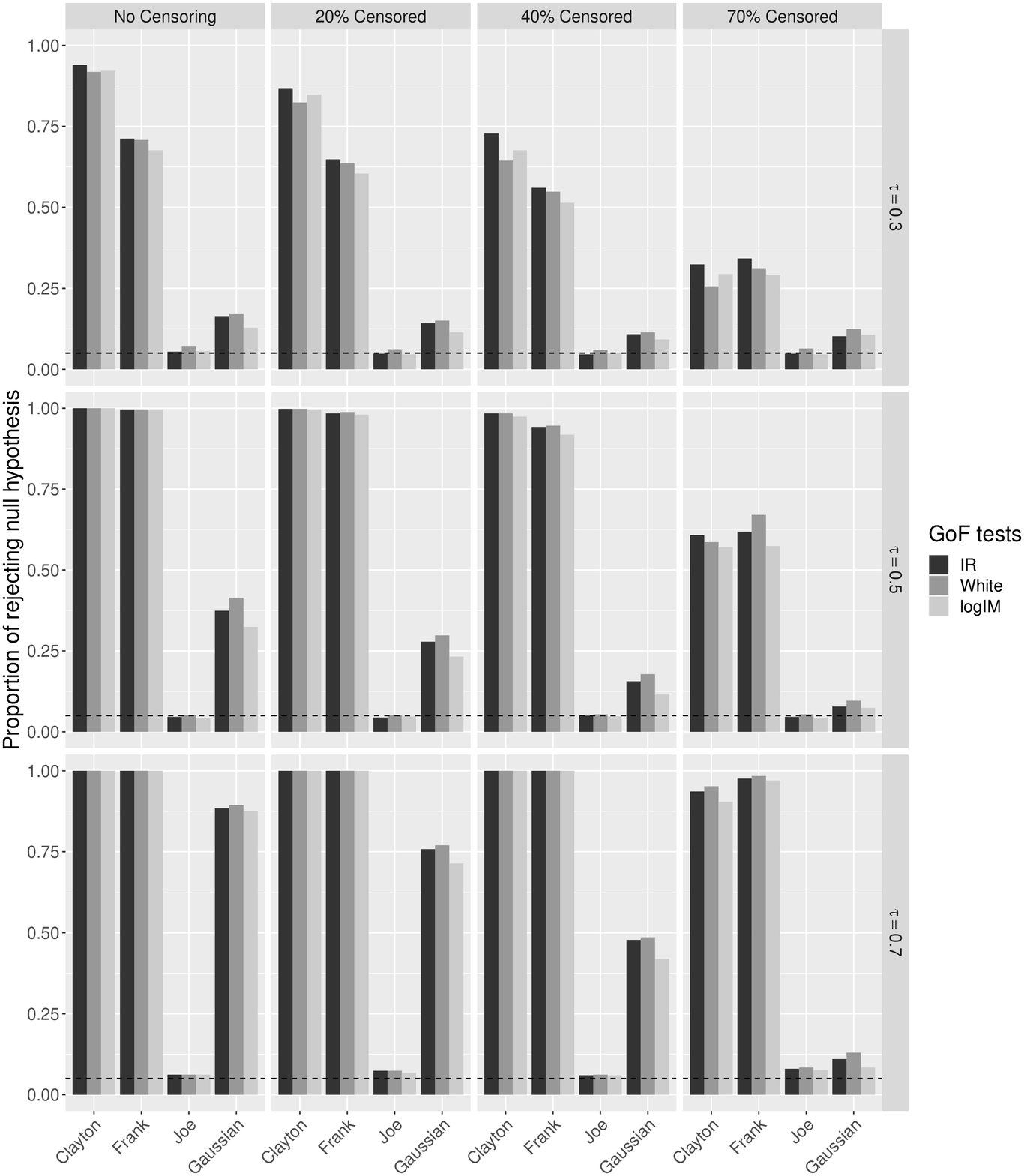}
\end{figure}

\begin{figure}
\centering
\caption{Simulation results: Proportions of rejecting {\bf Gaussian} when the true copula is Clayton, Frank, Joe, or Gaussian and the sample size is 600. The dashed lines represent the significance level 0.05.}\label{fig:rej_normal}
\vskip 0.3cm
\includegraphics[width=1\textwidth]{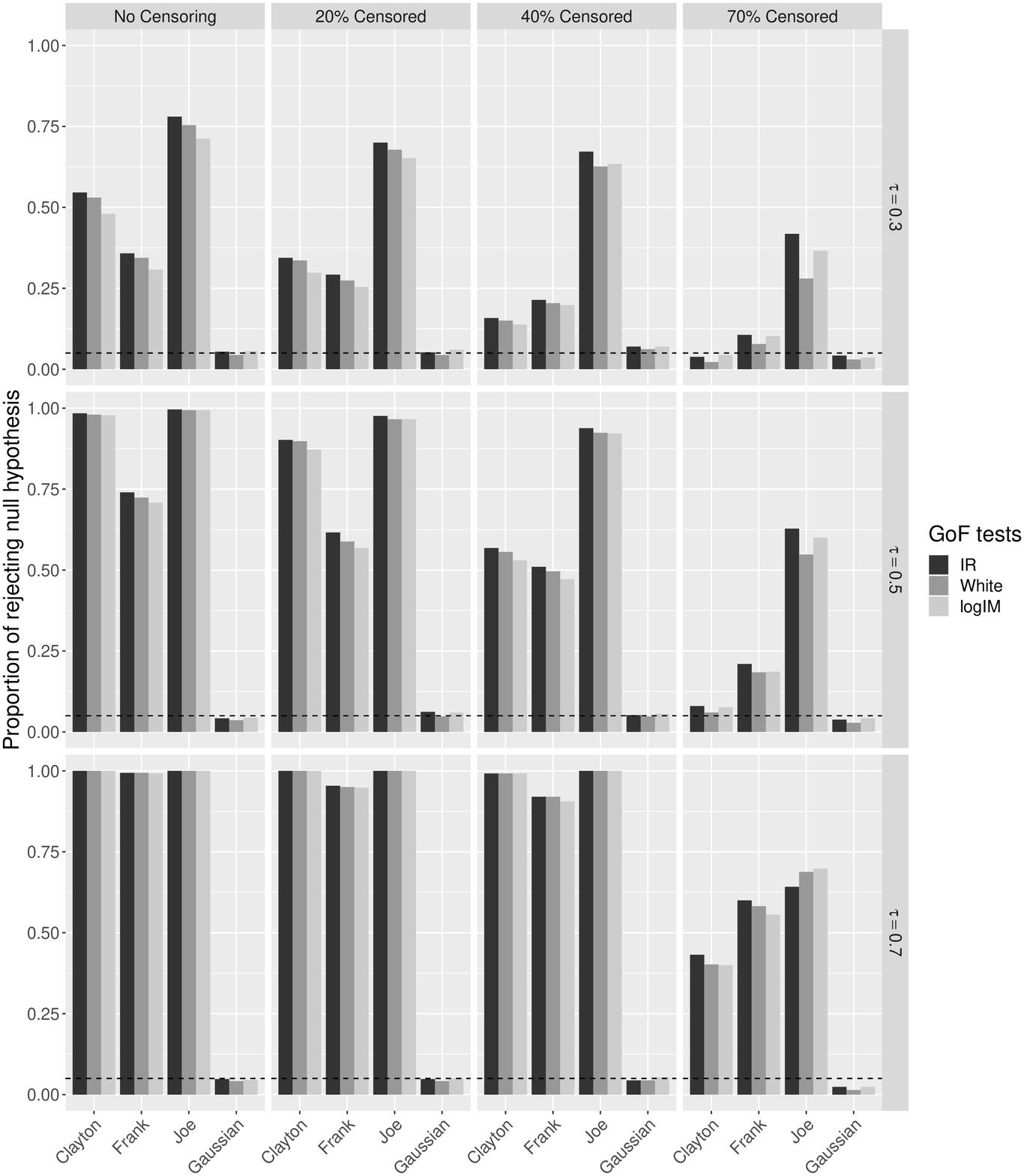}
\end{figure}

In this study, we investigate the type I error rate and power of the proposed IR test. The bivariate event times $(T_{i1},T_{i2})$ are generated from each of the four copula families; under a true copula, we test each of the four copula families as the null hypothesis. For example, in one scenario, $(T_{i1},T_{i2})$ is generated from Clayton,  i.e., the true copula is Clayton, and we test four different null copulas: Clayton, Frank, Joe, and Gaussian. We consider three different dependence levels: Kendall's $\tau=0.3,0.5,0.7$ and three sample sizes $n=100, 300, 600$.  Figures \ref{fig:rej_clayton} - \ref{fig:rej_normal} plots the proportion of rejecting the null hypothesis at the significance level 0.05 among 500 replications at sample size 600. For each simulation replication, the $P$-value is calculated from $B=500$ bootstrap resamples. The rejection proportions for sample size $n=100$ and $300$ are plotted in Figures 9 - 16 of the supplementary material. 

When the null copula is the same as the true copula, the rejection proportions are the empirical type I error rates, also extracted in Table \ref{tab:Type-I}. In most scenarios, the IR test can maintain the nominal test size, i.e., the empirical type I error rates are close to the significance level 0.05. When the null copula is different from the true copula, the rejection proportions are the empirical test power. The results indicate that Kendall's $\tau$, sample size, and censoring rate all affect the power. First, Kendall's $\tau$ reflects the strength of the dependence between the bivariate event times. When $\tau$ is large, i.e., the event times are highly dependent with each other, the true copula's distinct features such as tail dependence are more pronounced, and thus, our IR test is more powerful to detect deviations from the null copula. However, when the dependency is weak, copula families appear similar to each other (See Figures 1 and 3 in the supplementary material). Thus, the IR test has a lower power for a smaller Kendall's $\tau$.  Similar pattens are observed in \cite{genest2009Goodnessoffit} and \cite{zhang2016Goodnessoffit}. Second, as expected, when the sample size is larger or the censoring rate is lower or both, the data provide more information of the underlying true copula, and consequently, the IR test is more powerful.

We observe that when the censoring rate is 70\%, the proportion of rejecting Clayton when the true copula is Clayton is much lower than the significance level 0.05. In other words, the IR test is over conservative against Clayton when the event times are heavily censored. A possible explanation is that when the censoring time follows an exponential distribution, it is more likely to censor smaller event times, leading to insufficient information on the lower-tail dependence, which is a distinct feature of Clayton. As a result, the data exhibit minimal evidence against Clayton. With the same reason, when the true copula is Frank, it is difficult to tell apart from Clayton because they appear alike under heavy censoring (Figures 1 - 4 of the supplementary material). Thus, the proportion of rejecting Clayton when the true copula is Frank is low. Similarly, the proportion of rejecting Frank when the true copula is Clayton is also low. By contrast, since Joe has the upper-tail dependence, the IR test has a much higher power of rejecting Joe when the true copula is Clayton or Frank, or rejecting Clayton or Frank when the true copula is Joe. 

We also observe low proportions of rejecting Frank when the true copula is Gaussian for all sample sizes, Kendall's $\tau$ values, and censoring rates (Figure \ref{fig:rej_frank}).  It could be because both families have no dependence on either tails. However, when the true copula is Frank and the null copula is Gaussian, the test performs better (Figure \ref{fig:rej_normal}). It calls for more investigations.

As pointed out in the introduction, our IR test can be regarded as a specific form of comparing the two information matrices in the class of generalized IM tests \citep{prokhorov2019Generalized}. The other forms include the White test, determinant White test, trace White test, determinant IR test, log trace IM test, log GAIC IM test, log eigenspectrum IM test, and eigenvalue test. However, for the case of scaler parameter, i.e., $p=1$, some tests are equivalent. Specifically, the determinant IR and eigenvalue tests are the same as the IR test: $R_n=\widehat {\bS}_n(\bthetahatn)^{-1}\widehat {\bV}_n(\bthetahatn)$. The White, determinant White, and trace White are the same; they all take a difference: $T_n = \widehat {\bV}_n(\bthetahatn) - \widehat {\bS}_n(\bthetahatn)$.  The log trace IM, log GAIC IM, and log eigenspectrum IM are equivalent, given as $Z_n = \log[\widehat {\bS}_n(\bthetahatn)] - \log[\widehat {\bV}_n(\bthetahatn)]$. Thus, in this study, we compare our IR test $R_n$ with $T_n$, referred to as the White test, and $Z_n$, referred to as the log IM test. The $P$-values of these two tests are also obtained by the parametric bootstrap resampling procedure described in Section \ref{sec:bootstrap}. The results show that these three tests perform similarly for most scenarios.

\subsection{Copula Selection}\label{sec:copula_selection}

We also examine how well using the $P$-value of the IR test as the criterion can correctly select the true copula as the best among the four families. With each simulated data, we obtain the $P$-value for testing each of Clayton, Frank, Joe, and Gaussian as the null hypothesis. Following Section \ref{sec:selection}, we select the copula family with the largest $P$-value as the best. Figures 17 - 28 in the supplementary material report the percentage of choosing each family as the best among the 500 replications.  Consistent with our findings on the test power, when the sample size is larger or the dependence is stronger or the censoring rate is lower, the proportion of selecting the true copula as the best is higher. Copulas with similar properties are more difficult to tell apart. For example, when the true copula is Gaussian, Frank copula is a strong competitor, even when the sample size is 600, Kendall's $\tau=0.7$, and the event times are fully observed. In addition, our IR test performs similarly to the other two generalized IM tests.

\section{Data Example}\label{sec:data}

\begin{figure}
\centering
\caption{Data example: Scatter plot of estimated pseudo-observations $\widehat U_{i1}$ and $\widehat U_{i2}$.}\label{fig:data_U}
\vskip 0.1cm
\includegraphics[width=1\textwidth]{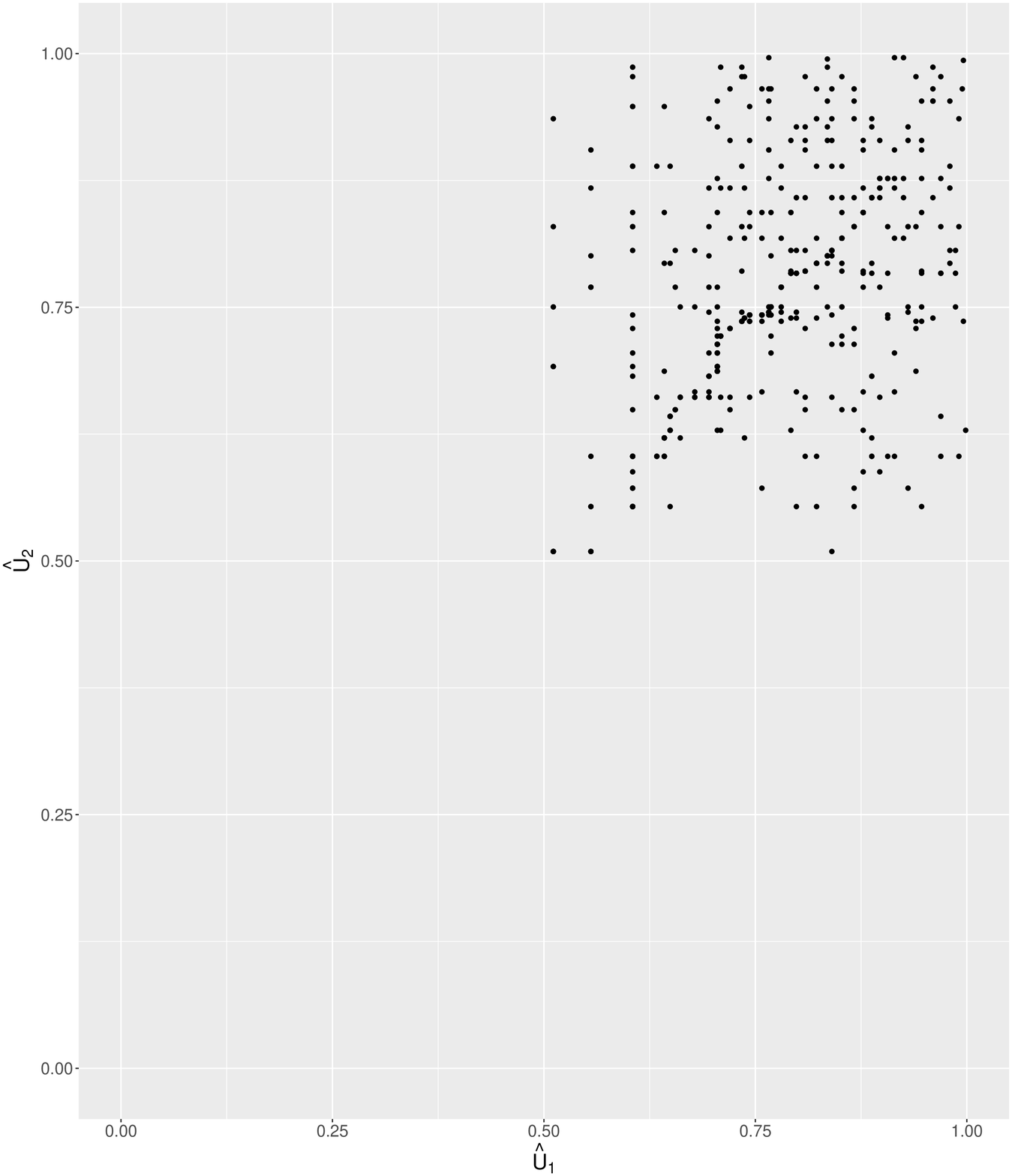}
\end{figure}

The data example is 748 dizygotic female twin pairs from the Australian NHMRC Twin Registry \citep{duffy1990Appendectomy}, and the bivariate event times $(T_1, T_2)$ are the ages at appendicectomy measured for each twin pair. For this data, the event times are heavily censored with the censoring rate of about 74\%. Among the 748 twin pairs, 82 (11\%) pairs have both event times observed, 222 (30\%) have one event time observed and the other censored, and 444 (59\%) have  both event times censored.  Figure \ref{fig:data_U} plots the estimated pseudo-observations $\{(\Uhat_{i1},\Uhat_{i2}), i =1,\cdots,n\}$. 

\cite{emura2010Goodnessoffit} analyzed this data and concluded that Gumbel provides the best fit over three other copula families: Clayton, Frank, and Log-copula. In this manuscript, we test for five copula families: Clayton, Frank, Gumbel, Joe, and Gaussian using our proposed IR test as well as the White test and log IM test.  Table \ref{tab:data} reports their test statistic values and $P$-values calculated using $B=1000$ bootstrapped resamples.  

Among the three tests, only the log IM test reaches the same conclusion as \cite{emura2010Goodnessoffit}: Gumbel is the best copula family with the $P$-value 0.307, and Clayton is the second best with  the $P$-value 0.273. In contrast, for both our IR test and White test, Clayton is the best copula and Gumbel is the second. However, under the IR test, the difference of the $P$-values between Clayton and Gumbel is tiny: the $P$-values is 0.296 for Clayton and 0.291 for Gumbel. It indicates that Gumbel's goodness-of-fit is comparable with Clayton. On the other hand, under the White test, the lead of Clayton over Gumbel is more substantial ($P$-values 0.379 for Clayton and 0.258 for Gumbel).

\section{Concluding Remarks}\label{sec:concluding}

Information matrix equivalence plays an important role in model diagnosis, and a number of GoF tests have been established based on this principle. However, this equivalence has not been verified for censored data. Thus, one major contribution of this work is to prove the equivalence of the two information matrices under a class of semiparametric copula models  for multivariate data in the presence of right censoring. The proof provides a framework which might be extended to other censoring schemes.

Based on this equivalence, we propose an IR test for the specification of the copula function via comparing consistent estimates of the two information matrices. This test is likelihood-based and depends on only the parametric form of the assumed copula function. Thus, it can be applied to all copula families, and do not rely on choices of weight functions, bandwidth, or smoothing parameters. In addition, the IR statistic is asymptotically equivalent to a class of PIOS test statistics, which provides a global measure of how the assumed model fits the data via the leave-one-out cross-validation.  Furthermore,  the IR test does not assume any parametric form of alternative copulas. It can be regarded as an omnibus test. 

In this manuscript, we derive the asymptotic properties of the IR statistic following similar arguments in \cite{chen2010Estimation}. They considered a more general distributional assumption for censoring: the joint distribution of the bivariate censoring times could be different across subjects. Under this relaxed assumption, the pseudo-true value of the copula parameter is defined as
$$
\bthetaast_n = \arg\max_{\btheta}  n^{-1} \sum_{i=1}^n \, \allowbreak \Ezero[\ell(\btheta,U_{i1}^\szero,U_{i2}^\szero)]. 
$$
This value depends on the sample size since the observed survival times might not be identically distributed due to non-identically distributed bivariate censoring times. Correspondingly, the definitions of the sensitivity and variability matrices can be modified as $\bSast(\btheta) = n^{-1}\sum_{i=1}^n \, \allowbreak  \Ezero[-\ell_{\btheta\btheta}]$ and $\bVast(\btheta) = n^{-1}\sum_{i=1}^n \Ezero[\ell_{\btheta}\ell_{\btheta}']$, which also depend on the sample size. It worths pointing out that the proof of Theorem \ref{thm:equivalence} is still valid, and thus, the information matrix equivalence still holds, and the IR test is still valid. However, for generating bootstrap resamples of censoring times, the Kaplan-Meier estimator of the censoring survival function is not appropriate when assuming heterogenous censoring distribution. Under this assumption, it would require some subject-specific covariates $Z_i$ to estimate the subject-specific censoring survival function $G_{ir}(t) = Pr(C_{ir}>t \mid Z_i)$ for $r=1,2$.

In general, if testing within Archimedean families, the GoF tests that target these families are expected to be more powerful than our proposed IR test because they utilize their distinct properties such as cross-ratio functions or Kendall distribution. On the other hand, our proposed IR test can compare copula families beyond Archimedean. In Section \ref{sec:selection}, we demonstrate how to use the $P$-value of the IR test to select the best copula family among several candidates. 

Equation (\ref{equ:interchange}) is the key step for proving the information matrix equivalence. It also implies that this equivalence holds for any censoring distribution when the assumed copula is correctly specified. However, when the assumed copula is misspecified, the difference between two information matrices depends on the censoring distribution. Our simulation study has shown that the censoring rate is one factor that affects the performance of the IR test. We hypothesize that besides the censoring rate, the shape of the censoring distribution might be another factor. For example, as discussed in our simulation, the shape of the exponential distribution for censoring leads to insufficient information on the lower-tail dependence. It causes the low power of differentiating between Clayton and Frank. It is our interest to conduct more studies to investigate other distributions for censoring, such as gamma, Weibull, or uniform distributions. 

In the simulation study and data example, we compare our IR test with two other forms of generalized IM tests, and they perform similarly. For example, all three tests exhibit a lower power for rejecting Frank when the true copula is Gaussian because they both have neither upper-tail or lower-tail dependence. Our studies focus on the case of scalar copula parameter, i.e., $p=1$, for which the class of generalized IM tests reduces to three forms of comparing IMs: ratio, difference, and difference of logarithm. However, if $p > 1$, the class would not be limited to only these three forms. In addition, different IM-based tests would perform more diversely. It worths further investigation for cases with $p > 1$.

\section{Supplementary Material}

In the supplementary material, we present the expressions of the copula function, and the derivatives of the log-likelihood function for Clayton, Frank, Joe, and Gaussian copulas. We also show more results of the simulation study, including (i) scatter plots (Figures 1 - 4) of estimated pseudo-observations $(\widehat U_{i1},\widehat U_{i2})$ from one simulated bivariate censored data, (ii) QQ plots (Figures 5 - 8) of the IR and PIOS statistics when the null copula is the true copula, (iii) bar plots (Figures 9 - 16) of proportions of rejecting the null hypothesis for sample sizes 100 and 300, and (iv) bar plots (Figures 17 - 28) of proportions of selecting different copula families as the best copula. 

\section{Acknowledgement}

The work is supported by the grant DMS-2210481 from the National Science Foundation.

\bibliographystyle{apalike}
\bibliography{GoF_BiSurv}

\clearpage
\newpage

\begin{table}
\centering
\small
\caption{Simulation results: Empirical type I errors for the IR test $R_n$, White test $T_n$, and log IM test $Z_n$.}\label{tab:Type-I}
\begin{tabular}{cc|ccc|ccc|ccc|ccc}
\toprule
 &  & \multicolumn{3}{c|}{No Cen.} & \multicolumn{3}{c|}{20\% Cen.} &  \multicolumn{3}{c|}{40\% Cen.} & \multicolumn{3}{c}{70\% Cen.} \\
$\tau$ & $n$ & $R_n$ & $T_n$& $Z_n$  &$R_n$ & $T_n$& $Z_n$  & I$R_n$ & $T_n$& $Z_n$ &$R_n$ & $T_n$& $Z_n$\\
 \hline
\multicolumn{14}{c}{Clayton Copula}\\
 & 100 & 0.038 & 0.018 & 0.044 & 0.034 & 0.014 & 0.048 & 0.024 & 0.002 & 0.026 & 0.000 & 0.000 & 0.078 \\
0.3 & 300 & 0.036 & 0.026 & 0.032 & 0.030 & 0.018 & 0.034 & 0.022 & 0.010 & 0.020 & 0.012 & 0.002 & 0.016 \\
 & 600 & 0.044 & 0.036 & 0.046 & 0.034 & 0.030 & 0.036 & 0.018 & 0.014 & 0.022 & 0.012 & 0.004 & 0.014 \\
\hline 
 & 100 & 0.048 & 0.028 & 0.038 & 0.032 & 0.022 & 0.036 & 0.032 & 0.012 & 0.040 & 0.022 & 0.000 & 0.040 \\
0.5 & 300 & 0.036 & 0.030 & 0.042 & 0.028 & 0.018 & 0.024 & 0.024 & 0.018 & 0.028 & 0.008 & 0.004 & 0.010 \\
 & 600 & 0.048 & 0.040 & 0.052 & 0.042 & 0.040 & 0.042 & 0.022 & 0.018 & 0.024 & 0.004 & 0.004 & 0.008 \\
\hline 
 & 100 & 0.054 & 0.026 & 0.058 & 0.026 & 0.014 & 0.026 & 0.020 & 0.010 & 0.022 & 0.016 & 0.002 & 0.016 \\
0.7 & 300 & 0.050 & 0.034 & 0.050 & 0.036 & 0.024 & 0.040 & 0.014 & 0.012 & 0.018 & 0.016 & 0.006 & 0.018 \\
 & 600 & 0.028 & 0.028 & 0.034 & 0.018 & 0.016 & 0.020 & 0.014 & 0.010 & 0.014 & 0.008 & 0.002 & 0.008 \\
\hline 
\multicolumn{14}{c}{Frank Copula}\\
 & 100 & 0.032 & 0.034 & 0.038 & 0.028 & 0.032 & 0.044 & 0.040 & 0.048 & 0.050 & 0.022 & 0.034 & 0.034 \\
0.3 & 300 & 0.050 & 0.052 & 0.052 & 0.048 & 0.046 & 0.052 & 0.046 & 0.044 & 0.042 & 0.058 & 0.060 & 0.050 \\
 & 600 & 0.036 & 0.040 & 0.030 & 0.040 & 0.040 & 0.034 & 0.038 & 0.038 & 0.036 & 0.040 & 0.040 & 0.036 \\
\hline 
 & 100 & 0.038 & 0.036 & 0.044 & 0.038 & 0.036 & 0.058 & 0.048 & 0.054 & 0.066 & 0.024 & 0.024 & 0.048 \\
0.5 & 300 & 0.054 & 0.054 & 0.048 & 0.038 & 0.036 & 0.040 & 0.040 & 0.040 & 0.048 & 0.030 & 0.034 & 0.048 \\
 & 600 & 0.038 & 0.028 & 0.028 & 0.038 & 0.042 & 0.040 & 0.036 & 0.038 & 0.034 & 0.038 & 0.038 & 0.034 \\
\hline 
 & 100 & 0.048 & 0.040 & 0.068 & 0.050 & 0.038 & 0.058 & 0.066 & 0.046 & 0.078 & 0.018 & 0.010 & 0.118 \\
0.7 & 300 & 0.066 & 0.060 & 0.058 & 0.054 & 0.046 & 0.048 & 0.054 & 0.040 & 0.058 & 0.030 & 0.030 & 0.052 \\
 & 600 & 0.040 & 0.040 & 0.034 & 0.042 & 0.042 & 0.034 & 0.038 & 0.038 & 0.038 & 0.042 & 0.042 & 0.064 \\
\hline 
\multicolumn{14}{c}{Joe Copula}\\
 & 100 & 0.054 & 0.064 & 0.066 & 0.046 & 0.062 & 0.062 & 0.052 & 0.074 & 0.060 & 0.034 & 0.066 & 0.038 \\
0.3 & 300 & 0.066 & 0.078 & 0.078 & 0.058 & 0.066 & 0.076 & 0.050 & 0.060 & 0.070 & 0.046 & 0.050 & 0.054 \\
 & 600 & 0.054 & 0.072 & 0.056 & 0.048 & 0.062 & 0.046 & 0.046 & 0.060 & 0.050 & 0.048 & 0.064 & 0.046 \\
\hline 
 & 100 & 0.050 & 0.044 & 0.052 & 0.052 & 0.046 & 0.054 & 0.040 & 0.034 & 0.050 & 0.048 & 0.036 & 0.050 \\
0.5 & 300 & 0.068 & 0.062 & 0.082 & 0.058 & 0.060 & 0.078 & 0.050 & 0.048 & 0.060 & 0.056 & 0.054 & 0.062 \\
 & 600 & 0.046 & 0.052 & 0.042 & 0.044 & 0.052 & 0.048 & 0.050 & 0.054 & 0.048 & 0.046 & 0.054 & 0.044 \\
\hline 
 & 100 & 0.050 & 0.038 & 0.046 & 0.054 & 0.040 & 0.052 & 0.044 & 0.034 & 0.038 & 0.046 & 0.028 & 0.042 \\
0.7 & 300 & 0.056 & 0.042 & 0.050 & 0.050 & 0.044 & 0.050 & 0.052 & 0.042 & 0.050 & 0.064 & 0.046 & 0.052 \\
 & 600 & 0.062 & 0.062 & 0.062 & 0.074 & 0.074 & 0.068 & 0.060 & 0.062 & 0.060 & 0.080 & 0.084 & 0.076 \\
\hline 
\multicolumn{14}{c}{Gaussian Copula}\\
 & 100 & 0.026 & 0.014 & 0.048 & 0.042 & 0.016 & 0.044 & 0.034 & 0.014 & 0.036 & 0.014 & 0.000 & 0.018 \\
0.3 & 300 & 0.040 & 0.028 & 0.046 & 0.040 & 0.034 & 0.044 & 0.052 & 0.042 & 0.052 & 0.024 & 0.012 & 0.034 \\
 & 600 & 0.054 & 0.044 & 0.056 & 0.052 & 0.044 & 0.060 & 0.070 & 0.062 & 0.070 & 0.042 & 0.030 & 0.036 \\
\hline 
 & 100 & 0.022 & 0.018 & 0.044 & 0.022 & 0.012 & 0.042 & 0.022 & 0.002 & 0.034 & 0.018 & 0.002 & 0.026 \\
0.5 & 300 & 0.046 & 0.042 & 0.042 & 0.040 & 0.034 & 0.048 & 0.054 & 0.036 & 0.054 & 0.024 & 0.010 & 0.032 \\
 & 600 & 0.042 & 0.036 & 0.044 & 0.062 & 0.048 & 0.060 & 0.052 & 0.048 & 0.056 & 0.038 & 0.028 & 0.042 \\
\hline 
 & 100 & 0.032 & 0.028 & 0.040 & 0.030 & 0.020 & 0.032 & 0.026 & 0.010 & 0.026 & 0.022 & 0.006 & 0.020 \\
0.7 & 300 & 0.048 & 0.042 & 0.044 & 0.044 & 0.040 & 0.040 & 0.046 & 0.046 & 0.042 & 0.012 & 0.008 & 0.016 \\
 & 600 & 0.048 & 0.042 & 0.046 & 0.048 & 0.042 & 0.050 & 0.044 & 0.044 & 0.054 & 0.024 & 0.014 & 0.024 \\
  \bottomrule
\end{tabular}
\end{table}

\begin{table}
\centering
\caption{Data example: The PMLE $\widehat \theta_n$ of the copula parameter, the test statistic with the $P$-value (in the paratheses) of the IR, White, and log IM tests for Clayton, Frank, Gumbel, Joe, and Gaussian.}\label{tab:data}
\vspace{0.2cm}
\begin{tabular}{c|cccc}
\toprule
Copula & $\widehat \theta_n$ & IR & White & log IM\\
\hline
Clayton & 0.750 & 1.085 (0.296) & 0.003 (0.379) & 0.081 (0.273) \\
Frank & 1.795 & 1.075 (0.039) & 0.001 (0.038) & 0.072 (0.046) \\
Gumbel & 1.162 & 1.060 (0.291) & 0.051 (0.258) & 0.058 (0.307) \\
Joe & 1.204 & 1.085 (0.215) & 0.045 (0.191) & 0.081 (0.238) \\
Gaussian & 0.304 & 1.083 (0.166) & 0.035 (0.188) & 0.079 (0.177) \\
\bottomrule
\end{tabular}
\end{table}

\clearpage
\newpage 

\clearpage
\newpage

\appendix

\section*{Appendix}

\section{Proof of Theorem \ref{thm:equivalence}}\label{app:Thm1}

The outline of the prove is as follows. First, we will show that for an assumed copula with a given $\btheta$, 
\begin{equation}
\bSast(\btheta) = \bVast(\btheta) - \bA(\btheta) ,  
\end{equation}
where $\bA(\btheta)$ is a $p\times p$ matrix. Second, we prove that if the assumed copula is correctly specified, $\bA(\btheta^*)={\bf 0}_{p\times p}$. 

\paragraph{Derive $\bSast(\btheta) = \bVast(\btheta)-\bA(\btheta)$.}

The sensitivity and variability matrices in Equation (\ref{equ:IM-censor}) are functions of $\ell_{\btheta} \ell_{\btheta}'$ and $\mathfrak \ell_{\btheta\btheta}$. We derive the expressions of these two quantities. By Equation (\ref{equ:log-likelihood}), we have 
\begin{align}
\label{equ:score}\ell_{\btheta} & = \delta_1\delta_2 \frac{\cbbm_{\btheta}}{\cbbm} + \delta_1(1-\delta_2) \frac{\cbbm_{1,\btheta}}{\cbbm_1} + (1-\delta_1)\delta_2\frac{\cbbm_{2,\btheta}}{\cbbm_2} + (1-\delta_1)(1-\delta_2) \frac{\Cbbm_{\btheta}}{\Cbbm}.
\end{align}
Thus,
\begin{align}
\label{equ:lthetaltheta}\ell_{\btheta}\mathfrak \ell_{\btheta}'= & \delta_1\delta_2 \frac{\cbbm_{\btheta}\cbbm_{\btheta}'}{\cbbm^2} + \delta_1(1-\delta_2) \frac{\cbbm_{1,\btheta}\cbbm_{1,\btheta}'}{\cbbm_1^2} + (1-\delta_1)\delta_2\frac{\cbbm_{2,\btheta}\cbbm_{2,\btheta}'}{\cbbm_2^2} + (1-\delta_1)(1-\delta_2) \frac{\Cbbm_{\btheta}\Cbbm_{\btheta}'}{\Cbbm^2},
\end{align}
and
\begin{align}
\nonumber \ell_{\btheta\btheta} & = \delta_1\delta_2 \left[\frac{\cbbm_{\btheta\btheta}}{\cbbm} - \frac{\cbbm_{\btheta}\cbbm_{\btheta}'}{\cbbm^2} \right] + \delta_1(1-\delta_2) \left[\frac{\cbbm_{1,\btheta\btheta}}{\cbbm_1} - \frac{\cbbm_{1,\btheta}\cbbm_{1,\btheta}'}{\cbbm_1^2} \right] \\
\label{equ:lthetatheta}& + (1-\delta_1)\delta_2 \left[\frac{\cbbm_{2,\btheta\btheta}}{\cbbm_2} - \frac{\cbbm_{2,\btheta}\cbbm_{2,\btheta}'}{\cbbm_2^2} \right] + (1-\delta_1)(1-\delta_2) \left[\frac{\Cbbm_{\btheta\btheta}}{\Cbbm} - \frac{\Cbbm_{\btheta}\Cbbm_{\btheta}'}{\Cbbm^2}\right].
\end{align}
Consequently, $-\ell_{\btheta\btheta} = \ell_{\btheta}\ell_{\btheta}' - \boldsymbol{\Delta}$, where $\boldsymbol{\Delta}$ is a $p\times p$ matrix with the $(j,k)$-th element
\begin{align*}
\Delta_{jk}(\btheta,u_1,u_2) & = \delta_1\delta_2 \frac{\cbbm_{\theta_j\theta_k}(u_1,u_2;\btheta)}{\cbbm(u_1,u_2;\btheta)} + \delta_1(1-\delta_2)\frac{\cbbm_{1,\theta_j\theta_k}(u_1,u_2;\btheta)}{\cbbm_1(u_1,u_2;\btheta)} \\
& + (1-\delta_1)\delta_2\frac{\cbbm_{2,\theta_j\theta_k}(u_1,u_2;\btheta)}{\cbbm_2(u_1,u_2;\btheta)} + (1-\delta_1)(1-\delta_2)\frac{\Cbbm_{\theta_j\theta_k}(u_1,u_2;\btheta)}{\Cbbm(u_1,u_2;\btheta)},
\end{align*}
Note that, for simplicity, we suppress $(\delta_1,\delta_2)$ from the $\Delta$ function.  Thus, $\bA(\btheta)$ is a $p\times p$ matrix with the $(j,k)$-th element $A_{jk}(\btheta)=\Ezero[\Delta_{jk}(\btheta,U_1^\szero,U_2^\szero)]$. \smallskip

To derive the expression of $A_{jk}(\btheta)$, we invoke the double expectation theorem by conditioning on $(C_1,C_2)$, i.e., 
 $A_{jk}(\btheta)=\Ezero_{(C_1,C_2)}\left\{\Ezero_{(T_1,T_2)}[\Delta_{jk}|C_1,C_2]\right\}$,  where $\Ezero_{(T_1,T_2)}$ and $\Ezero_{(C_1,C_2)}$ denote the expectations w.r.t. the true distributions of $(T_1,T_2)$ and $(C_1,C_2)$, respectively. In Remark \ref{remark:log-likelihood}, we stated that given that the true marginal survival functions $H_r^\szero(\cdot)$ are known, the copula $\Cbbm$ can be regarded as the joint CDF of $(Y_1,Y_2)$ with $Y_r=H_r^\szero(T_r)$, $r=1,2$, which are uniformly distributed on $(0,1)$. Thus, the expectation w.r.t. $(T_1,T_2)$ is equivalent to the expectation w.r.t. $(Y_1, Y_2)$, which gives
\begin{equation}\label{equ:Ajk}
A_{jk}(\btheta)=\Ezero_{(C_1,C_2)}\left\{\Ezero_{(Y_1,Y_2)}[\Delta_{jk}(\btheta; U_1^\szero, U_2^\szero)|C_1,C_2]\right\}.  \end{equation}
In addition, $U_r^\szero$ can be expressed as $U_r^\szero=\max\{Y_r,H_r^\szero(C_r)\}$ and $\delta_r = I\left(Y_r\geq H_r^\szero(C_r)\right)$.

\paragraph{Proof of $A_{jk}(\btheta^*)=0$ under correct specification.}  If the assumed copula $\Cbbm$ is correctly specified, $\Ezero_{(Y_1,Y_2)}$ is taken w.r.t. $\Cbbm$ as follows:
$$
\Ezero_{(Y_1,Y_2)}[\Delta_{jk}|C_1,C_2] = \iint  \Delta_{jk}(\btheta; U_1^\szero, U_2^\szero) \cbbm(u_1,u_2;\btheta) du_1du_2.
$$
Given the regularity conditions R1 - R6, by the law of total probability, we can show that this conditional expectation $\Ezero_{(Y_1,Y_2)}[\Delta_{jk}|C_1,C_2] = \mathcal A_{(1,1)}+\mathcal A_{(1,0)}+\mathcal A_{(0,1)}+\mathcal A_{(0,0)}$ where $ \Acal_{(1,1)}, \Acal_{(1,0)}, \Acal_{(0,1)}$, and $\Acal_{(0,0)}$ correspond to each censoring scenario $(\delta_1,\delta_2)$, given as
\begin{align}
\label{equ:A11}\Acal_{(1,1)}(\btheta,C_1,C_2) & = \iint_{[H_1^\szero(C_1),1]\times[H_2^\szero(C_2),1]} \cbbm_{\theta_j\theta_k}(u_1,u_2;\btheta) du_1du_2,\\
\label{equ:A10}\Acal_{(1,0)} (\btheta,C_1,C_2)  & = \iint_{[H_1^\szero(C_1),1] \times [0,H_2^\szero(C_2)]} \cbbm_{\theta_j\theta_k}(u_1,u_2;\btheta) du_1du_2,\\
\label{equ:A01} \Acal_{(0,1)} (\btheta,C_1,C_2)  & = \iint_{[0,H_1^\szero(C_1)]\times [H_2^\szero(C_2),1]} \cbbm_{\theta_j\theta_k}(u_1,u_2;\btheta) du_2du_2,\\
\label{equ:A00}\Acal_{(0,0)}(\btheta,C_1,C_2) & =   \iint_{[0,H_1^\szero(C_1)]\times [0,H_2^\szero(C_2)]} \cbbm_{\theta_j\theta_k}(u_1,u_2;\btheta) du_2du_2.
\end{align}

\begin{itemize}
\item Under the scenario with $\delta_1=1$ and $\delta_2=1$, $\Delta_{jk}=\cbbm_{\theta_j\theta_k}/\cbbm$ with  both $U_1^\szero=Y_1$ and $U_2^\szero=Y_2$ being random variables. In addition, the integral under this scenario is taken over the region of $(Y_1,Y_2)$: $\Omega_{11}=[H_1^\szero(C_1),1]\times[H_2^\szero(C_2),1]$. Thus, the conditional expectation is 
$$
\iint_{\Omega_{11}} \frac{\cbbm_{\theta_j\theta_k} (u_1,u_2;\btheta)}{\cbbm(u_1,u_2;\btheta)}  \cbbm(u_1,u_2;\btheta) du_1du_2 =  \iint_{\Omega_{11}} \cbbm_{\theta_j\theta_k} (u_1,u_2;\btheta) du_1du_2,
$$
which results in $A_{11}$ in Equation (\ref{equ:A11}).

\item Under the scenario with $\delta_1=1$ and $\delta_2=0$, $\Delta=\cbbm_{1,\theta_j\theta_k}/\cbbm_1$ with  $U_2^\szero=H_2^\szero(C_2)$ as a fixed number and $U_1^\szero=Y_1$ as the only random variable. In addition, the integral under this scenario is taken over the region of $(Y_1,Y_2)$: $\Omega_{10}=[H_1^\szero(C_1),1]\times[0,H_2^\szero(C_2)]$. Thus, the conditional expectation is 
\begin{align*}
& \int_{[H_1^\szero(C_1),1]} \frac{\cbbm_{1,\theta_j\theta_k} (u_1,H_2^\szero(C_2);\btheta)}{\cbbm_1(u_1,H_2^\szero(C_2);\btheta)}  \left\{\int_{[0,H_2^\szero(C_2)]} \cbbm (u_1,u_2;\btheta) du_2\right\} du_1\\
=  &  \int_{[H_1^\szero(C_1),1]} \cbbm_{1,\theta_j\theta_k} (u_1,H_2^\szero(C_2);\btheta) du_1,
\end{align*}
since $\int_{[0,H_2^\szero(C_2)]} \cbbm (u_1,u_2;\btheta) du_2 = \cbbm_{1} (u_1,H_2^\szero(C_2);\btheta) $. The regularity condition R6 ensures that functions $\cbbm_{\theta_j\theta_k}$, $j,k=1,\cdots,p$, are dominated by integrable function w.r.t. $(u_1,u_2)$ for all $\btheta$.  It allows the following interchangeability between the integral and derivation in $\cbbm_{1,\theta_j\theta_k} $: 
$$ 
\frac{\partial^2}{\partial \theta_j\partial \theta_k}  \int_{[0,H_2^\szero(C_2)]} \cbbm(u_1,u_2;\btheta)du_2 =  \int_{[0,H_2^\szero(C_2)]} \cbbm_{\theta_j\theta_k}(u_1,u_2;\btheta)du_2.
$$
Thus, the conditional expectation is $ \iint_{\Omega_{10}}  \cbbm _{\theta_j\theta_k}(u_1,u_2;\btheta) du_1du_2$, which is $A_{10}$ in Equation (\ref{equ:A10}).

\item Under the scenario with $\delta_1=0$ and $\delta_2=1$, $\Delta=\cbbm_{2,\theta_j\theta_k}/\cbbm_2$ with  $U_1^\szero=H_1^\szero(C_1)$ as a fixed number and $U_2^\szero=Y_2$ as the only random variable. In addition, the integral under this scenario is taken over the region of $(Y_1,Y_2)$: $\Omega_{01}=[0,H_1^\szero(C_1)] \times [H_2^\szero(C_2),1]$. Thus, the conditional expectation is 
\begin{align*}
& \int_{[H_2^\szero(C_2),1]} \frac{\cbbm_{2,\theta_j\theta_k} (H_1^\szero(C_1),u_2;\btheta)}{\cbbm_2(H_1^\szero(C_1),u_2;\btheta)}  \left\{\int_{[0,H_1^\szero(C_1)]} \cbbm (u_1,u_2;\btheta) du_1\right\} du_2\\
 = & \int_{[H_2^\szero(C_2),1]}\cbbm_{2,\theta_j\theta_k} (H_1^\szero(C_1),u_2;\btheta) du_2,
\end{align*}
since $\int_{[0,H_1^\szero(C_1)]} \cbbm (u_1,u_2;\btheta) du_1 = \cbbm_2(H_1^\szero(C_1),u_2;\btheta)$. Similarly,  in $\cbbm_{2,\theta_j,\theta_k}$, the integral and derivation are interchangeable as follows:
$$
 \frac{\partial^2}{\partial \theta_j\partial \theta_k} \int_{[0,H_1^\szero(C_1)]} \cbbm(u_1,u_2;\btheta) du_1=\int_{[0,H_1^\szero(C_1)]}  \cbbm_{\theta_j\theta_k}(u_1,u_2;\btheta) du_1.
$$
Thus, the conditional expectation is $ \iint_{\Omega_{01}}  \cbbm_{\theta_j\theta_k} (u_1,u_2;\btheta) du_1du_2$, which is $A_{01}$ in Equation (\ref{equ:A01}).

\item Under the scenario with $\delta_1=0$ and $\delta_2=0$, $\Delta=\Cbbm_{\theta_j\theta_k}/\Cbbm$ with both $U_1^\szero=H_1^\szero(C_1)$ and $U_2^\szero=H_2^\szero(C_2)$ as fixed numbers. In addition, the integral under this scenario is taken over the region of $(Y_1,Y_2)$: $\Omega_{00}=[0,H_1^\szero(C_1)] \times [0,H_2^\szero(C_2)]$. Thus, the conditional expectation is 
\begin{align*}
 & \frac{\Cbbm_{\theta_j\theta_k} (H_1^\szero(C_1),H_2^\szero(C_2);\btheta)}{\Cbbm(H_1^\szero(C_1),H_2^\szero(C_2);\btheta)}   \left\{\iint_{[0,H_1^\szero(C_1)]\times [0,H_2^\szero(C_2)]} \cbbm (u_1,u_2;\btheta) du_1 du_2\right\} \\
= &\Cbbm_{\theta_j\theta_k} (H_1^\szero(C_1),H_2^\szero(C_2);\btheta).
\end{align*}
Again, in $\Cbbm_{\theta_j\theta_k} $, the integral and derivations can be interchangeable as follows:
\begin{align*}
& \frac{\partial^2}{\partial \theta_j\partial \theta_k} \iint_{[0,H_1^\szero(C_1)]\times[0,H_2^\szero(C_2)]}\cbbm(u_1,u_2;\btheta)du_1du_2 \\
=  & \iint_{[0,H_1^\szero(C_1)]\times[0,H_2^\szero(C_2)]}\cbbm_{\theta_j\theta_k}(u_1,u_2;\btheta)du_1du_2.
\end{align*}
Thus, the conditional expectation is $A_{00}$ in Equation (\ref{equ:A00}). 
\end{itemize}

Combining all the four censoring statuses, we have
\begin{align*}
\Ezero_{(T_1,T_2)}[\Delta_{jk}|C_1,C_2] & =\iint_{\Omega_{11}\bigcup \Omega_{10}\bigcup \Omega_{01} \bigcup \Omega_{00}}  \cbbm_{\theta_j\theta_k}(u_1,u_2;\btheta) du_1du_2 .
\end{align*}
Again, due to the interchangeability between the integral and derivation, we have
$$
\Ezero_{(T_1,T_2)}[\Delta_{jk}|C_1,C_2]  =\frac{\partial^2}{\partial \theta_j\partial \theta_k}  \iint_{\Omega_{11}\bigcup \Omega_{10}\bigcup \Omega_{01} \bigcup \Omega_{00}}  \cbbm(u_1,u_2;\btheta) du_1du_2.
$$
Since $\Omega_{11}\bigcup \Omega_{10}\bigcup \Omega_{01} \bigcup \Omega_{00} =[0,1]^2$ and $\iint_{[0,1]^2}  \cbbm(u_1,u_2;\btheta) du_1du_2=1$ for any $\btheta$, we can show that  for any $\btheta$, $\Ezero_{(Y_1,Y_2)}[\Delta_{jk}|C_1,C_2]  = 0$ for $j,k=1,\cdots,p$. It implies that $A_{jk}(\btheta) = \Ezero_{(C_1,C_2)}\left\{\Ezero_{(Y_1,Y_2)}[\Delta_{jk}|C_1,C_2]\right\} = 0$. Furtermore, evaluating at $\btheta=\btheta^*$, we have $A_{jk}(\btheta^*)=0$, which proves Theorem \ref{thm:equivalence}.

\medskip
\paragraph{Expression of $A_{jk}(\btheta^*)$ under copula misspecification.} Again, we use the double expectation theorem in Equation (\ref{equ:Ajk}). If the assumed copula is misspecified, this conditional expectation $\Ezero_{(Y_1,Y_2)}[\Delta_{jk}|C_1,C_2]$ is taken w.r.t. $\Cbbm^\szero$.  Let $\cbbm_r^\szero=\frac{\partial \Cbbm^\szero(u_1,u_2)}{\partial u_r}$, $r=1,2$, and $\cbbm^\szero=\frac{\partial^2\Cbbm^\szero(u_1,u_2)}{\partial u_1\partial u_2}$. Following the above derivations under correct copula specification, we can show that
\begin{align*}
& \Ezero_{(Y_1,Y_2)}[\Delta_{jk}|C_1,C_2]  = \iint_{\Omega_{11}} \cbbm_{\theta_j\theta_k}(u_1,u_2;\btheta) w_{11}(u_1,u_2;\btheta) du_1du_2 \\
&  \quad \quad + \iint_{\Omega_{10}} \cbbm_{\theta_j\theta_k}(u_1,H_2^\szero(c_2);\btheta) w_{10}(u_1,H_2^\szero(c_2);\btheta) du_1\\
& \quad \quad + \iint_{\Omega_{01}} \cbbm_{\theta_j\theta_k}(H_1^\szero(c_1),u_2;\btheta) w_{10}(H_1^\szero(c_1),u_2;\btheta) du_2\\
& \quad \quad +  \iint_{\Omega_{00}}  \cbbm_{\theta_j\theta_k}(H_1^\szero(c_1),H_2^\szero(c_2);\btheta)w_{00}(H_1^\szero(c_1),H_2^\szero(c_2);\btheta) du_1du_2,
\end{align*}
where $w_{11}(u_1,u_2;\btheta)=\cbbm^\szero(u_1,u_2)/\cbbm(u_1,u_2;\btheta)$, $w_{10}(u_1,u_2;\btheta)=\cbbm_1^\szero(u_1,u_2)/\cbbm_1(u_1,u_2;\btheta)$, $w_{01}(u_1,u_2;\btheta)=\cbbm_2^\szero(u_1,u_2)/\cbbm_2(u_1,u_2;\btheta)$, and $w_{00}(u_1,u_2;\btheta)=\Cbbm^\szero(u_1,u_2)/\Cbbm(u_1,u_2;\btheta)$. \smallskip

By Definition \ref{defn:correct-spec}, when the assumed copula is misspecified, there exists some $(u_1,u_2)$ such that $w_{d_1,d_2}(u_1,u_2;\btheta^*)\neq 1$ for some $d_1,d_2=0,1$. Thus, for some $j,k=1,\cdots,p$,
$$
 \Ezero_{(Y_1,Y_2)}[\Delta_{jk}(\btheta^*,U_1^\szero,U_2^\szero)|C_1,C_2] \neq \frac{\partial^2}{\partial \theta_j\theta_k}\iint_{[0,1]^2} \cbbm(u_1,u_2;\btheta^*) du_1du_2.
 $$
Consequently, $A_{jk}(\btheta^*)\neq 0$ for some $j,k=1,\cdots,p$. 

\section{Proof of Theorem \ref{thm:convergence}}\label{app:Thm2}

To show $\left|R_n-tr\left[\bSast(\bthetaast)^{-1}\bVast(\bthetaast)\right]\right|=o_p(1)$, we need to first prove the consistency of  $\bShatn(\bthetahatn)$ and $\bVhatn(\bthetahatn)$. \cite{chen2010Estimation} has shown the consistency of $\bShatn(\bthetahatn)$ which requires conditions A2 and A4 (i) \& (ii) listed in their paper.  To prove the consistency of $\bVhatn(\bthetahatn)$, our conditions R5 and C5 are analogous to Chen et al.'s those two conditions, respectively. Thus, following the same arguments in their paper,  we can show that $\sup_{\btheta \in \Theta} n^{-1}\sum_{i=1}^n \, \allowbreak  \|\ell_{\btheta}(\btheta,\Uhat_{i1},\Uhat_{i2}) \ell_{\btheta}(\btheta,\Uhat_{i1},\Uhat_{i2})'I(X_{ir}\leq \eta)\|$ is asymptotically ignorable as $\eta \rightarrow 0$. This together with the continuity of $\ell_{\btheta}(\btheta,u_1,u_2)$ (our condition R5), and the consistency of the Kaplan-Meier estimate $\Hhat_r$ and the PMLE $\bthetahatn$, leads to $\|\bVhatn(\bthetahatn)-\bVast(\bthetaast)\|=o_p(1)$. 

Our condition C3 (ii) (which is equivalent to Condition A1 (ii) of \cite{chen2010Estimation}) ensures that $\bSast(\bthetaastn)$ is finite and non-singular. Thus, by Slutsky’s Theorem, it implies $tr\left[\bShatn(\bthetahatn)^{-1}\bVhatn(\bthetahatn)\right]$ converges $tr\left[\bSast(\bthetaast)^{-1}\bVast(\bthetaast)\right]$ in probability as $n\rightarrow \infty$.\\
\medskip
\noindent \underline{The proof of Theorem \ref{thm:convergence} ends.}

\section{Proof of Theorem \ref{thm:normal}}\label{app:Thm3}

To prove this theorem, we need to prove the following lemma:
\begin{lem}\label{lem:expansion-SV}
Under Conditions R1 - R6 and C1 - C7, 
\begin{enumerate}[(1)]
\item $\sqrt{n}\left\{\bShatn(\bthetahatn) - \bSast(\bthetaast) \right\} = \frac{1}{\sqrt{n}}\sumin {\bf h}_S(\bthetaast,X_{i1},X_{i2},\delta_{i1},\delta_{i2})+o_p(1)$, where ${\bf h}_S(\bthetaast,X_{i1},\,\allowbreak X_{i2},\delta_{i1},\delta_{i2})$ is a $p\times p$ matrix with the $(j,k)$-th element $h_{S_{jk}}(\bthetaast, X_{i1},X_{i2},\delta_{i1},\delta_{i2})$ given in Equation (\ref{equ:hS}) being independent random variables with mean 0.
\item $\sqrt{n}\left\{\bVhatn(\bthetahatn) - \bVast(\bthetaast) \right\} = \frac{1}{\sqrt{n}}\sumin {\bf h}_V(\bthetaast, X_{i1},X_{i2},\delta_{i1},\delta_{i2})$, where ${\bf h}_V(\bthetaast, X_{i1},X_{i2}, \, \allowbreak \delta_{i1},\delta_{i2})$ is a $p\times p$ matrix with the $(j,k)$-th element $h_{V_{jk}}(\bthetaast,X_{i1},X_{i2},\delta_{i1},\delta_{i2})$ given in Equation (\ref{equ:hV}) being independent random variables with mean 0. 
\end{enumerate}
\end{lem}

\paragraph{Proof of Lemma \ref{lem:expansion-SV}.} Let $\bSast_{jk}(\bthetaast)=\Ezero\left[-\ell_{\theta_j\theta_k}(\bthetaast,U_1,U_2)\right]$ denote the $(j,k)$-th element of $\bSast(\bthetaast)$. Similarly,  let $\bShatnjk(\bthetahatn)=- n^{-1}\sum_{i=1}^n \ell_{\theta_j\theta_k}(\bthetahatn,\Uhat_{i1},\Uhat_{i2})$ denote the $(j,k)$-th element of $\bShatn(\bthetahatn)$, $j,k=1,\cdots,p$.  By the mean-value theorem, we have
\begin{align*}
\bShatnjk(\bthetahatn)& = -n^{-1}\sum_{i=1}^n \ell_{\theta_j\theta_k}(\bthetaast,\Uhat_{i1},\Uhat_{i2}) + \left[-n^{-1}\sum_{i=1}^n \ell_{\theta_j\theta_k,\btheta}(\tilde \btheta,\Uhat_{i1},\Uhat_{i2})\right]'(\bthetahatn - \bthetaast),
\end{align*}
where $\tilde \btheta$ lies on the linear segment between $\bthetaast$ and $\bthetahatn$. 

Using the same arguments for the consistency of $\bShatn(\bthetahatn)$, by condition C6 (i) (analogous to Chen et al.'s condition A2) and condition C6 (ii) \& (iii) (analogous to Chen et al.'s Condition A4), we can show $\sup_{\btheta \in \Theta} n^{-1}\sum_{i=1}^n \|\ell_{\theta_j\theta_k,\btheta}(\btheta,\Uhat_{i1},\Uhat_{i2})I(X_{ir}\leq \eta)\|$ is asymptotically ignorable as $\eta \rightarrow 0$. This together with the continuity of $\ell_{\theta_j\theta_k,\btheta}$ (in our condition C6 (i)) and the consistency of the Kaplan-Meier estimate and PMLE $\bthetahatn$, we can show that $\|n^{-1}\sum_{i=1}^n \ell_{\theta_j\theta_k,\btheta}(\tilde \btheta,\Uhat_{i1},\Uhat_{i2})-\Ezero\left[\ell_{\theta_j\theta_k,\btheta}(\bthetaast,U_{1}^\szero,U_{2}^\szero)\right]\|=o_p(1)$. Let $\bM_{jk}(\bthetaast) = \Ezero\left[\ell_{\theta_j\theta_k,\btheta}(\bthetaast,U_{1}^\szero,U_{2}^\szero)\right]$ (a $p\times1$ vector). Because $\bthetahatn$ is $\sqrt{n}$-consistent, we have
\begin{equation}\label{equ:S-theta}
\bShatn(\bthetahatn)_{jk} =  -n^{-1}\sum_{i=1}^n \ell_{\theta_j\theta_k}(\bthetaast,\Uhat_{i1},\Uhat_{i2}) - \bM_{jk}(\bthetaast)'(\bthetahatn-\bthetaast) + o_p(n^{-1/2}).
\end{equation}
Again applying the mean-value theorem on Equation (\ref{equ:S-theta}), we have
\begin{align}
\nonumber & \bShatnjk(\bthetahatn) - \bSast_{jk}(\bthetaast) \\
\nonumber = & n^{-1}\sum_{i=1}^n \left[-\ell_{\theta_j\theta_k}(\bthetaast,U_{i1}^\szero,U_{i2}^\szero) -\bSast_{jk}(\bthetaast)\right]  - n^{-1}\sum_{r=1}^2\sum_{i=1}^n \ell_{\theta_j\theta_k,u_r}(\bthetaast,\tilde U_{i1},\tilde U_{i2}) (\Uhat_{ir}-U_{ir}^\szero) \\
\label{equ:S-exp}& \quad - \bM_{jk}(\bthetaast)'(\bthetahatn - \bthetaast) + o_p(n^{-1/2})
\end{align}
where $(\tilde U_{i1},\tilde U_{i2})$ lies on the linear segment between $(\Uhat_{i1},\Uhat_{i2})$ and $(U_{i1}^\szero,U_{i2}^\szero)$. 

Based on the expansion of $\bthetahatn$ around $\bthetaast$ in \cite{chen2010Estimation}, we have
\begin{align}
\nonumber&  \bthetahatn-\bthetaast  \\
\label{equ:theta-exp}  = & \bSast(\bthetaast)^{-1}\frac{1}{n}\sumin \left[\ell_{\btheta}(\bthetaast,U_{i1}^\szero,U_{i2}^\szero) + W_1(\bthetaast,X_{i1},\delta_{i1}) + W_2(\bthetaast,X_{i2},\delta_{i2})\right]+ o_p(n^{-1/2}) \end{align}
where for $r=1,2$, 
\begin{equation}\label{equ:Wr}
W_r(\bthetaast,X_{ir},\delta_{ir}) = \Ezero\left[\ell_{\btheta, u_r}(\bthetaast, U_1^\szero,U_2^\szero)I_{ir}(X_r)\mid X_{ir},\delta_{ir}\right] 
\end{equation}
with
$$
I_{ir}(X_r) = - H_r(X_r)\left[\int_{-\infty}^{X_{r}}\frac{dN_{ir}(u)}{P_{n,r}(u)} - \int_{-\infty}^{X_r}\frac{I(X_{ir}\geq u)d\Lambda_r^\szero(u)}{P_{n,r}(u)}\right]
$$
with $\Lambda^\szero_r(u) = -\log H_r^\szero(u)$, the true cumulative hazard function of $T_{ir}$, $N_{ir}(u)=\delta_{ir}I(X_{ir}\leq u)$, $dN_{ir}(u) = N_{ir}(u) - N_{ir}(u-)$, and $P_{n,r}(u) = n^{-1}\sum_{k=1}^n Pr(X_{kr}\geq u) $. Using similar arguments for obtaining Equation (\ref{equ:theta-exp}), under our condition R6 (iii) and C7 (i) (analogous to Chen et al.'s condition A3 (i) \& (ii)) and condition C6 (ii)  (analogous to Chen et al.'s Condition A4 (i)), we can show that 
\begin{equation}\label{equ:U-exp}
n^{-1}\sum_{i=1}^n \ell_{\theta_j\theta_k,u_r}(\bthetaast,\tilde U_{i1},\tilde U_{i2}) (\Uhat_{ir}-U_{ir}^\szero) =  n^{-1}\sumin h_{S_{jk},r}(\bthetaast,X_{ir},\delta_{ir})
+ o_p(n^{-1/2}),
\end{equation}
where $h_{S_{jk},r}(\bthetaast,X_{ir},\delta_{ir}) = \Ezero\left[\ell_{\theta_j\theta_k,u_r}(\bthetaast,U_{1}^\szero,U_{2}^\szero)I_{ir}(X_{r})\mid X_{ir},\delta_{ir} \right]$.

From Equations (\ref{equ:S-exp}), (\ref{equ:theta-exp}), and (\ref{equ:U-exp}), we have 
$$
\sqrt{n}\left\{\bShatnjk(\bthetahatn) - \bSast_{jk}(\bthetaast) \right\} = \frac{1}{\sqrt{n}}\sumin h_{S_{jk}}(\bthetaast,X_{i1},X_{i2},\delta_{i1},\delta_{i2})+o_p(1),
$$ 
where
\begin{align}
\nonumber & h_{S_{jk}}(\bthetaast,X_{i1},X_{i2},\delta_{i1},\delta_{i2})\\
\nonumber = & \left[-\ell_{\theta_j\theta_k}(\bthetaast,U_{i1}^\szero,U_{i2}^\szero) -\bSast_{jk}(\bthetaast)\right]  - h_{S_{jk},1}(\bthetaast,X_{i1},\delta_{i1}) - h_{S_{jk},2}(\bthetaast,X_{i2},\delta_{i2}) \\
\label{equ:hS}
 & \quad -\bM_{jk}(\bthetaast)'\bSast(\bthetaast)^{-1}\left[\ell_{\btheta}(\bthetaast,U_{i1},U_{i2}) + W_1(\bthetaast,X_{i1},\delta_{i1}) + W_2(\bthetaast,X_{i2},\delta_{i2})\right]
\end{align}

Let $\bVast_{jk}(\bthetaast)$ and $\bVhatnjk(\bthetaast)$ denote the $(j,k)$-th element of $\bVast(\bthetaast)$ and $\bVhatn(\bthetaast)$. 
We apply the same techniques above, with our condition C5 (i) (analogous to Chen et al.'s condition A4 (i)) and condition C7 (ii) \& (iii) (analogous to Chen et al.'s condition A3 (i) \& (ii)), we can show
$$
\sqrt{n}\left\{\bVhatnjk(\bthetahatn) - \bVast_{jk}(\bthetaast) \right\} = \frac{1}{\sqrt{n}}\sumin h_{V_{jk}}(\bthetaast,X_{i1},X_{i2},\delta_{i1},\delta_{i2})+o_p(1),
$$ 
where 
\begin{align}
\nonumber & h_{V_{jk}}(\bthetaast,X_{i1},X_{i2},\delta_{i1},\delta_{i2})\\
\nonumber = & \left[\ell_{\theta_j}(\bthetaast,U_{i1}^\szero,U_{i2}^\szero)\ell_{\theta_k}(\bthetaast,U_{i1}^\szero,U_{i2}^\szero)' - \bVast_{jk}(\bthetaast)\right] + h_{V_{jk},1}(\bthetaast,X_{i1},\delta_{i1}) + h_{V_{jk},2}(\bthetaast,X_{i2},\delta_{i2}) \\
\label{equ:hV}
 & \quad + \bP_{jk}(\bthetaast)'\bSast(\bthetaast)^{-1}\left[\ell_{\btheta}(\bthetaast,U_{i1}^\szero,U_{i2}^\szero) + W_1(\bthetaast,X_{i1},\delta_{i1}) + W_2(\bthetaast,X_{i2},\delta_{i2})\right]
\end{align}
with 
\begin{align*}
h_{V_{jk},r}(\bthetaast,X_{ir},\delta_{ir})  & = \Ezero\left\{\left[\ell_{\theta_j,u_r}(\bthetaast,U_{1}^\szero,U_{2}^\szero)\ell_{\theta_k}(\bthetaast,U_{1}^\szero,U_{2}^\szero) \right. \right. \\
& \quad +\left.\left. \ell_{\theta_k,u_r}(\bthetaast,U_{1}^\szero,U_{2}^\szero)\ell_{\theta_j}(\bthetaast,U_{1}^\szero,U_{2}^\szero)\right] \, \allowbreak \ast I_{ir}(X_{r})\mid X_{ir},\delta_{ir} \right\}
\end{align*}
and $\bP_{jk}(\bthetaast) = \Ezero[\ell_{\theta_j,\btheta}(\bthetaast,U_{1}^\szero,U_{2}^\szero)\ell_{\theta_k}(\bthetaast,U_{1}^\szero,U_{2}^\szero) + \ell_{\theta_k,\btheta}(\bthetaast,U_{1}^\szero,U_{2}^\szero)\ell_{\theta_j}(\bthetaast,\, \allowbreak U_{1}^\szero,U_{2}^\szero)]$. \medskip

\noindent \underline{The proof of Lemma \ref{lem:expansion-SV} ends.}
\medskip

\noindent \textbf{Proof of Theorem \ref{thm:normal}}:
Under the null hypothesis that the assumed copula function is correctly specified, $\bR^*(\btheta^*)=\bSast(\btheta^*)^{-1}\bVast(\btheta^*)=I_p$ due to Theorem \ref{thm:equivalence}, and consequently, by Theorem \ref{thm:convergence}, $R_n\rightarrow p=tr\left[I_p\right]$ in probability as $n\rightarrow \infty$. In addition, $R_n-p$ can be expressed as $R_n-p=tr[\bShatn(\bthetahatn)^{-1}\bVhatn(\bthetahatn)-\bSast(\bthetaast)^{-1}\bVast(\bthetaast)]$. With algebraic derivations, we have 
\begin{align}
\nonumber & \sqrt{n}(R_n-p) = \sqrt{n}tr\left[\bShatn(\bthetahatn)^{-1}\bVhatn(\bthetahatn) -\bSast(\bthetaast)^{-1}\bVast(\bthetaast)\right]\\
\nonumber = & tr\left[\bSast(\bthetaast)^{-1}\sqrt{n}\left\{\bVhatn(\bthetahatn)-\bVast(\bthetaast)\right\}\right] \\
\nonumber & +   tr\left[\bSast(\bthetaast)^{-1}\bVast(\bthetaast)\bSast(\bthetaast)^{-1}\sqrt{n}\left\{\bSast(\bthetaast) - \bShatn(\bthetahatn)\right\}\right] \\
\label{equ:Rn-exp}& + tr\left[\left\{\bShatn(\bthetahatn)^{-1}\bVhatn(\bthetahatn) - \bSast(\bthetaast)^{-1}\bVast(\bthetaast)\right\}\bSast(\bthetaast)^{-1}\sqrt{n}\left\{\bSast(\bthetaast) - \bShatn(\bthetahatn)\right\}\right]
\end{align}%
Under the null hypothesis, $\bSast(\bthetaast)=\bVast(\bthetaast)$, the second term in Equation (\ref{equ:Rn-exp}) becomes $tr[\bSast(\bthetaast)^{-1}\sqrt{n}\left\{\bSast(\bthetaast) - \bShatn(\bthetahatn)\right\}]$. The third term in Equation (\ref{equ:Rn-exp}) is $o_p(1)$ because $\|\bShatn(\bthetahatn)^{-1} \bVhatn(\bthetaast) - \bSast(\bthetaast)^{-1}\,\allowbreak\bVast(\bthetaast)\|=o_p(1)$ shown in the  proof of Theorem \ref{thm:convergence} (Appendix \ref{app:Thm2}) and $\|\bShatn(\bthetahatn) -\bSast(\bthetaast) \| =O_p(n^{-1/2})$ by Lemma \ref{lem:expansion-SV}. Thus, we can write 
$$
\sqrt{n}(R_n-p) = \frac{1}{\sqrt{n}}\sumin h_R(X_{i1},X_{i2},\delta_{i1},\delta_{i2},\btheta) + o_p(1),
$$
where 
\begin{align}
\label{equ:hR}h_R(\bthetaast,X_{i1},X_{i2},\delta_{i1},\delta_{i2}) & = tr\left[\bSast(\bthetaast)^{-1}\left\{h_V(\bthetaast, X_{i1},X_{i2},\delta_{i1},\delta_{i2}) -  h_S(\bthetaast, X_{i1},X_{i2},\delta_{i1},\delta_{i2})\right\}\right].
\end{align}
By Central Limit Theorem for independent random variables, we can show that $\sqrt{n}(R_n-p)$ converges in distribution to a normal random variable with mean 0 and variance $\sigma^2=Var[h_R(X_{i1},X_{i2},\delta_{i1},\delta_{i2},\btheta)]$.\\
\medskip
\underline{The proof of Theorem \ref{thm:normal} ends.}

\section{Proof of Theorem \ref{thm:IR-PIOS}}\label{app:Thm4}

To prove this theorem, we need to first prove the following lemma:

\begin{lem}\label{lem:theta-os}
Under Condition R1 - R6 and C1 - C4, $\sup_{1\leq i \leq n}\|\bthetahatn - \bthetahat_{(-i)}\| = O_p(n^{-1})$.
\end{lem}

\paragraph{Proof of Lemma \ref{lem:theta-os}}
The "out-of-sample" PMLE $\bthetahat_{(-i)}$ is obtained by maximizing $\sum_{s=1,s\neq i}^n \ell(\btheta,\, \allowbreak \Uhat_{s1},\Uhat_{s2})$, i.e., $\sum_{s=1,s\neq i}^n \ell_{\btheta}(\bthetahat_{-(i)},\Uhat_{s1},\Uhat_{s2})=0$. Apply the mean-value theorem, we have
\begin{align*}
0 & = \sum_{s=1,s\neq i}^n \ell_{\btheta}(\bthetahat_{-(i)},\Uhat_{s1},\Uhat_{s2})  = \sum_{s=1,s\neq i}^n \ell_{\btheta}(\bthetahatn,\Uhat_{s1},\Uhat_{s2}) +  \sum_{s=1,s\neq i}^n \ell_{\btheta\btheta}(\widetilde \btheta,\Uhat_{s1},\Uhat_{s2})(\bthetahat_{(-i)}-\bthetahat_n) \\
& = \sum_{s=1}^n \ell_{\btheta}(\bthetahatn,\Uhat_{s1},\Uhat_{s2}) - \ell_{\btheta}(\bthetahatn,\Uhat_{i1},\Uhat_{i2})+  \sum_{s=1,s\neq i}^n \ell_{\btheta\btheta}(\widetilde \btheta,\Uhat_{s1},\Uhat_{s2})(\bthetahat_{(-i)}-\bthetahat_n)
\end{align*}
where $\widetilde\btheta$ lies in the linear segment between $\bthetahat_{(-i)}$ and $\bthetahatn$. Since $ \sum_{s=1}^n \ell_{\btheta}(\bthetahatn,\Uhat_{s1},\Uhat_{s2}) =0$ ($\bthetahatn$ is the PMLE using all the observations), we have
\begin{equation}\label{equ:theta-os-exp}
\bthetahatn - \bthetahat_{(-i)} = \widehat{\bS}_{(-i)}(\widetilde\btheta)^{-1} n^{-1} \ell_{\btheta}(\bthetahatn,\Uhat_{i1},\Uhat_{i2})
\end{equation}
where $\widehat{\bS}_{(-i)}(\widetilde\btheta)=-n^{-1}\sum_{s=1,s\neq i}^n \ell_{\btheta\btheta}(\widetilde \btheta,\Uhat_{s1},\Uhat_{s2})$. Thus,
$$
\sup_{1\leq i \leq n}\|\bthetahatn - \bthetahat_{(-i)}\| \leq n^{-1} \sup_{1\leq i \leq n} \left\|\widehat{\bS}_{(-i)}(\widetilde\btheta)^{-1}\right\| \times \sup_{1\leq i \leq n} \left\|\ell_{\btheta}(\bthetahatn,\Uhat_{i1},\Uhat_{i2})\right\|.
$$

Using the same arguments for proving the consistency of $\bShatn(\bthetahat)$, we can prove that as $n\rightarrow \infty$, $
\widehat{\bS}_{(-i)}(\widetilde\btheta)\rightarrow \bSast(\bthetaast)$ in probability. Our condition C3 (ii) (equivalent to Chen et al.'s Condition A1 (ii))  assumes the boundedness for the eigenvalues of $\bSast(\bthetaast)$, which ensures that $\sup_{1\leq i \leq n} \left\|\widehat{\bS}_{(-i)}(\widetilde\btheta)^{-1}\right\|<\infty$. In addition, our condition R6 (i) \& (ii) (equivalent to Chen et al.'s Condition A3) ensures that $\sup_{1\leq i \leq n} \|\ell_{\btheta}(\bthetahatn,\Uhat_{i1},\Uhat_{i2})\| = O_p(1)$, and thus, $\sup_{1\leq i\leq n} \|n^{-1} \ell_{\btheta}(\bthetahatn,\, \allowbreak \Uhat_{i1},\Uhat_{i2})\|=O_p(1)$. It leads to  $\sup_{1\leq i \leq n}\|\bthetahatn - \bthetahat_{(-i)}\| = O_p(n^{-1})$.\\
\\
\noindent \underline{The proof of Lemma \ref{lem:theta-os} ends.} 

\paragraph{Proof of Theorem \ref{thm:IR-PIOS}.} Recall that the PIOS test statistic is defined as 
$$
T_n = \sum_{i=1}^n \ell(\bthetahatn,\Uhat_{i1},\Uhat_{i2}) - \sum_{i=1}^n \ell(\bthetahat_{(-i)},\Uhat_{i1},\Uhat_{i2}). 
$$
Applying the mean value theorem on $\ell(\bthetahat_{(-i)},\Uhat_{i1},\Uhat_{i2})$, we have
\begin{align*}
T_n & =- \sum_{i=1}^n \ell_{\btheta}(\bthetahatn,\Uhat_{i1},\Uhat_{i2})' \left(\bthetahat_{(-i)}-\bthetahatn\right) -\frac{1}{2} \sum_{i=1}^n \ell_{\btheta\btheta}(\breve{\btheta},\Uhat_{i1},\Uhat_{i2}) \left(\bthetahat_{(-i)}-\bthetahatn\right)^2,
\end{align*}
where $\breve{\btheta}$ lies on the linear segment between $\bthetahat_{(-i)}$ and $\bthetahatn$. Plugging in Equation (\ref{equ:theta-os-exp}), we have
\begin{align*}
T_n &= n^{-1}  \sum_{i=1}^n \ell_{\btheta}(\bthetahatn,\Uhat_{i1},\Uhat_{i2}) '\left\{\widehat{\bS}_{(-i)}(\widetilde\btheta)\right\}^{-1} \ell_{\btheta}(\bthetahatn,\Uhat_{i1},\Uhat_{i2})  \\
& \quad  \quad \quad -\frac{1}{2} \sum_{i=1}^n \ell_{\btheta\btheta}(\breve{\btheta},\Uhat_{i1},\Uhat_{i2}) \left(\bthetahat_{(-i)}-\bthetahatn\right)^2\\
& = tr\left[\widehat{\bS}_{(-i)}(\widetilde\btheta)^{-1} \left\{n^{-1}\sum_{i=1}\ell_{\btheta}(\bthetahatn,\Uhat_{i1},\Uhat_{i2}) \ell_{\btheta}(\bthetahatn,\Uhat_{i1},\Uhat_{i2})'\right\}\right] \\
& \quad  \quad \quad -\frac{1}{2} \left\{n^{-1}\sum_{i=1}^n \ell_{\btheta\btheta} (\breve{\btheta},\Uhat_{i1},\Uhat_{i2}) \right\}n\left(\bthetahat_{(-i)}-\bthetahatn\right)^2.
\end{align*}
Thus, 
\begin{align}
\nonumber & T_n - R_n \\
\label{equ:Tn-Rn} = & tr\left[\left\{\widehat{\bS}_{(-i)}(\widetilde\btheta)^{-1}-\bShatn(\bthetahatn)^{-1}\right\}\bVhatn(\bthetahat)\right]-\frac{1}{2} \left\{n^{-1}\sum_{i=1}^n \ell_{\btheta\btheta} (\breve{\btheta},\Uhat_{i1},\Uhat_{i2}) \right\}n\left(\bthetahat_{(-i)}-\bthetahatn\right)^2.
\end{align}
In the proof of  Lemma \ref{lem:theta-os}, we have shown that $\|\widehat{\bS}_{(-i)}(\widetilde\btheta)-\bSast(\bthetaast)\|=o_p(1)$. In addition, because $\|\bShatn(\bthetahatn)-\bSast(\bthetaast)\|=o_p(1)$, we have $\|\widehat{\bS}_{(-i)}(\widetilde\btheta)-\bShatn(\bthetahatn)\|=o_p(1)$, and consequently, the first term in Equation (\ref{equ:Tn-Rn}) is $o_p(1)$. For the second term, following similar arguments, we can show $\|n^{-1}\sum_{i=1}^n \ell_{\btheta\btheta} (\breve{\btheta},\Uhat_{i1},\Uhat_{i2})-\bSast(\bthetaast)\|=o_p(1)$. Together with $\sup_{1\leq i \leq n}\|\bthetahatn - \bthetahat_{(-i)}\| = O_p(n^{-1})$, the second term is $O_p(n^{-1})$. Combining the two terms, we have $|T_n - R_n|=o_p(1)$.\\
\\
\noindent \underline{The proof of Theorem \ref{thm:IR-PIOS} ends}.

\end{document}



\title{Supplementary Material of \\
Information matrix equivalence in the presence of censoring: A goodness-of-fit test for semiparametric copula models with multivariate survival data}
\author{Qian M. Zhou\\
\small Department of Mathematics and Statistics, Mississippi State University,\\
\small MS, USA, qz70@msstate.edu}
\date{}
\def \bS{\textbf{S}}
\def \bV{\textbf{V}}
\def \bSast{\bS^\ast}
\def \bVast{\bV^\ast}
\def \szero{{\mbox{\tiny 0}}}
\def \Ezero{\mathbbm{E}^\szero}

\maketitle

\section{Simulation Results}

In this section, we present the following results of the simulation study (Section 5 of the manuscript).

\begin{figure}
\centering
\caption{Estimated pseudo-observations $(\widehat U_{i1}, \widehat U_{i2})$ from one simulated bivariate censored survival data of sample size $n=100$ generated from each of the four copula families: Clayton, Frank, Joe, and Gaussian with Kendall's $\tau=0.3$.}
\label{fig:U_n100_tau0.3}
\vskip 0.3cm
\includegraphics[width=1\textwidth]{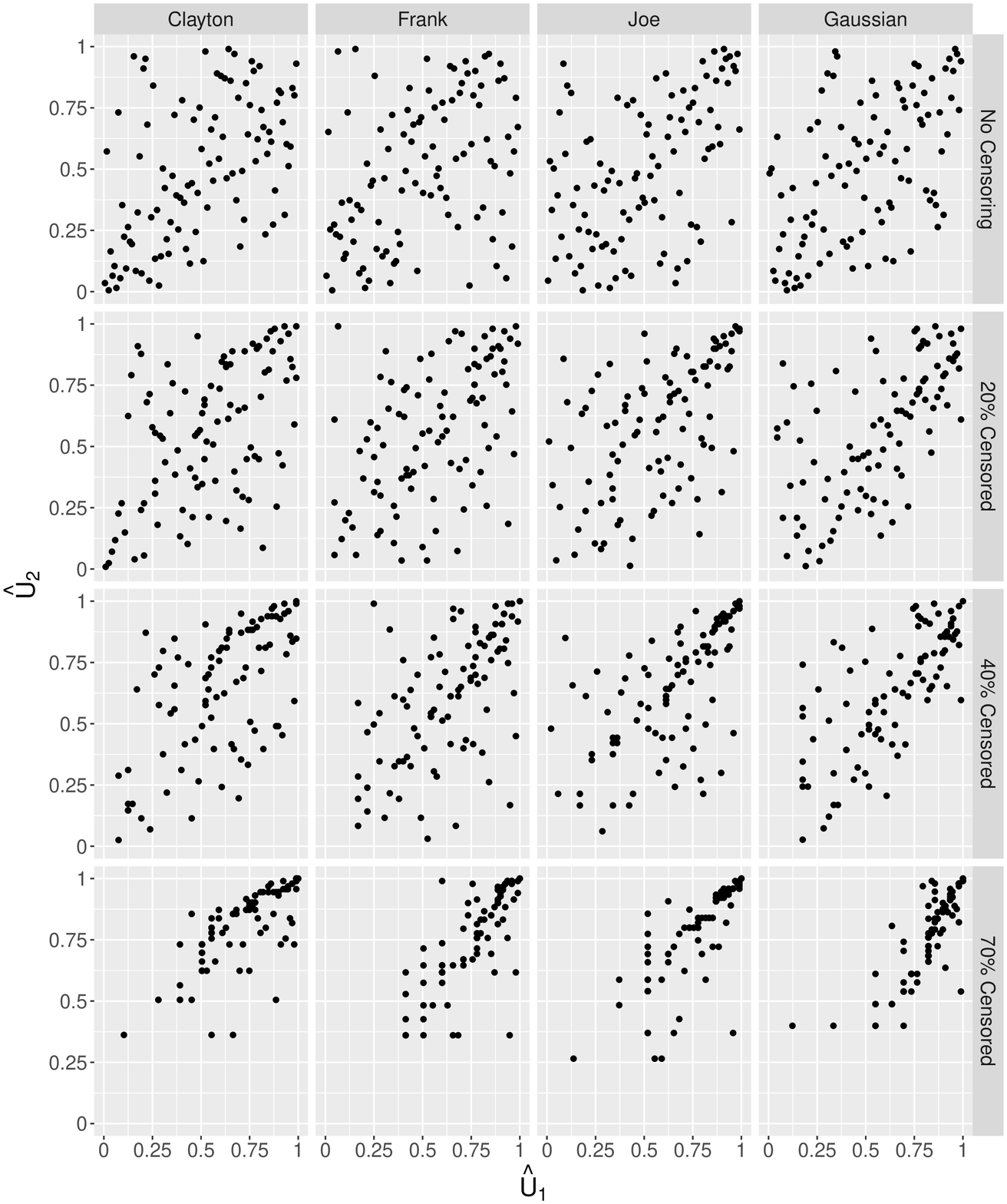}
\end{figure}

\begin{figure}
\centering
\caption{Estimated pseudo-observations $(\widehat U_{i1}, \widehat U_{i2})$ from one simulated bivariate censored survival data of sample size $n=100$ generated from each of the four copula families: Clayton, Frank, Joe, and Gaussian with Kendall's $\tau=0.7$.}
\label{fig:U_n100_tau0.7}
\vskip 0.3cm
\includegraphics[width=1\textwidth]{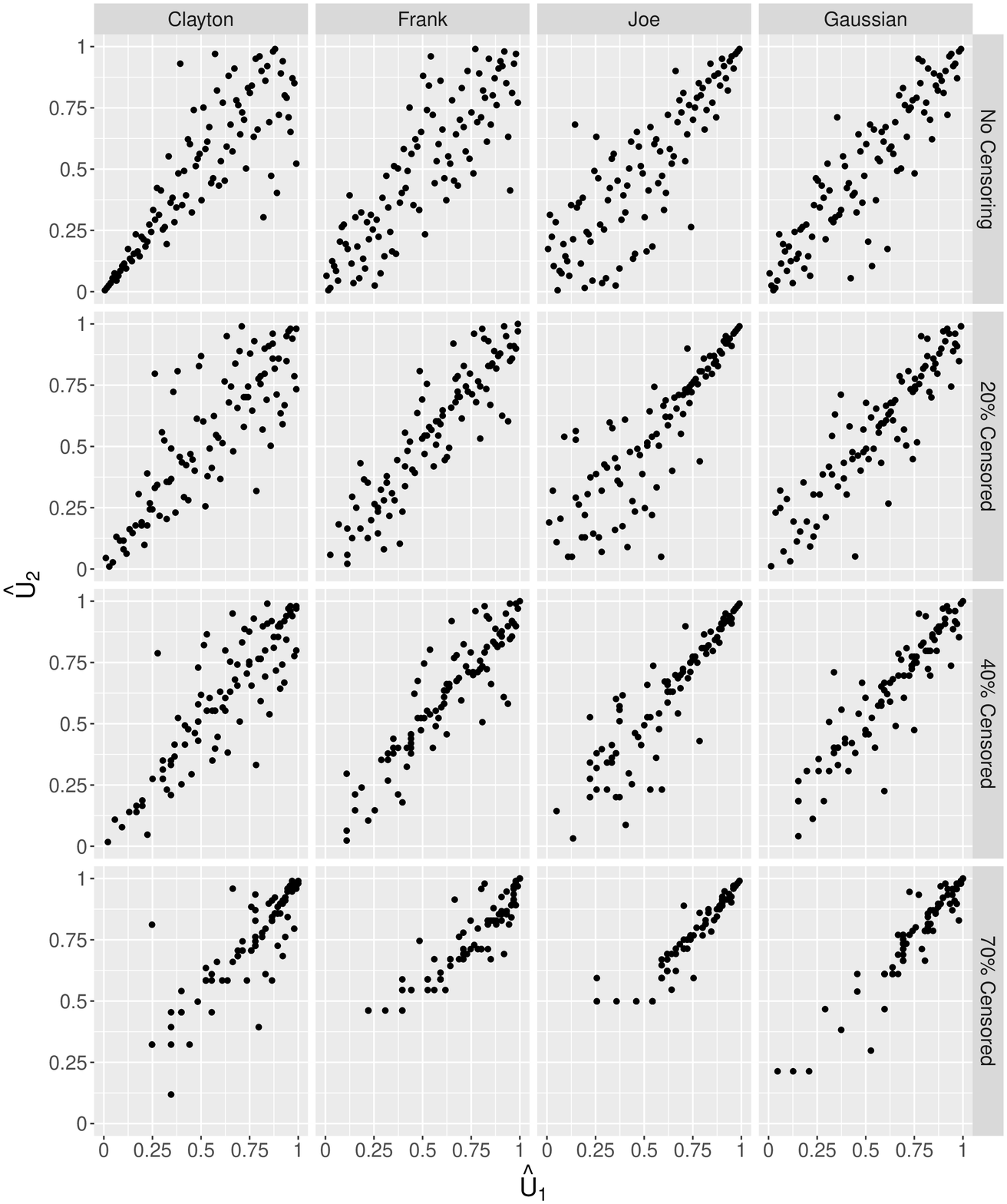}
\end{figure}

\begin{figure}
\centering
\caption{Estimated pseudo-observations $(\widehat U_{i1}, \widehat U_{i2})$ from one simulated bivariate censored survival data of sample size $n=600$ generated from each of the four copula families: Clayton, Frank, Joe, and Gaussian with Kendall's $\tau=0.3$.}
\label{fig:U_n600_tau0.3}
\vskip 0.3cm
\includegraphics[width=1\textwidth]{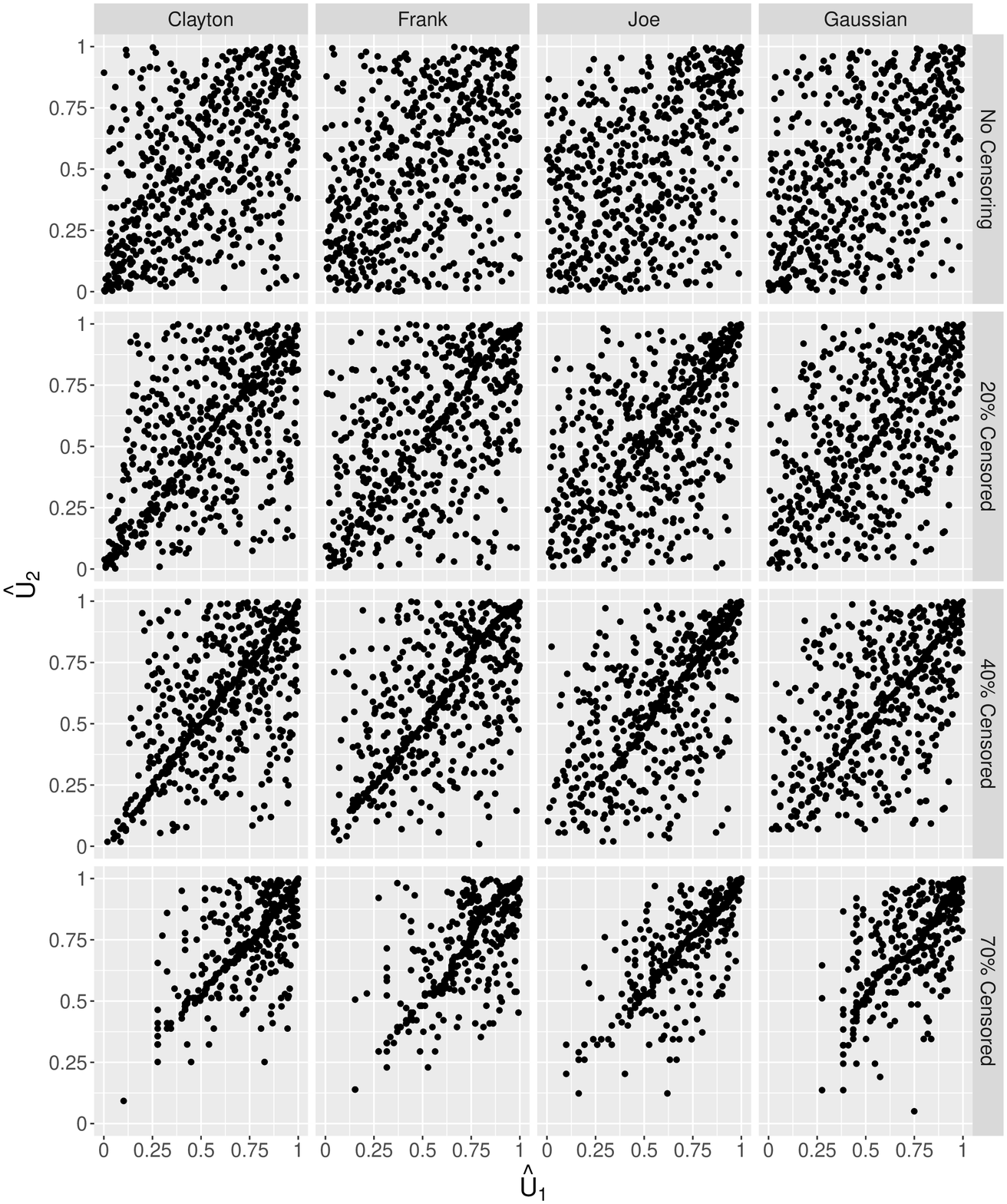}
\end{figure}

\begin{figure}
\centering
\caption{Estimated pseudo-observations $(\widehat U_{i1}, \widehat U_{i2})$ from one simulated bivariate censored survival data of sample size $n=600$ generated from each of the four copula families: Clayton, Frank, Joe, and Gaussian with Kendall's $\tau=0.7$.}
\label{fig:U_n600_tau0.7}
\vskip 0.3cm
\includegraphics[width=1\textwidth]{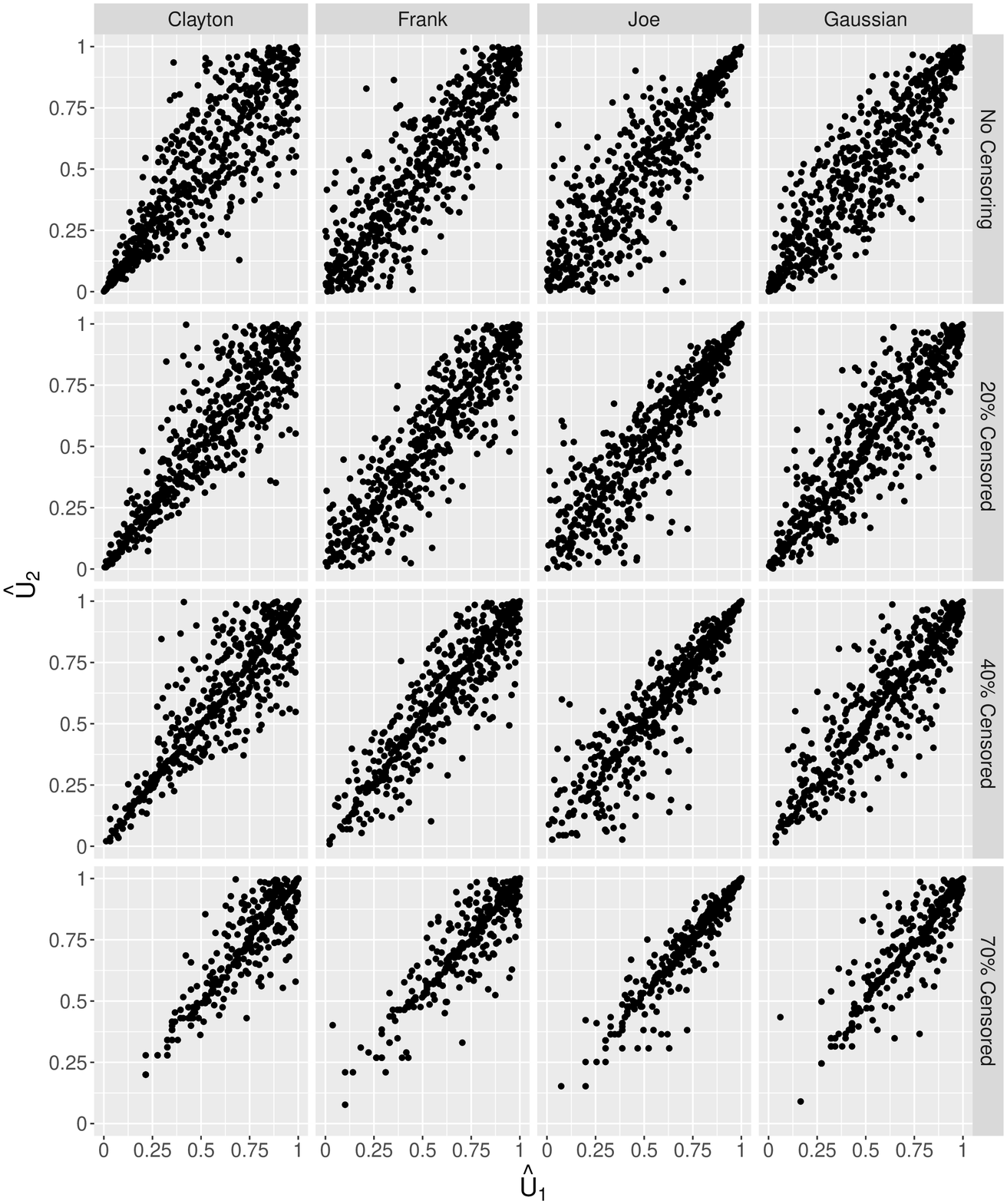}
\end{figure}

\begin{figure}
\centering
\caption{Normal QQ plots of 500 replications of the IR and PIOS statistics when the true and null copulas are both \textbf{Clayton} with Kendall's $\tau=0.5$.}\label{fig:QQ_clayton}
\vskip 0.3cm
\includegraphics[width=1\textwidth]{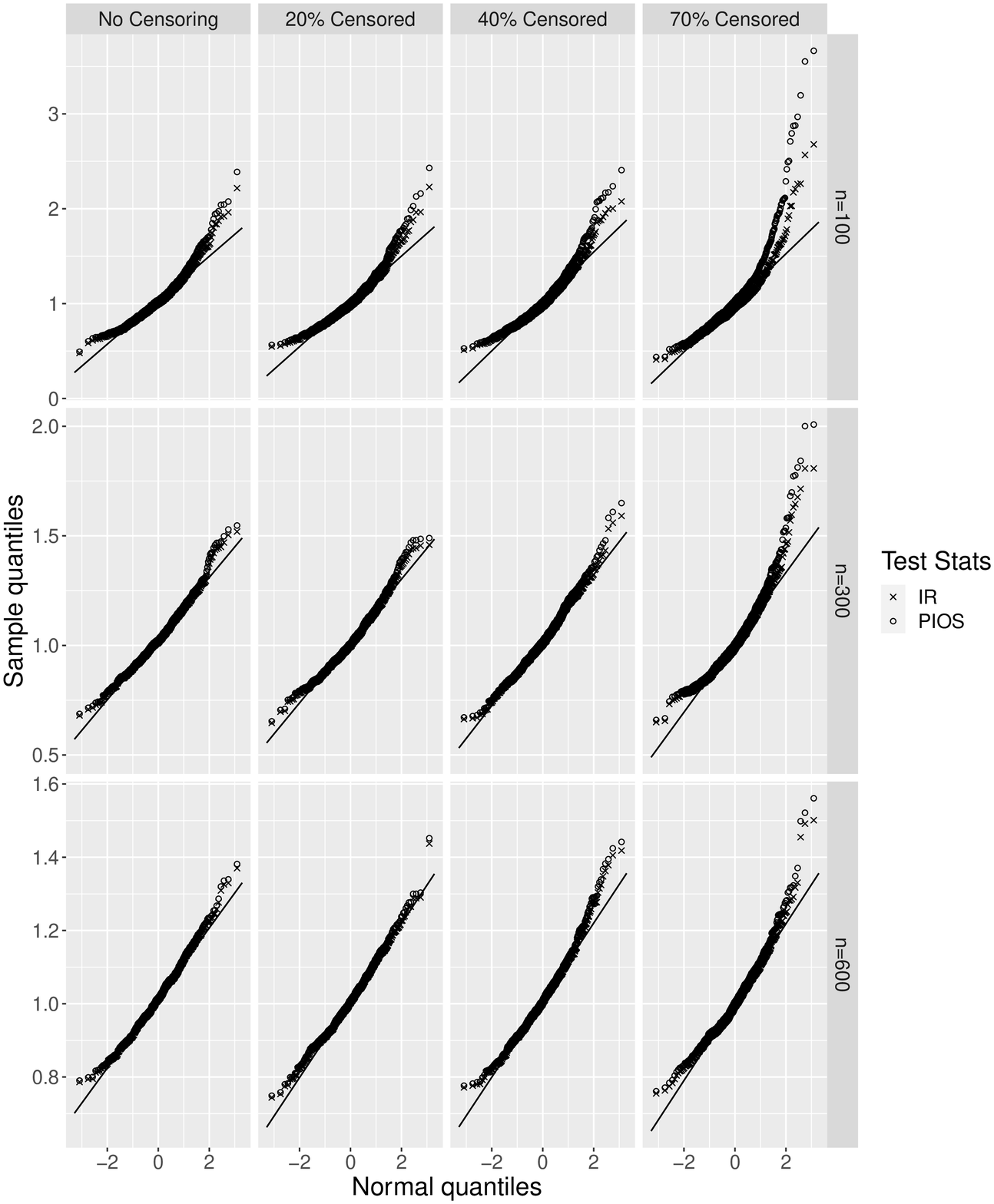}
\end{figure}

\begin{figure}
\centering
\caption{Normal QQ plots of 500 replications of the IR and PIOS statistics when the true and null copulas are both \textbf{Frank} with Kendall's $\tau=0.5$.}\label{fig:QQ_frank}
\vskip 0.3cm
\includegraphics[width=1\textwidth]{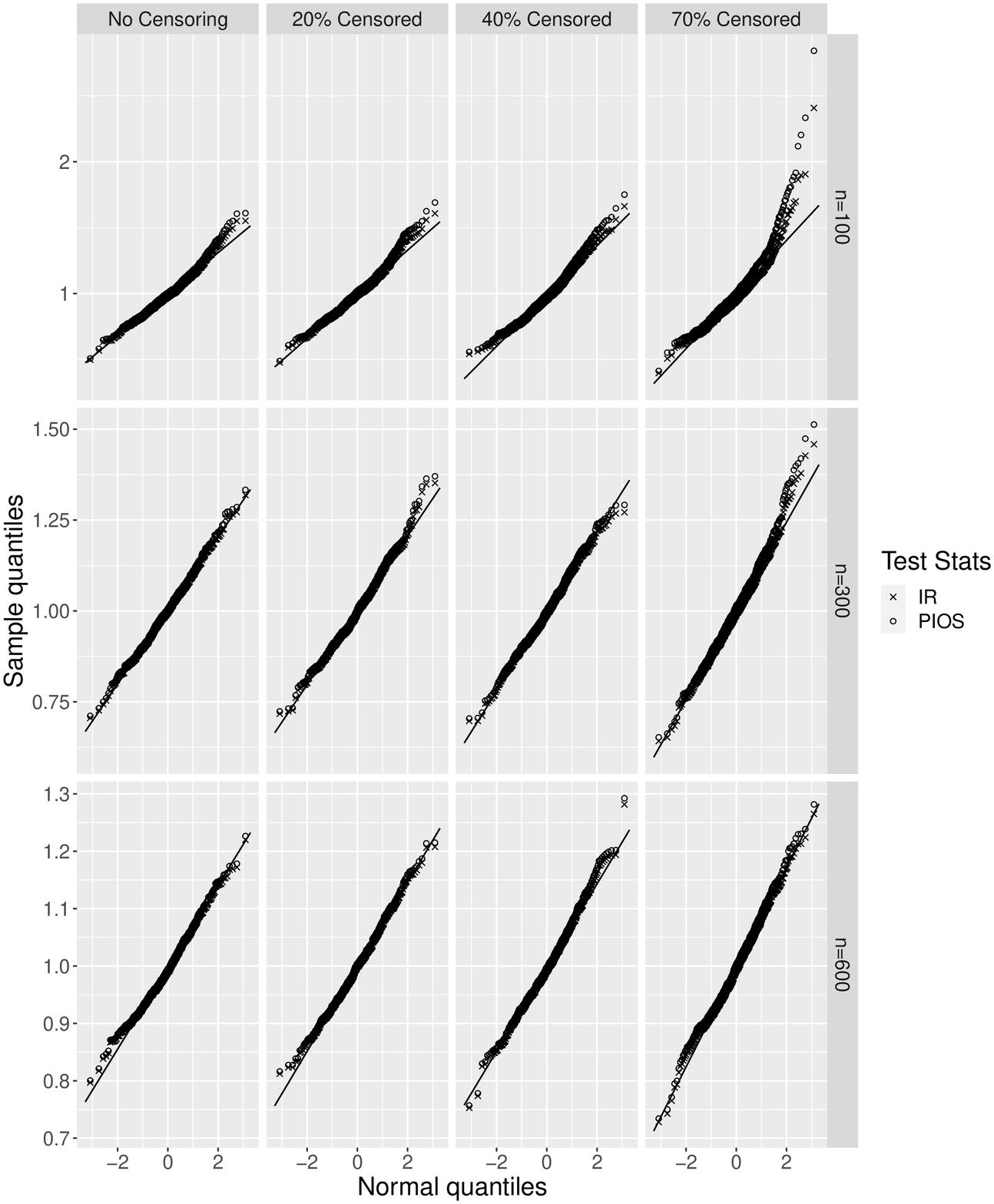}
\end{figure}

\begin{figure}
\centering
\caption{Normal QQ plots of 500 replications of the IR and PIOS statistics when the true and null copulas are both \textbf{Joe} with Kendall's $\tau=0.5$.}\label{fig:QQ_joe}
\vskip 0.3cm
\includegraphics[width=1\textwidth]{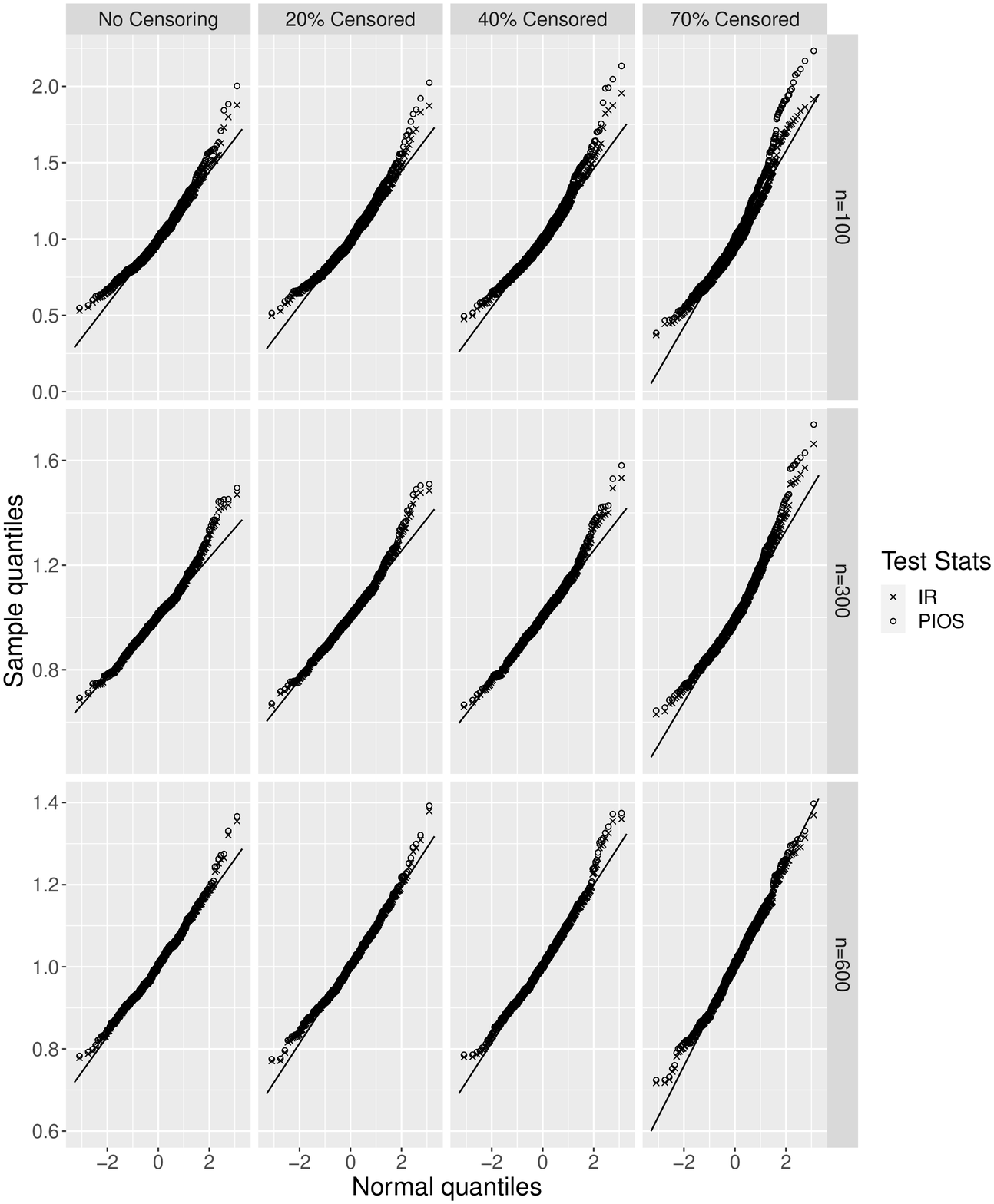}
\end{figure}

\begin{figure}
\centering
\caption{Normal QQ plots of 500 replications of the IR and PIOS statistics when the true and null copulas are both \textbf{Gaussian} with Kendall's $\tau=0.5$.}\label{fig:QQ_normal}
\vskip 0.3cm
\includegraphics[width=1\textwidth]{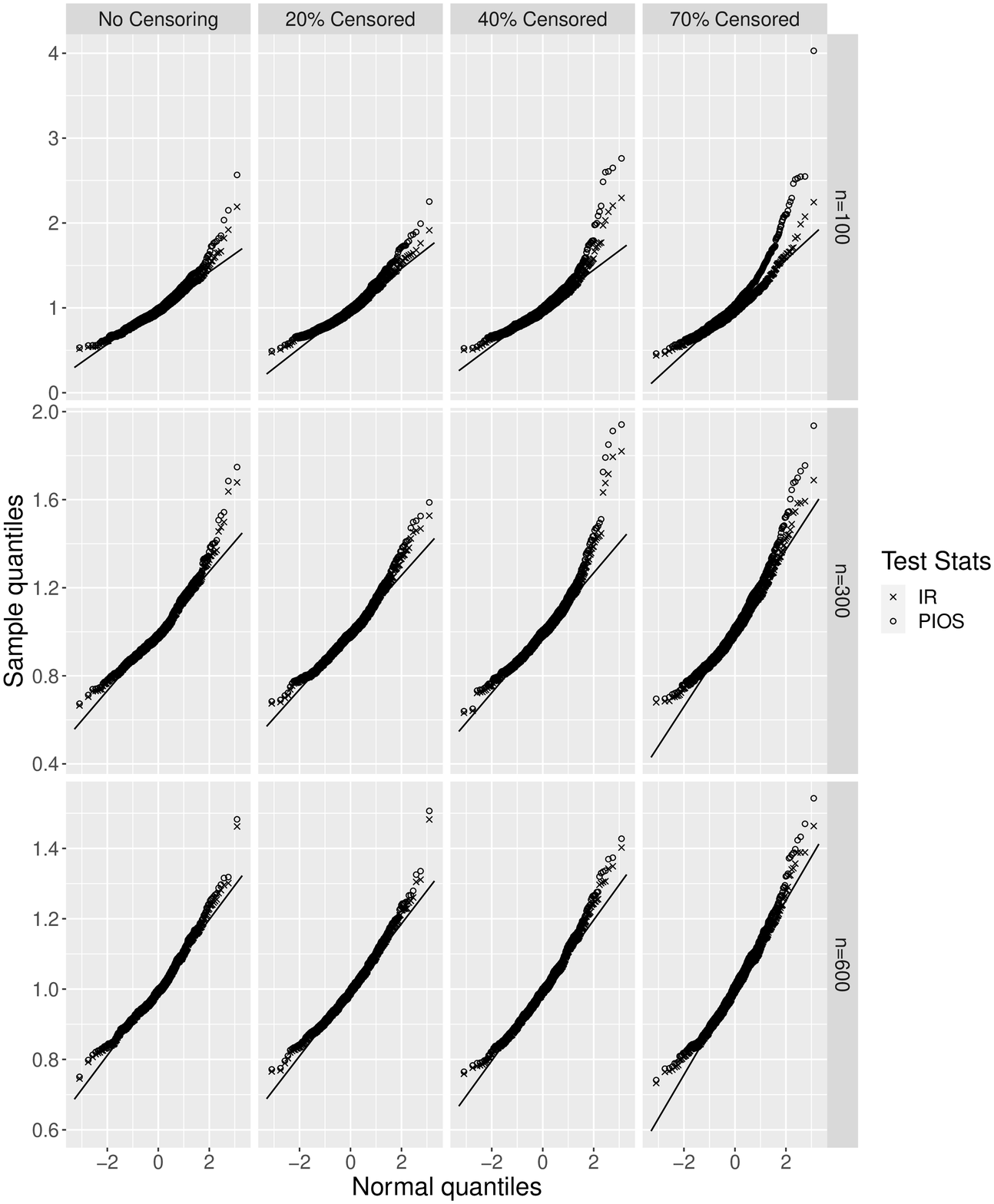}
\end{figure}

\begin{figure}
\centering
\caption{Simulation results: Proportions of rejecting {\bf Clayton} when the true copula is 
Clayton, Frank, Joe, or Gaussian and the sample size is 100. The dashed lines represent the significance level 0.05.}\label{fig:rej_clayton_n100}
\vskip 0.3cm
\includegraphics[width=1\textwidth]{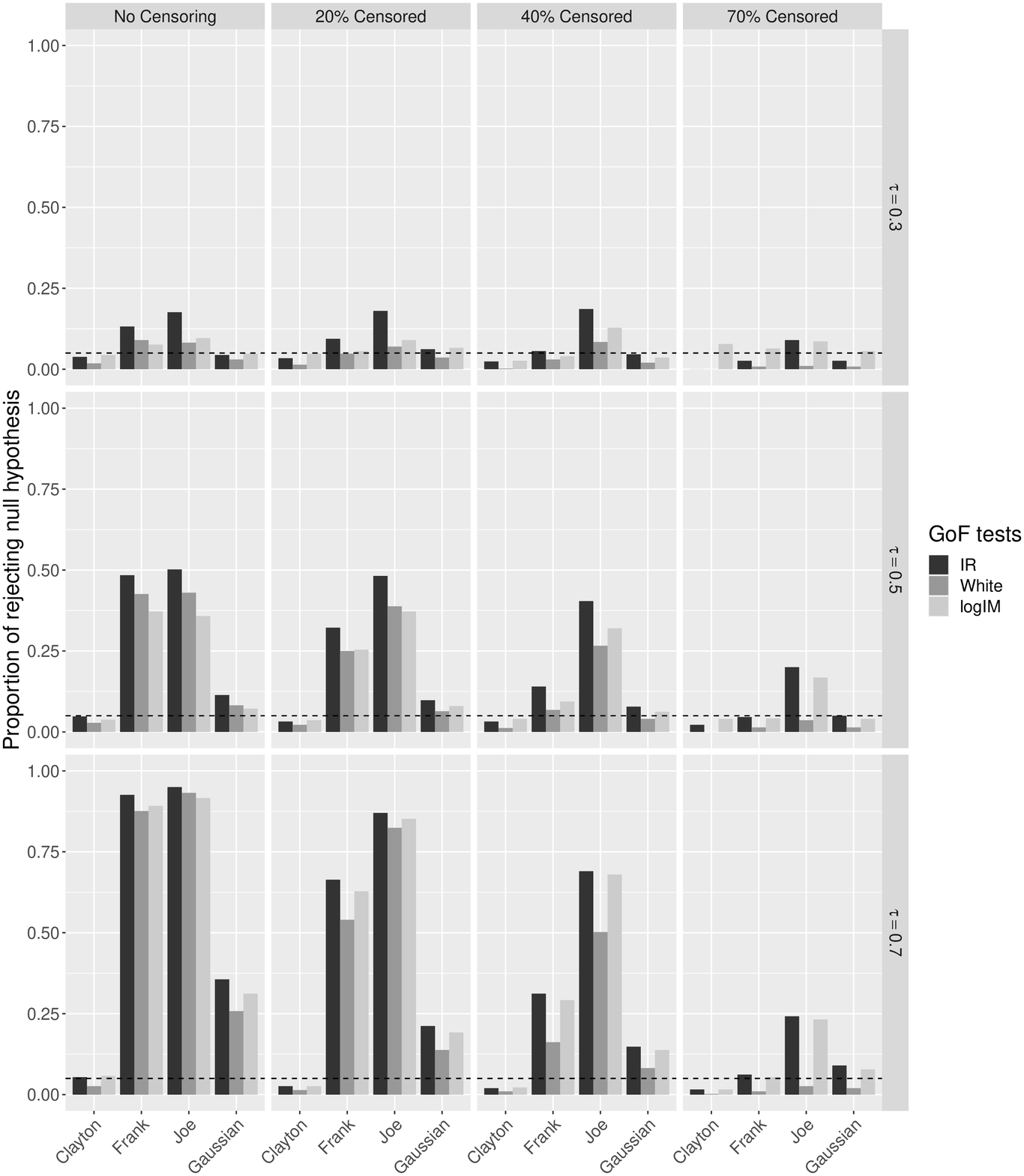}
\end{figure}

\begin{figure}
\centering
\caption{Simulation results: Proportions of rejecting {\bf Frank} when the true copula is 
Clayton, Frank, Joe, or Gaussian and the sample size is 100. The dashed lines represent the significance level 0.05.}\label{fig:rej_frank_n100}
\vskip 0.3cm
\includegraphics[width=1\textwidth]{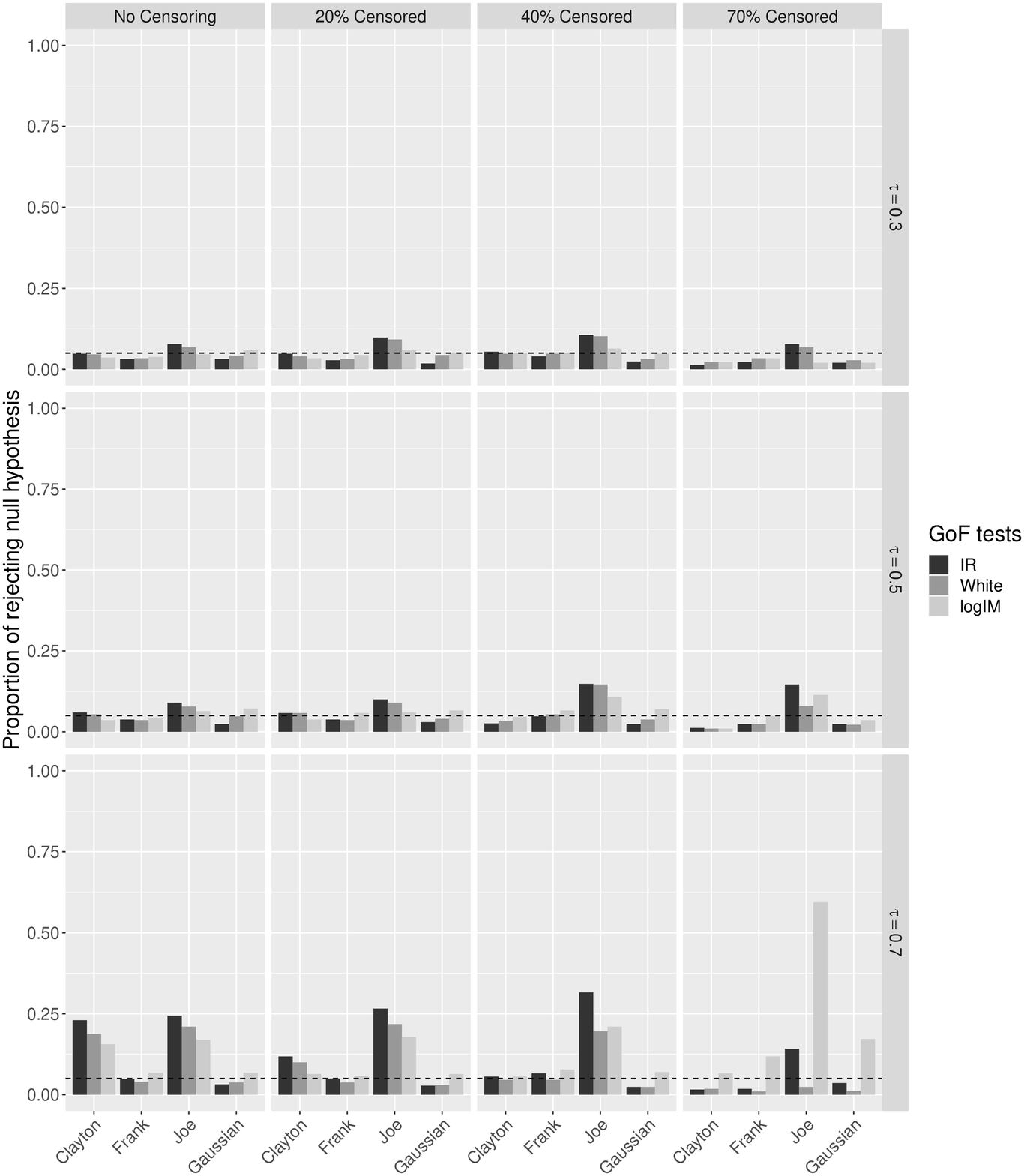}
\end{figure}

\begin{figure}
\centering
\caption{Simulation results: Proportions of rejecting {\bf Joe} when the true copula is 
Clayton, Frank, Joe, or Gaussian and the sample size is 100. The dashed lines represent the significance level 0.05.}\label{fig:rej_joe_n100}
\vskip 0.3cm
\includegraphics[width=1\textwidth]{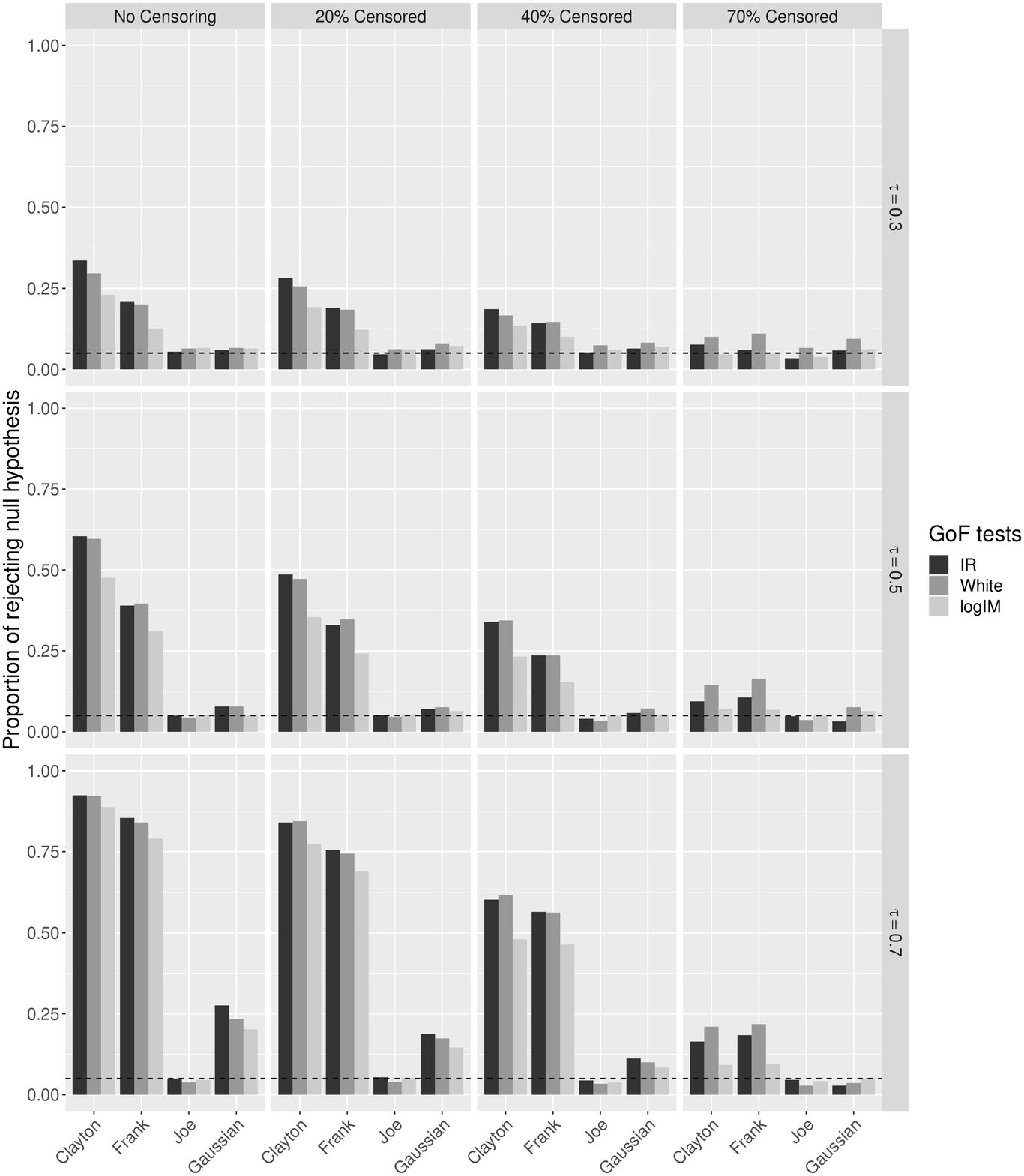}
\end{figure}

\begin{figure}
\centering
\caption{Simulation results: Proportions of rejecting {\bf Gaussian} when the true copula is Clayton, Frank, Joe, or Gaussian and the sample size is 100. The dashed lines represent the significance level 0.05.}\label{fig:rej_normal_n100}
\vskip 0.3cm
\includegraphics[width=1\textwidth]{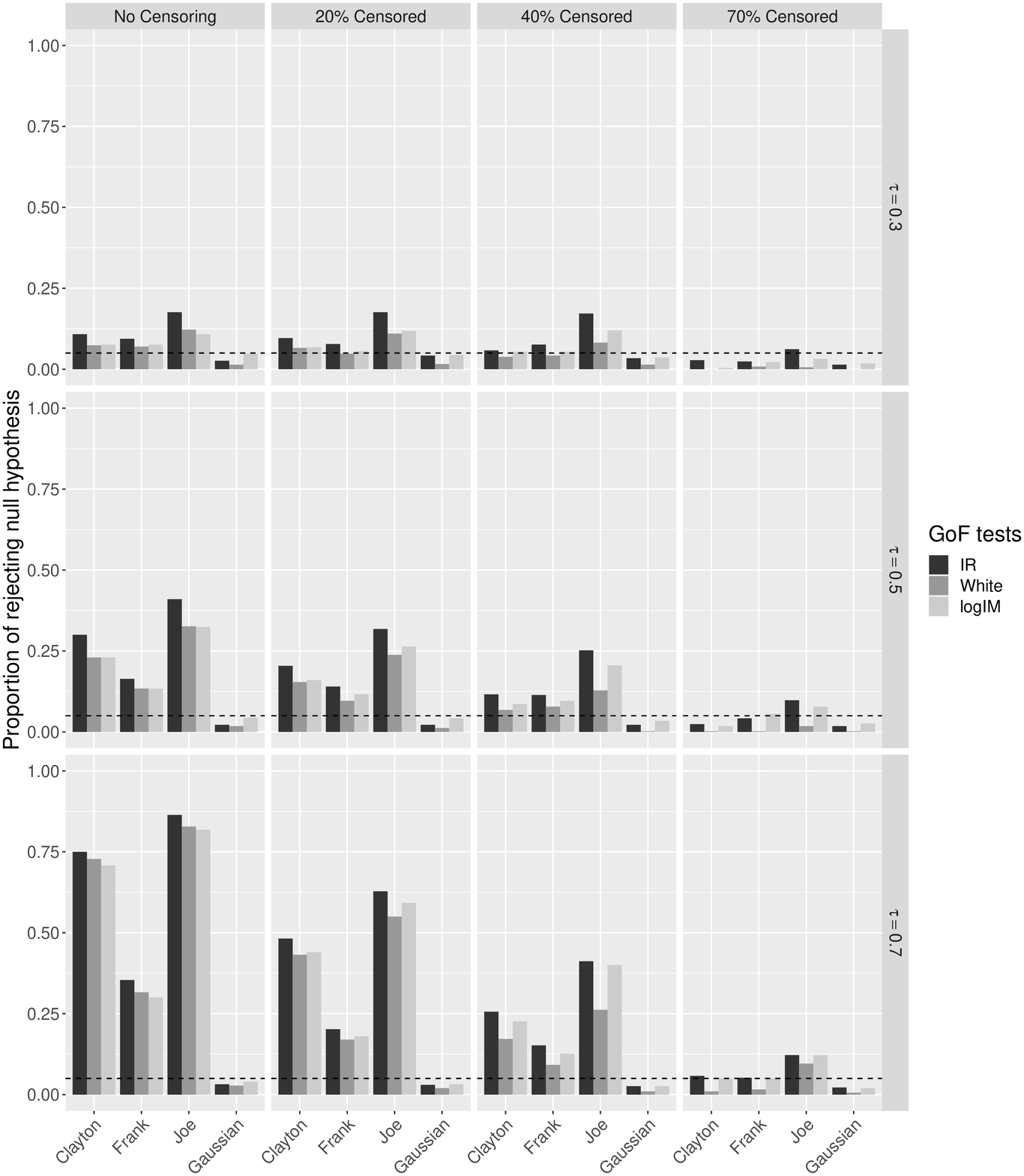}
\end{figure}

\begin{figure}
\centering
\caption{Simulation results: Proportions of rejecting {\bf Clayton} when the true copula is 
Clayton, Frank, Joe, or Gaussian and the sample size is 300. The dashed lines represent the significance level 0.05.}\label{fig:rej_clayton_n300}
\vskip 0.3cm
\includegraphics[width=1\textwidth]{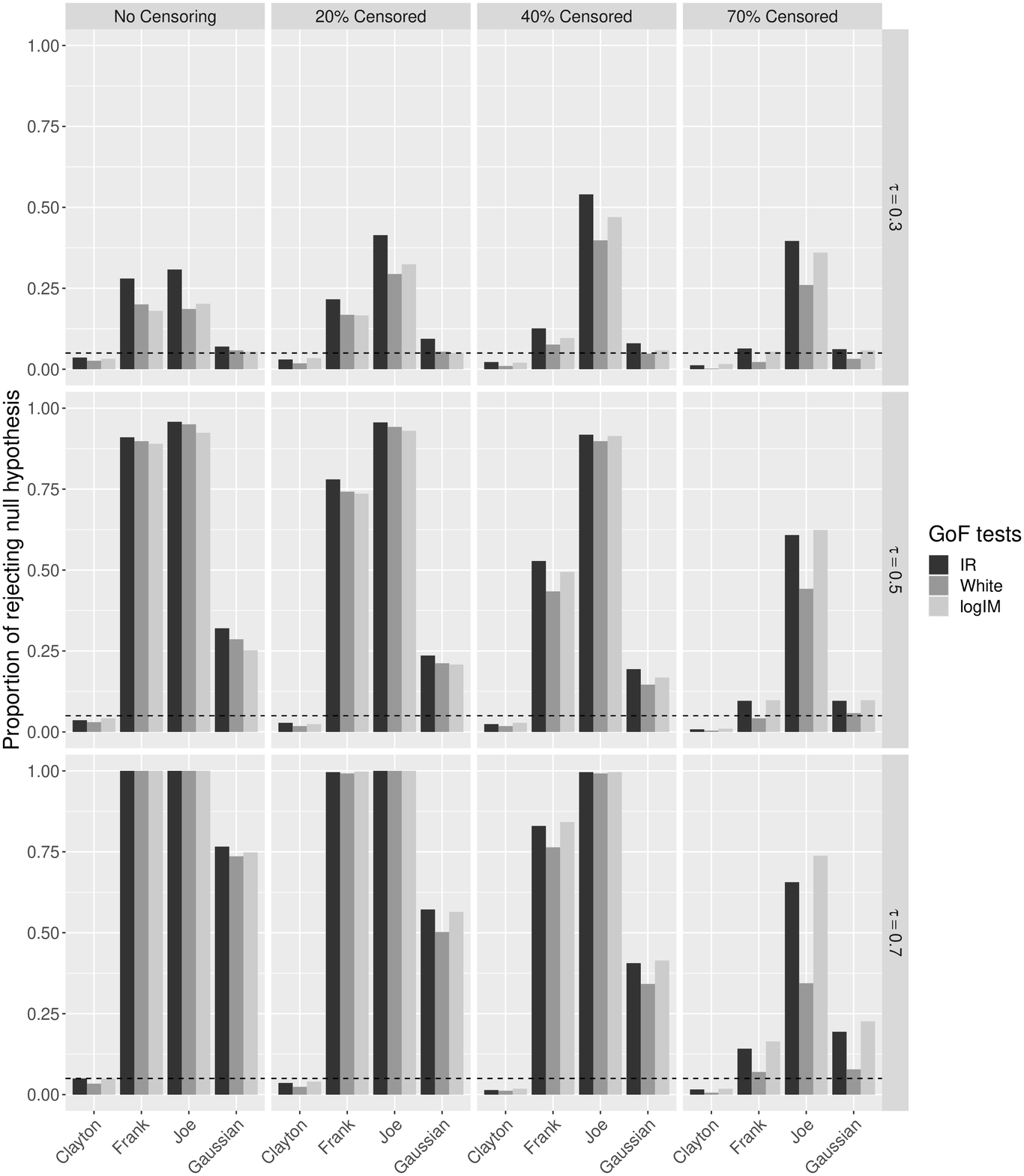}
\end{figure}

\begin{figure}
\centering
\caption{Simulation results: Proportions of rejecting {\bf Frank} when the true copula is 
Clayton, Frank, Joe, or Gaussian and the sample size is 300. The dashed lines represent the significance level 0.05.}\label{fig:rej_frank_n300}
\vskip 0.3cm
\includegraphics[width=1\textwidth]{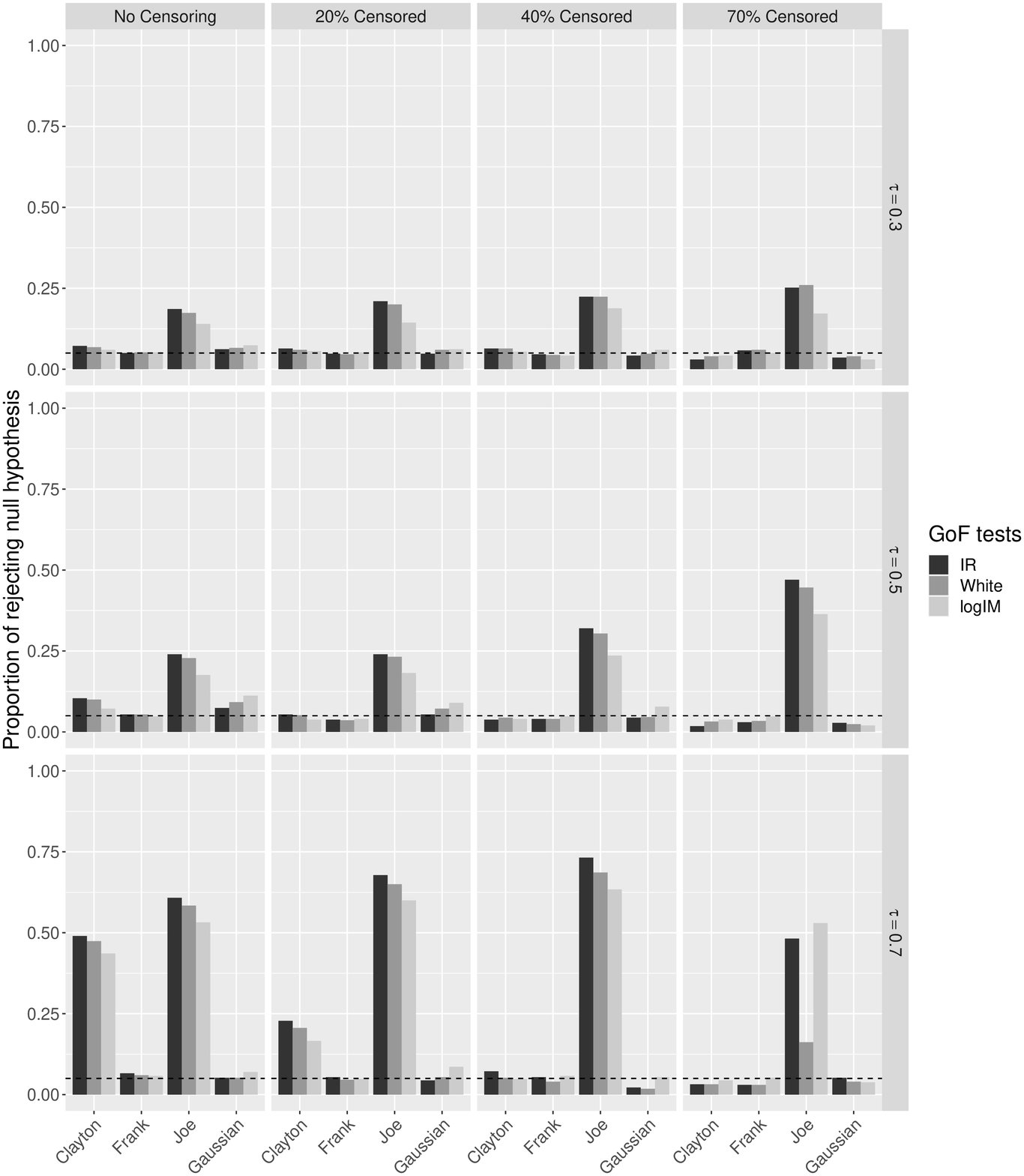}
\end{figure}

\begin{figure}
\centering
\caption{Simulation results: Proportions of rejecting {\bf Joe} when the true copula is 
Clayton, Frank, Joe, or Gaussian and the sample size is 300. The dashed lines represent the significance level 0.05.}\label{fig:rej_joe_n300}
\vskip 0.3cm
\includegraphics[width=1\textwidth]{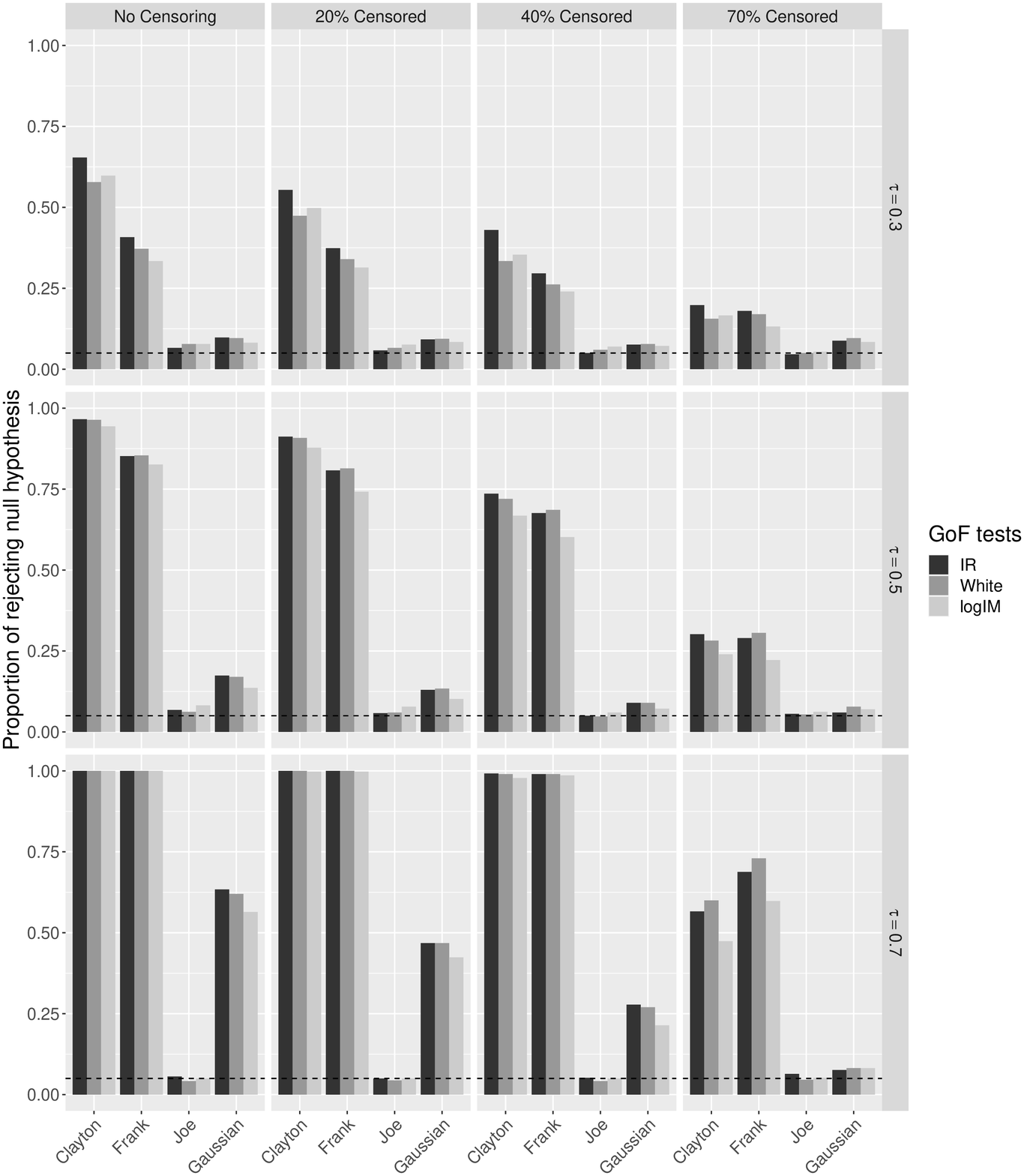}
\end{figure}

\begin{figure}
\centering
\caption{Simulation results: Proportions of rejecting {\bf Gaussian} when the true copula is Clayton, Frank, Joe, or Gaussian and the sample size is 300. The dashed lines represent the significance level 0.05.}\label{fig:rej_normal_n300}
\vskip 0.3cm
\includegraphics[width=1\textwidth]{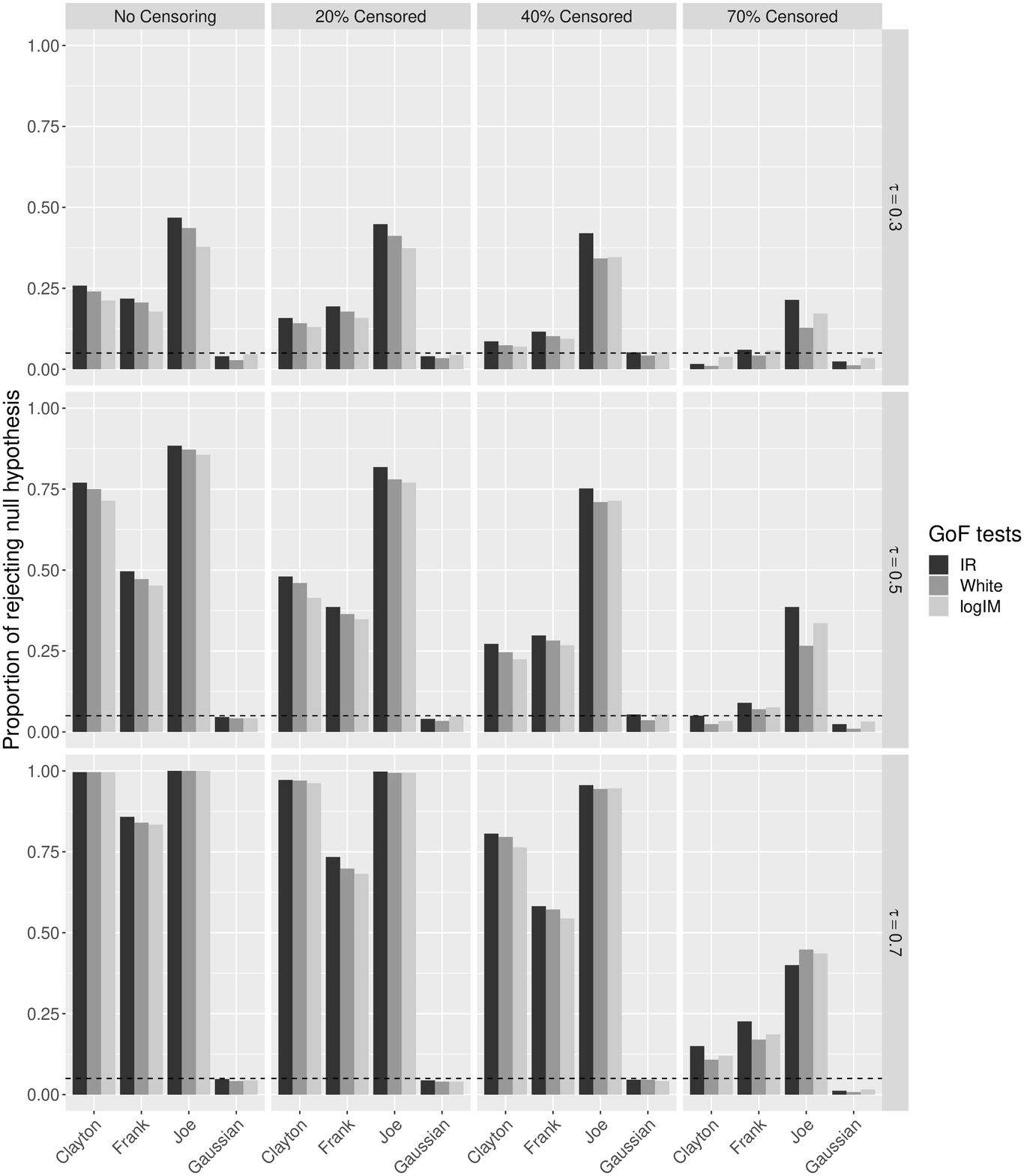}
\end{figure}

\begin{figure}
\centering
\caption{Proportions of selecting each of the four copula families: Clayton, Frank, Joe, and Gaussian as the best copula when the true copula is \textbf{Clayton} and the sample size is 100.}\label{fig:selection_clayton_n100}
\vskip 0.3cm
\includegraphics[width=1\textwidth]{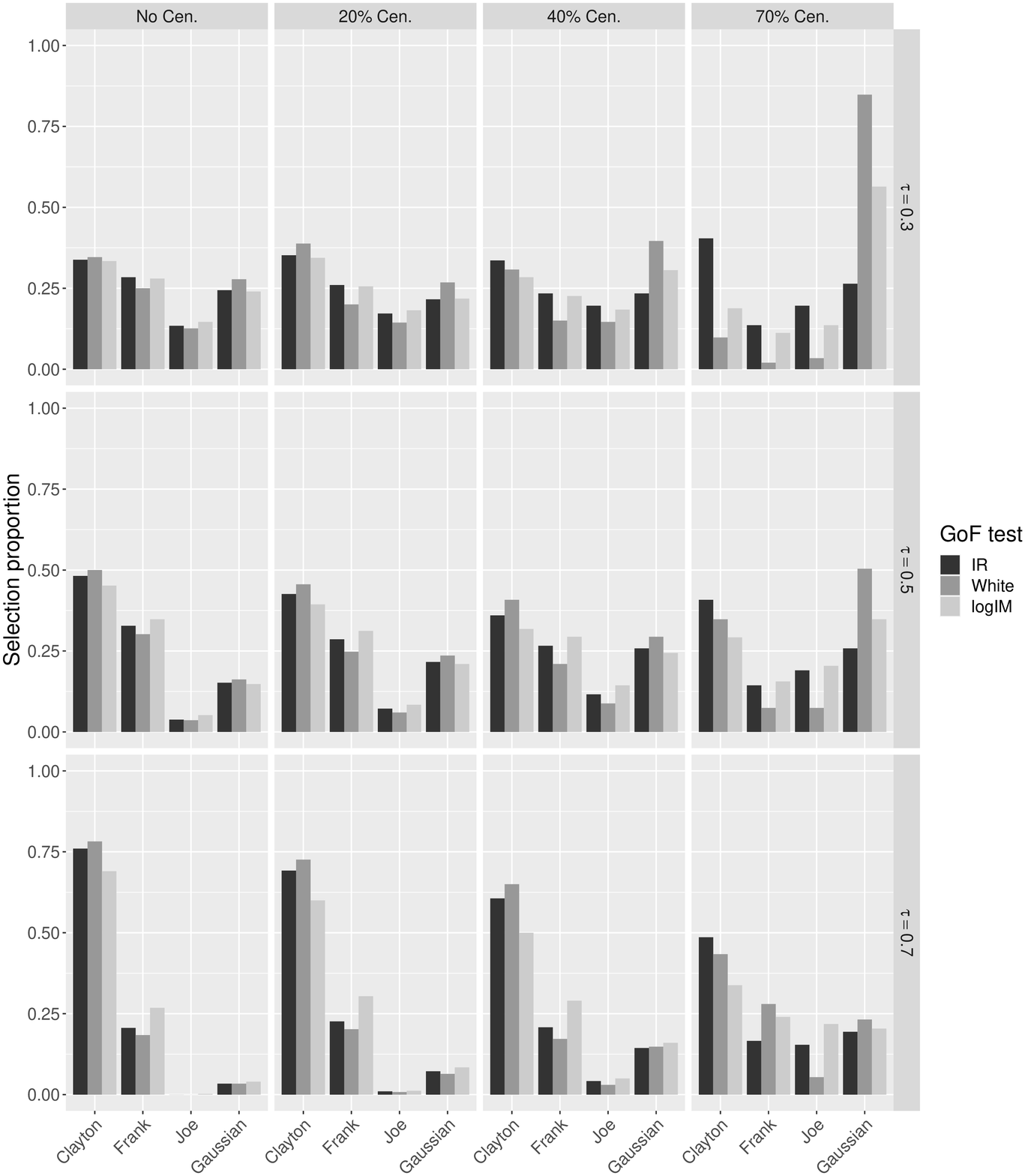}
\end{figure}

\begin{figure}
\centering
\caption{Proportions of selecting each of the four copula families: Clayton, Frank, Joe, and Gaussian as the best copula when the true copula is \textbf{Frank} and the sample size is 100.}\label{fig:selection_frank_n100}
\vskip 0.3cm
\includegraphics[width=1\textwidth]{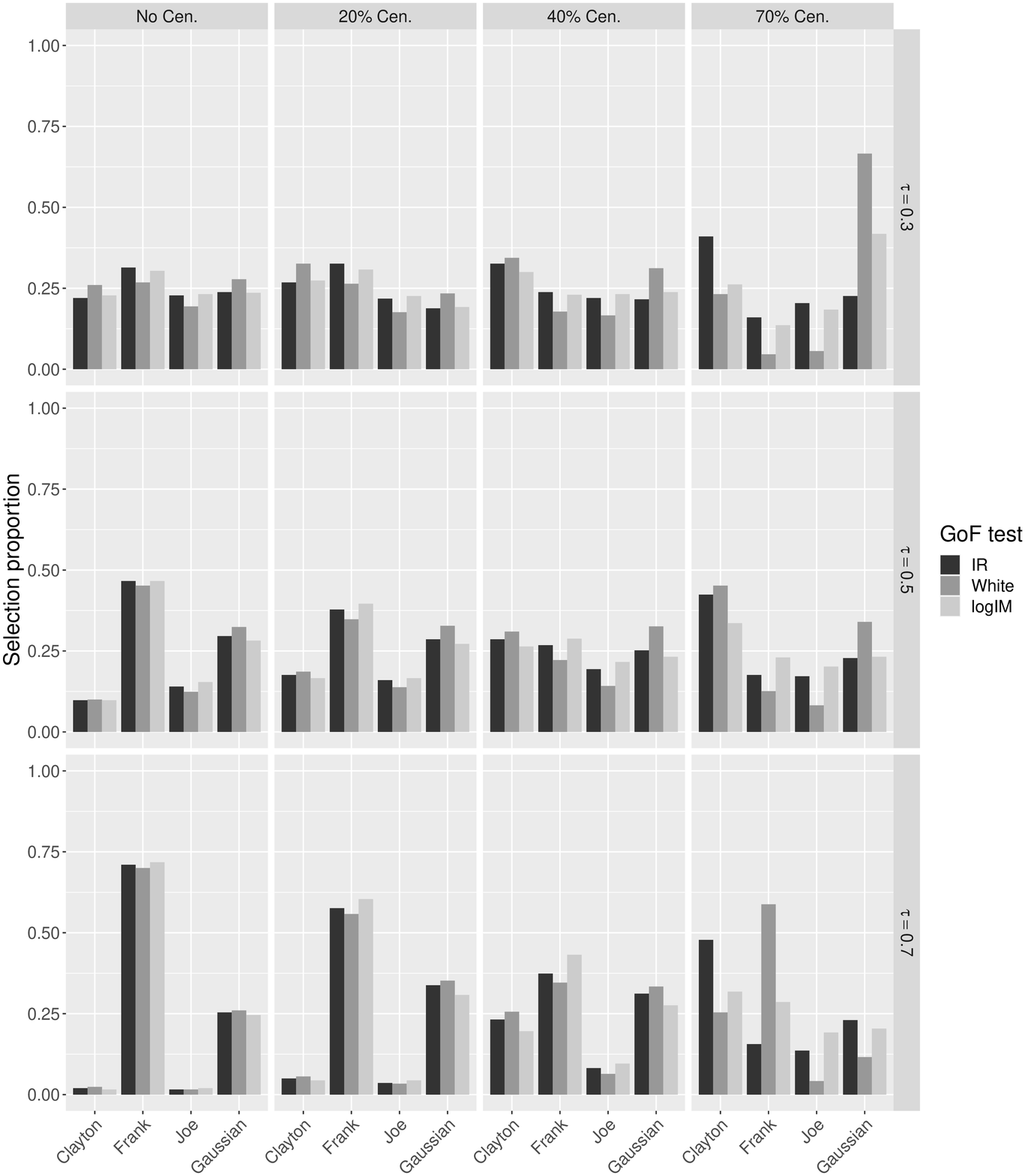}
\end{figure}

\begin{figure}
\centering
\caption{Proportions of selecting each of the four copula families: Clayton, Frank, Joe, and Gaussian as the best copula when the true copula is \textbf{Joe} and the sample size is 100.}\label{fig:selection_joe_n100}
\vskip 0.3cm
\includegraphics[width=1\textwidth]{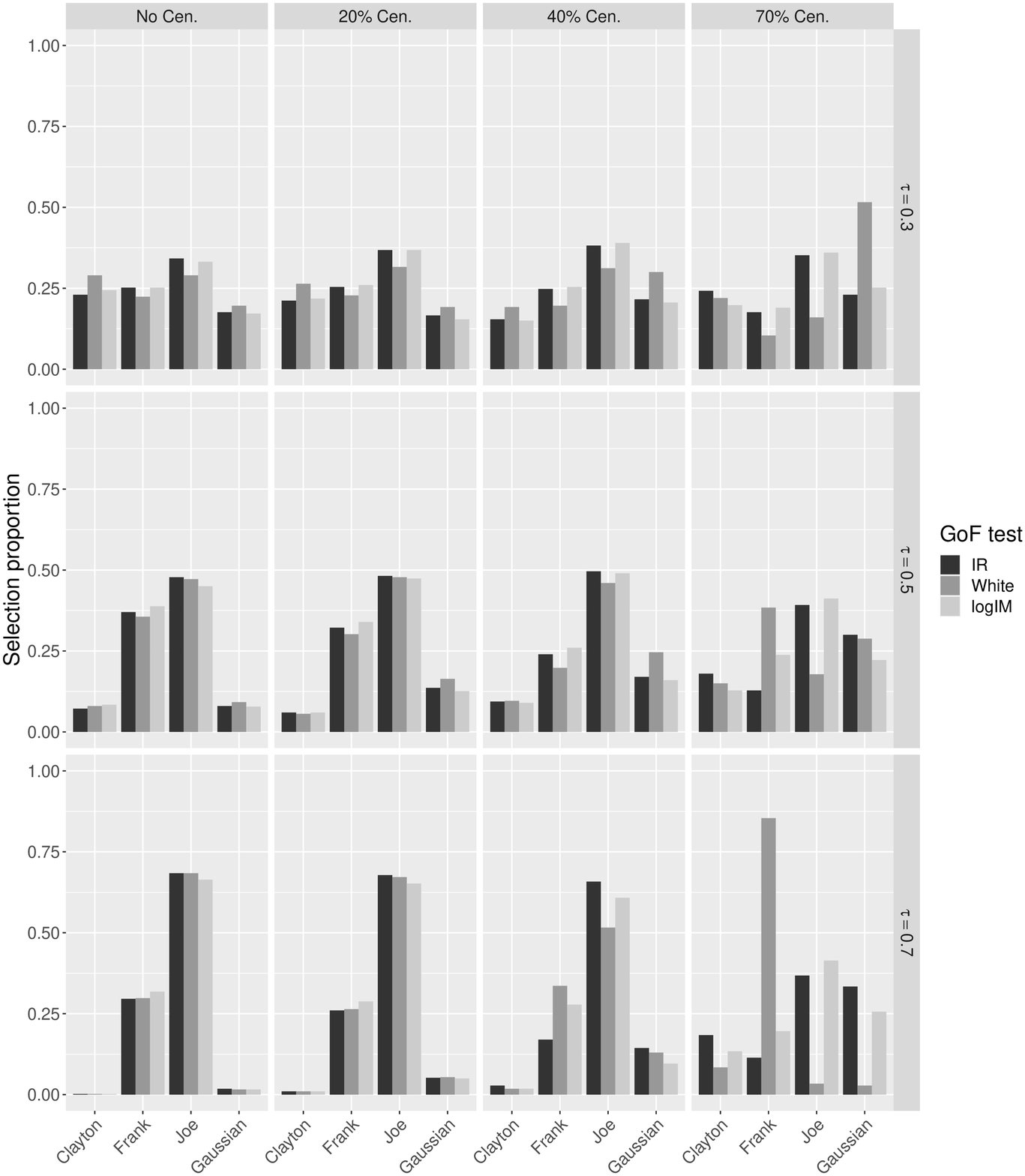}
\end{figure}

\begin{figure}
\centering
\caption{Proportions of selecting each of the four copula families: Clayton, Frank, Joe, and Gaussian as the best copula when the true copula is \textbf{Gaussian} and the sample size is 100.}\label{fig:selection_gaussian_n100}
\vskip 0.3cm
\includegraphics[width=1\textwidth]{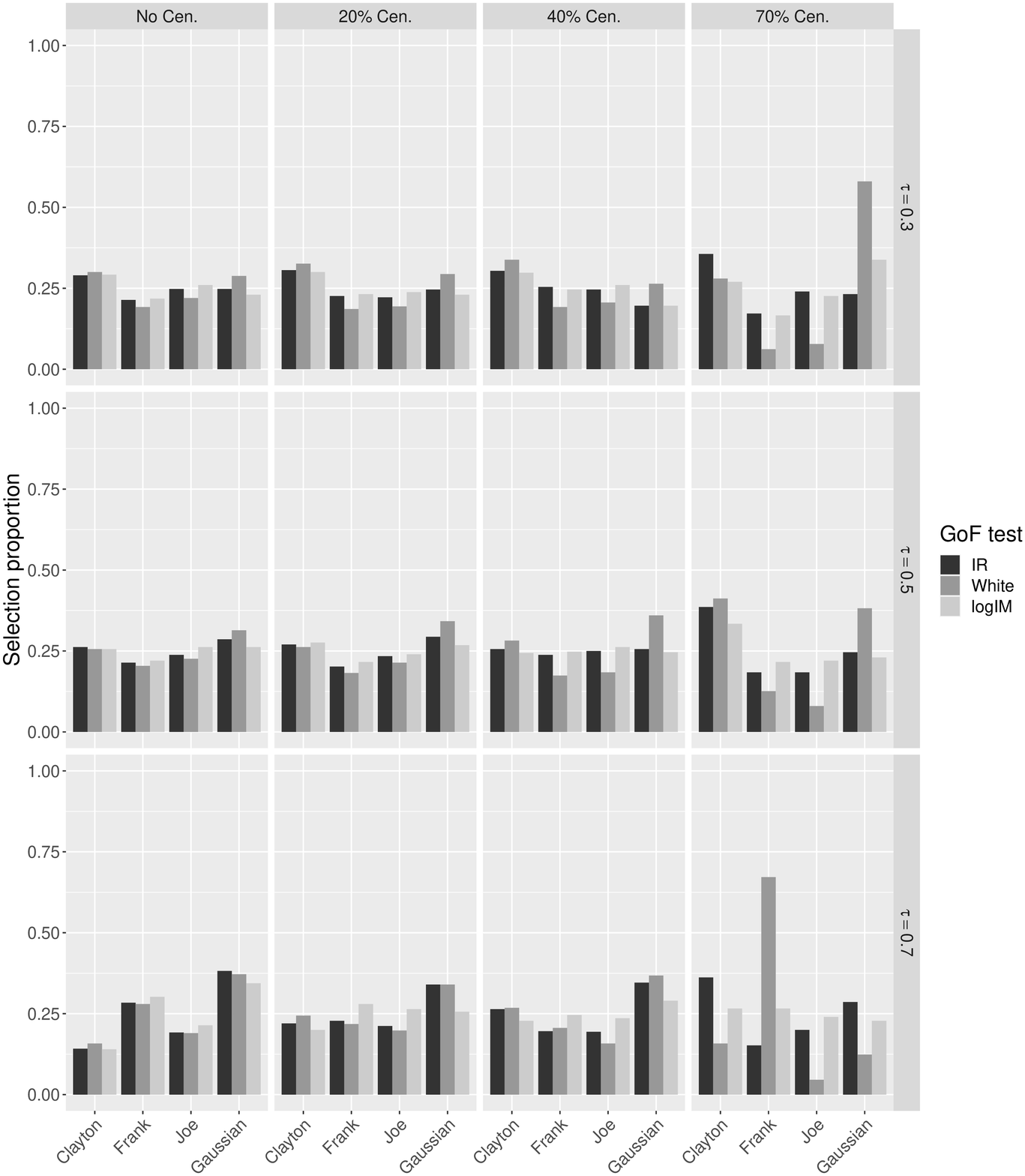}
\end{figure}

\begin{figure}
\centering
\caption{Proportions of selecting each of the four copula families: Clayton, Frank, Joe, and Gaussian as the best copula when the true copula is \textbf{Clayton} and the sample size is 300.}\label{fig:selection_clayton_n300}
\vskip 0.3cm
\includegraphics[width=1\textwidth]{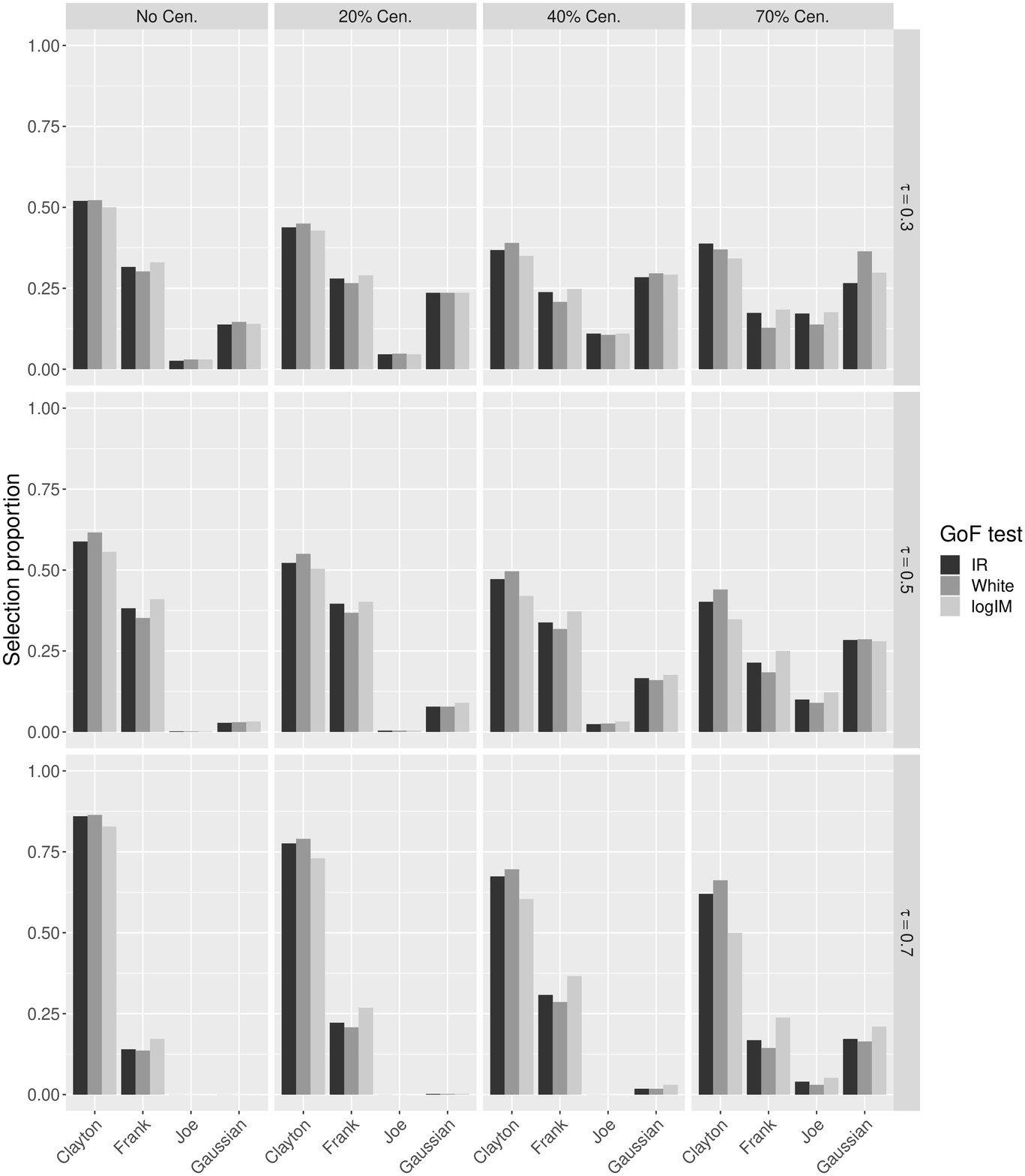}
\end{figure}

\begin{figure}
\centering
\caption{Proportions of selecting each of the four copula families: Clayton, Frank, Joe, and Gaussian as the best copula when the true copula is \textbf{Frank} and the sample size is 300.}\label{fig:selection_frank_n300}
\vskip 0.3cm
\includegraphics[width=1\textwidth]{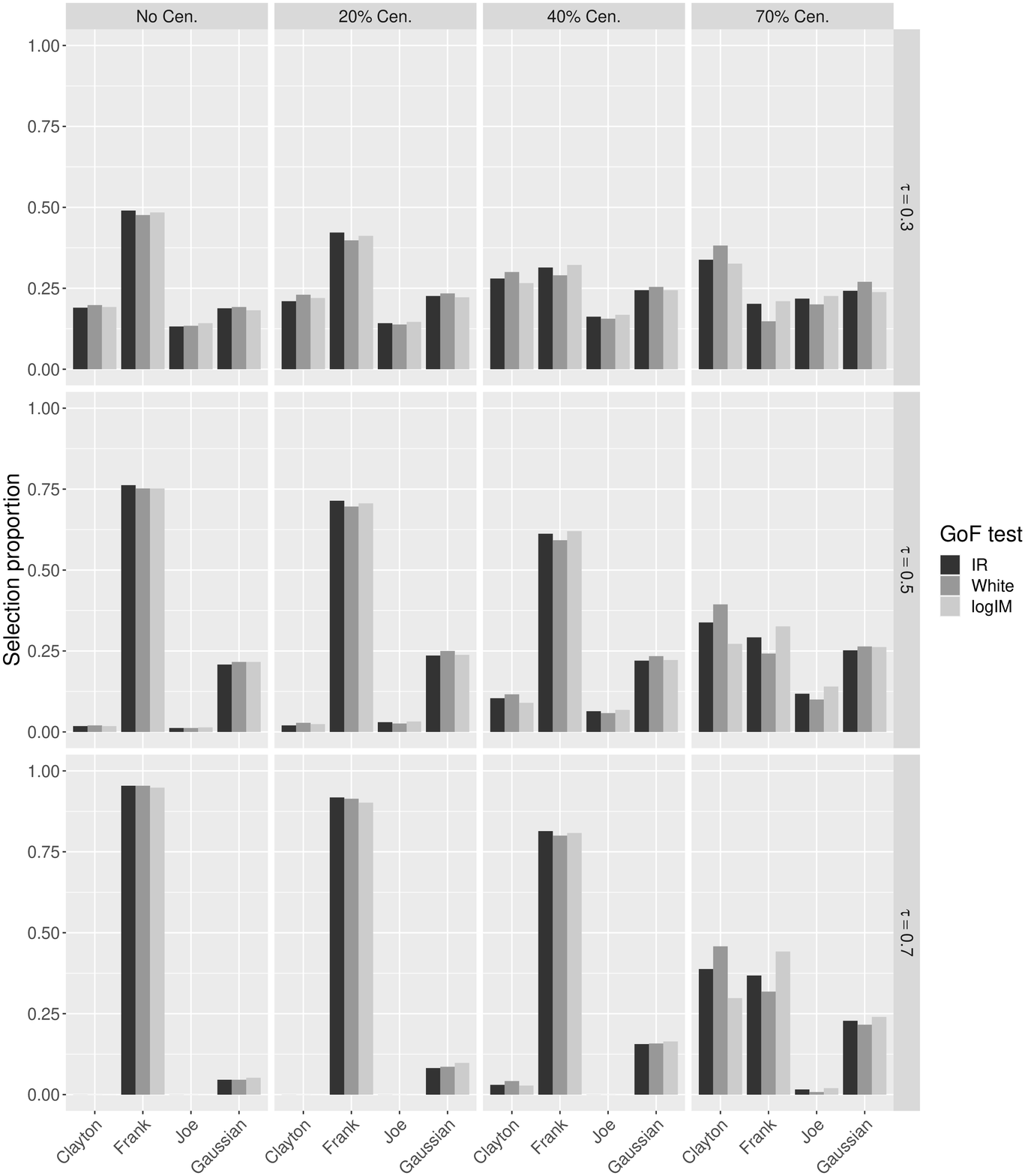}
\end{figure}

\begin{figure}
\centering
\caption{Proportions of selecting each of the four copula families: Clayton, Frank, Joe, and Gaussian as the best copula when the true copula is \textbf{Joe} and the sample size is 300.}\label{fig:selection_joe_n300}
\vskip 0.3cm
\includegraphics[width=1\textwidth]{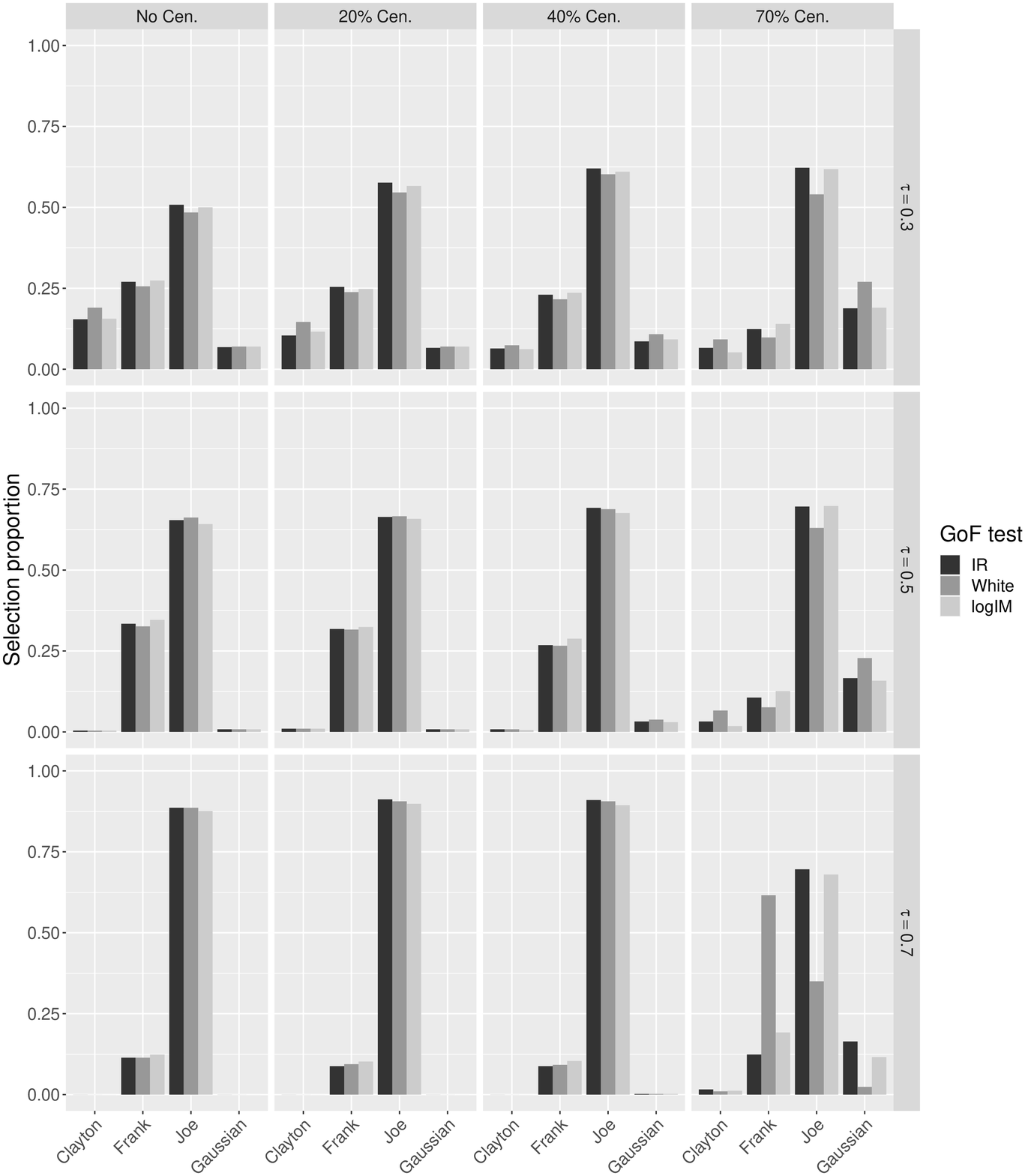}
\end{figure}

\begin{figure}
\centering
\caption{Proportions of selecting each of the four copula families: Clayton, Frank, Joe, and Gaussian as the best copula when the true copula is \textbf{Gaussian} and the sample size is 300.}\label{fig:selection_gaussian_n300}
\vskip 0.3cm
\includegraphics[width=1\textwidth]{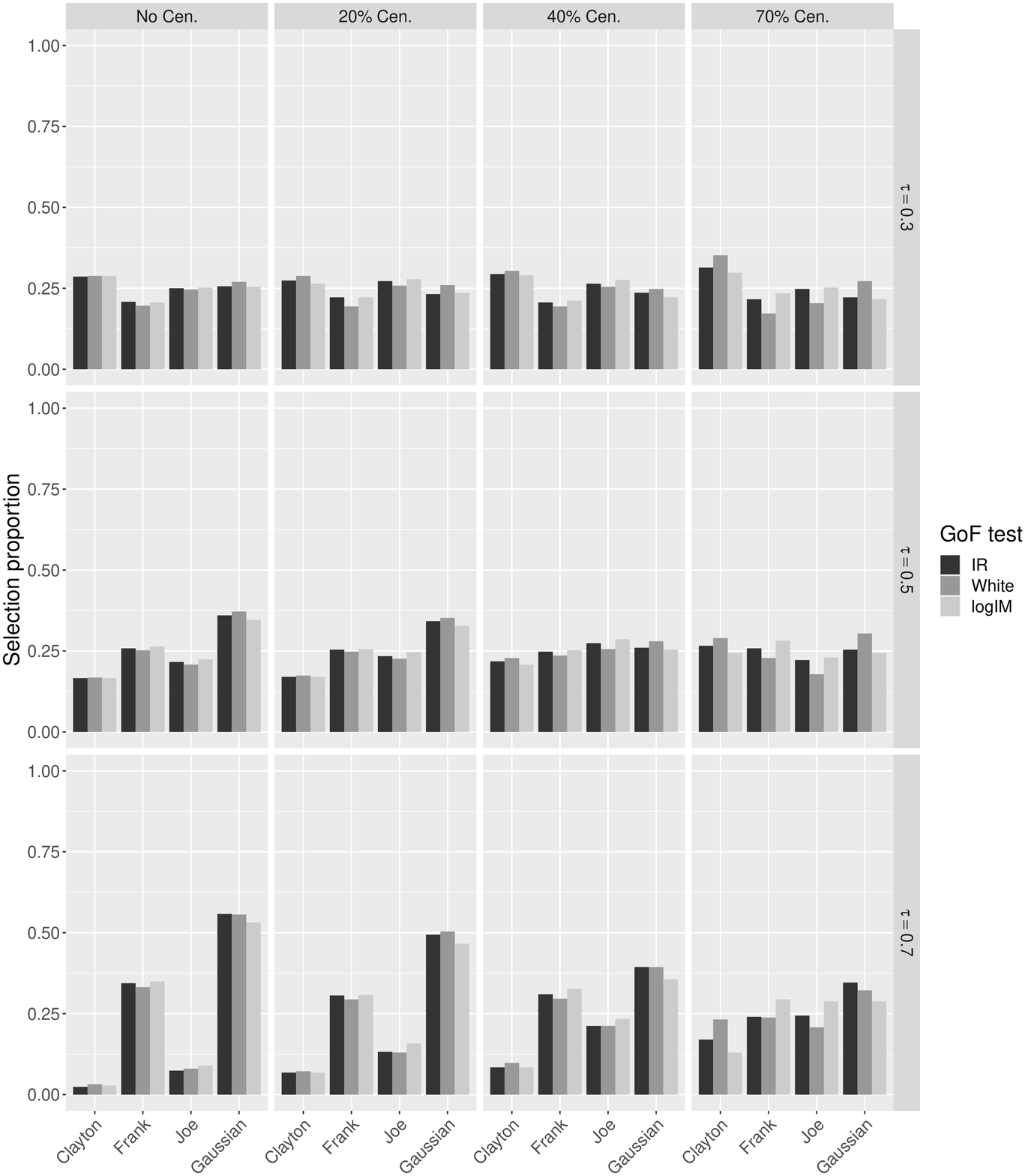}
\end{figure}

\begin{figure}
\centering
\caption{Proportions of selecting each of the four copula families: Clayton, Frank, Joe, and Gaussian as the best copula when the true copula is \textbf{Clayton} and the sample size is 600.}\label{fig:selection_clayton_n600}
\vskip 0.3cm
\includegraphics[width=1\textwidth]{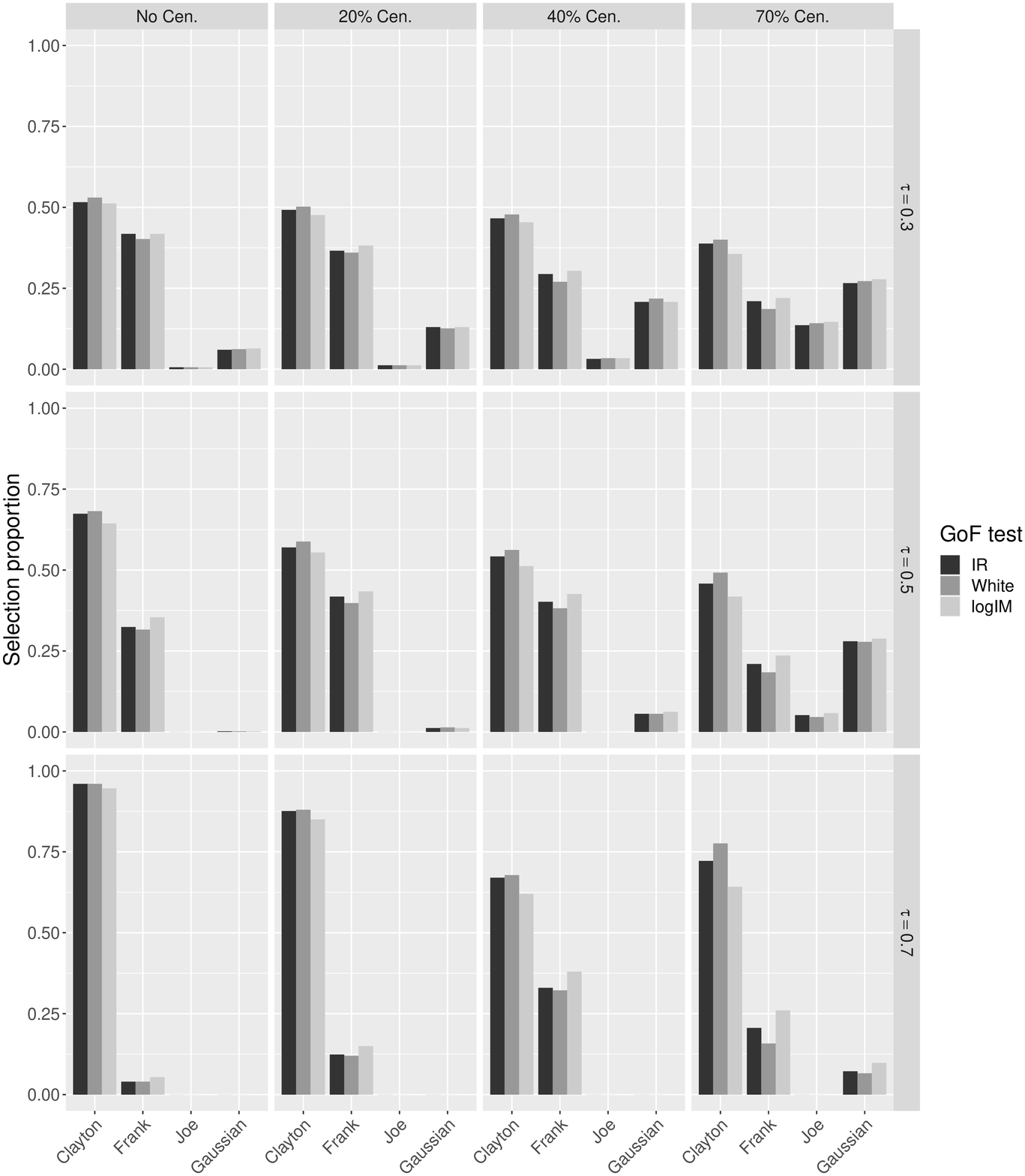}
\end{figure}

\begin{figure}
\centering
\caption{Proportions of selecting each of the four copula families: Clayton, Frank, Joe, and Gaussian as the best copula when the true copula is \textbf{Frank} and the sample size is 600.}\label{fig:selection_frank_n600}
\vskip 0.3cm
\includegraphics[width=1\textwidth]{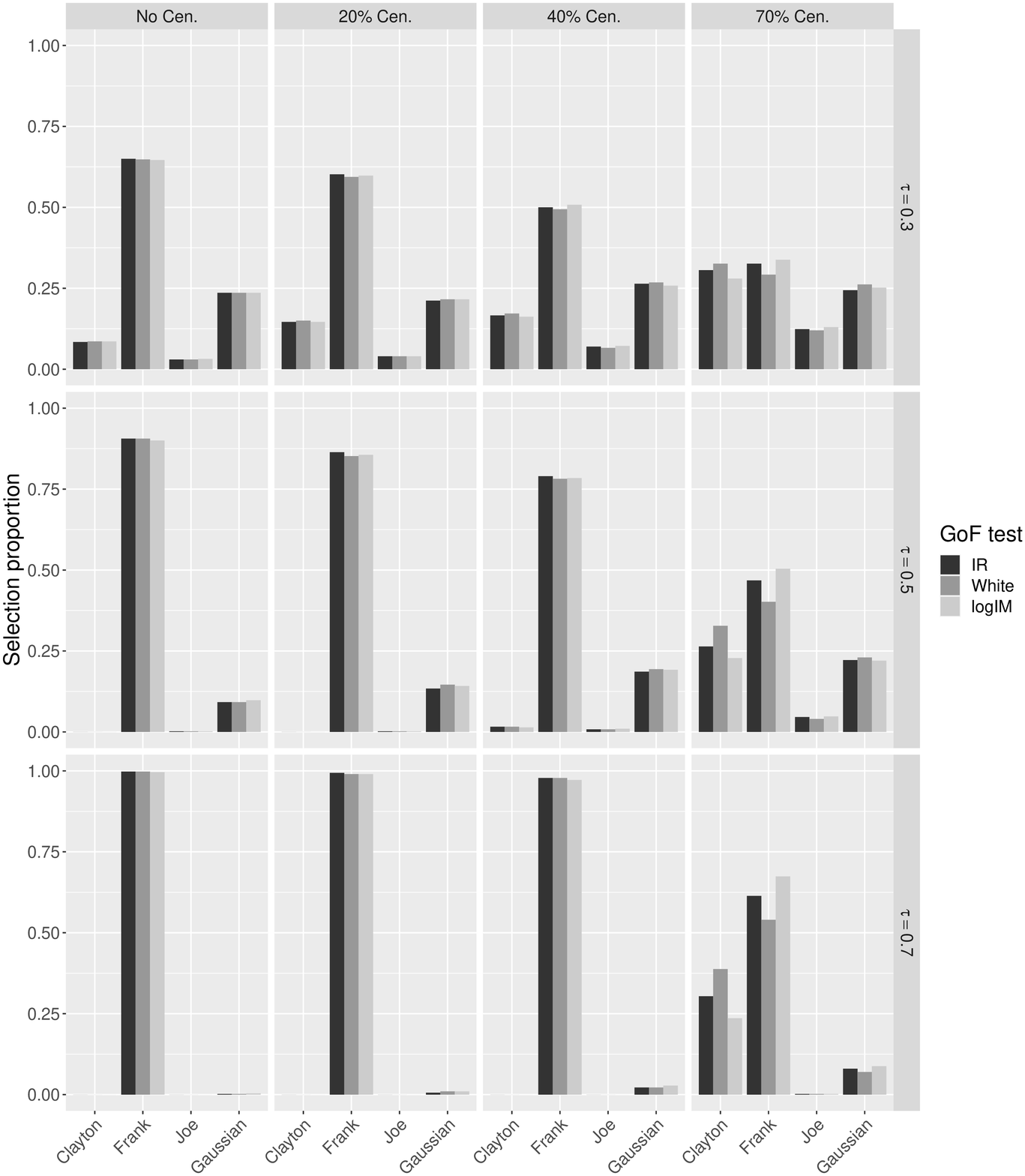}
\end{figure}

\begin{figure}
\centering
\caption{Proportions of selecting each of the four copula families: Clayton, Frank, Joe, and Gaussian as the best copula when the true copula is \textbf{Joe} and the sample size is 600.}\label{fig:selection_joe_n600}
\vskip 0.3cm
\includegraphics[width=1\textwidth]{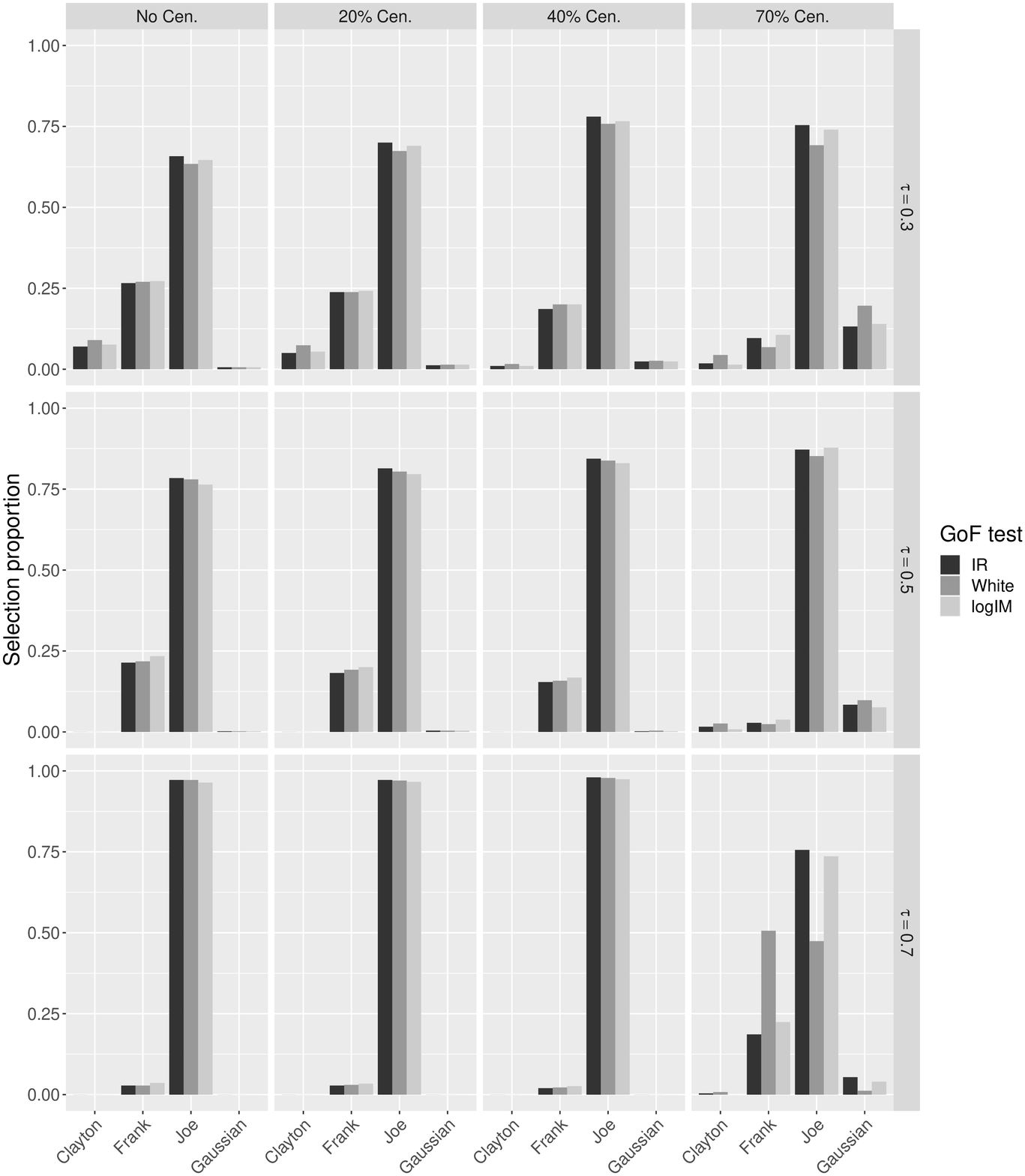}
\end{figure}

\begin{figure}
\centering
\caption{Proportions of selecting each of the four copula families: Clayton, Frank, Joe, and Gaussian as the best copula when the true copula is \textbf{Gaussian} and the sample size is 600.}\label{fig:selection_gaussian_n600}
\vskip 0.3cm
\includegraphics[width=1\textwidth]{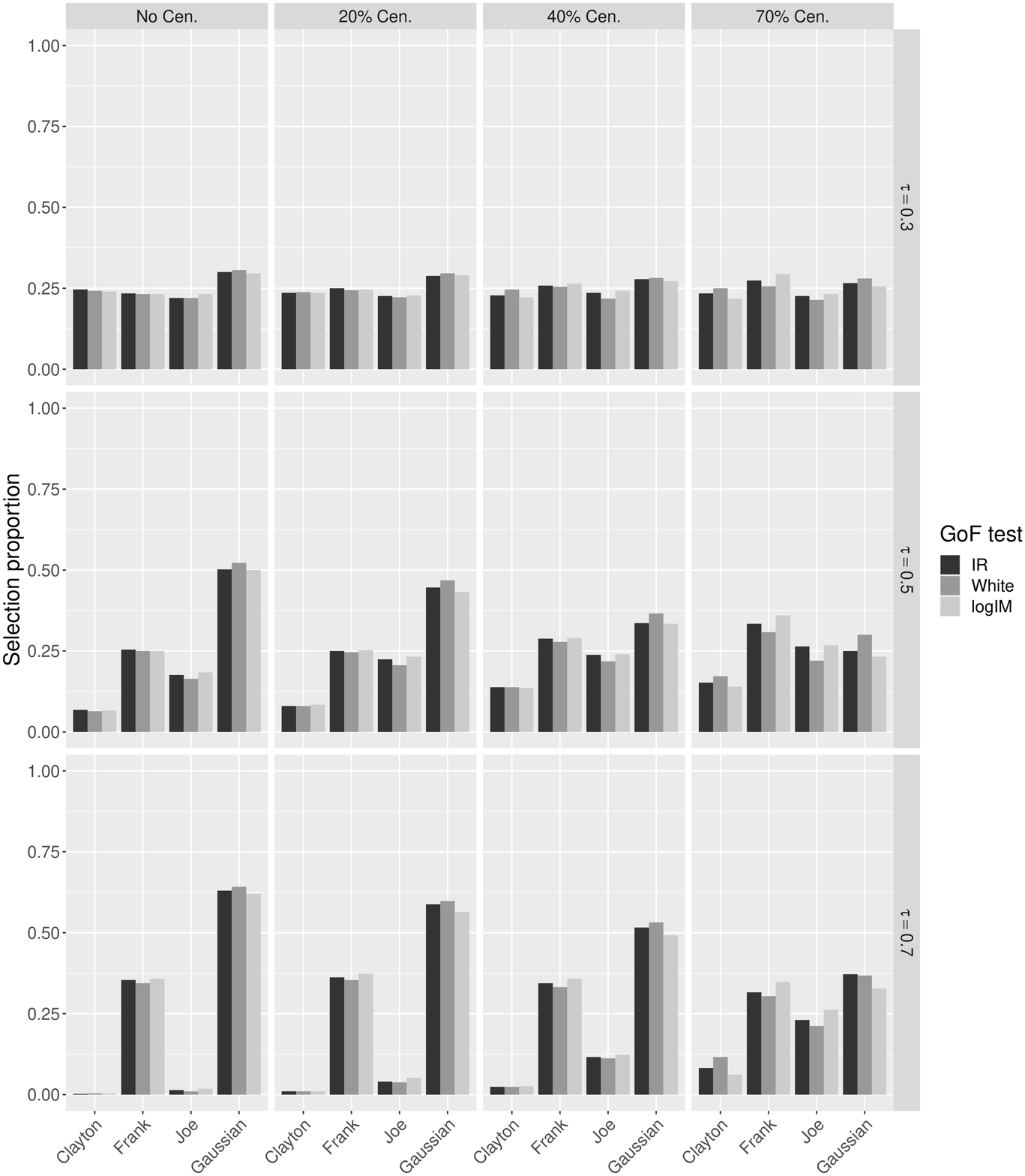}
\end{figure}

\begin{itemize}
\item Figures \ref{fig:U_n100_tau0.3} - \ref{fig:U_n600_tau0.7} plot the estimated pseudo-observations $(\widehat U_{i1}, \widehat U_{i2})$ obtained from one replication of the simulated bivariate censored survival times generated from one of the four copula families: Clayton, Frank, Joe, and Gaussian with Kendall's $\tau=0.3, 0.7$ and sample size $n=100, 600$.
\item Figures \ref{fig:QQ_clayton} - \ref{fig:QQ_normal} plot the normal quantile-quantile (QQ) plots of 500 replications of the IR and PIOS statistics under each of four copula families: Clayton, Frank, Joe, and Gaussian with Kendall's $\tau=0.5$ and sample size $n=100,300,600$. 
\item Figures \ref{fig:rej_clayton_n100} - \ref{fig:rej_normal_n300} plot the proportions of rejecting the null hypothesis at sample size $n=100$ or $300$ when the true copula is one of the four copula families: Clayton, Frank, Joe, and Gaussian with Kendall's $\tau=0.3, 0.5, 0.7$. 
\item Figures \ref{fig:selection_clayton_n100} - \ref{fig:selection_gaussian_n600} report the percentages of choosing each copula family as the best among the 500 replications when the true copula is one of the four copula families: Clayton, Frank, Joe, and Gaussian with Kendall's $\tau=0.3, 0.5, 0.7$ and sample size $n=100, 300, 600$. 
\end{itemize}

\section{Clayton, Frank, Joe, and Gaussian Copulas}\label{app:copula}

All the four families are determined by a scalar parameter, denoted by $\theta$. Here, we present the expressions of the copula function of each family and the corresponding expressions of $\ell_{\theta}$ and $\ell_{\theta\theta}$: the first-order and second-order partial derivatives of the log-likelihood function with respect to (w.r.t.) $\theta$. We also present the relationship between Kendall's $\tau$ and the copula parameter $\theta$.

Recall that $\Cbbm$ denotes the copula function with $\cbbm_r = \partial \Cbbm/\partial u_r$, $r=1,2$, and $\cbbm = \partial^2 \Cbbm/\partial u_1\partial u_2$. Let $\widetilde \Cbbm = \log \Cbbm$, $\widetilde \cbbm_1 = \log \cbbm_1$, $\widetilde \cbbm_2 = \log \cbbm_2$, and $\widetilde \cbbm = \log \cbbm$.

\subsection{Clayton copula} 
The Clayton copula function is given as
$$
\Cbbm(u_1,u_2;\theta) = (u_1^{-\theta}+u_2^{-\theta}-1)^{-1/\theta}.
$$
The relationship between Kendall's $\tau$ and the copula parameter $\theta$ is
$$
\tau = \frac{\theta}{\theta+2}.
$$
Based on the expression of $\Cbbm$, we have
\begin{align*}
\cbbm_1(u_1,u_2;\theta) & = u_1^{-(1+\theta)}  (u_1^{-\theta}+u_2^{-\theta}-1)^{-1/\theta-1} ,\\ 
\cbbm_2(u_1,u_2;\theta) & = u_2^{-(1+\theta)}  (u_1^{-\theta}+u_2^{-\theta}-1)^{-1/\theta-1},\\
\cbbm(u_1,u_2;\theta) & =  (1+\theta)u_1^{-(1+\theta)} u_2^{-(1+\theta)}  (u_1^{-\theta}+u_2^{-\theta}-1)^{-1/\theta-2}.
\end{align*}
Furthermore, let $\psi(u_1,u_2;\theta)= \log (u_1^{-\theta}+u_2^{-\theta}-1)$.  We have
\begin{align*}
\widetilde \Cbbm & = \log \Cbbm =  -(1/\theta) \psi(u_1,u_2;\theta)\\
\widetilde \cbbm_1 & = \log \cbbm_1 = -(1+\theta)\log u_1 +  (-1/\theta-1) \psi(u_1,u_2;\btheta),\\ 
\widetilde \cbbm_2 & =\log \cbbm_2 =  -(1+\theta)\log u_2 +  (-1/\theta-1) \psi(u_1,u_2;\btheta),\\
\widetilde \cbbm & = \log \cbbm = \log(1+\theta) -(1+\theta)\log u_1  -(1+\theta)\log u_2 -(1/\theta+2) \psi(u_1,u_2;\theta).
\end{align*}
The log-likelihood function can be expressed as
\begin{align*}
\ell = & \delta_1\delta_2 \widetilde \cbbm + \delta_1(1-\delta_2) \widetilde \cbbm_1 + (1-\delta_1)\delta_2 \widetilde \cbbm_2 + (1-\delta_1)(1-\delta_2) \widetilde \Cbbm\\
= & \, \delta_1\delta_2\log(1+\theta)- \delta_1(1+\theta)\log u_1 - \delta_2(1+\theta)\log u_2 - (1/\theta)\psi(u_1,u_2;\btheta)  \\
& - (\delta_1+\delta_2) \psi(u_1,u_2;\theta).
\end{align*}
Thus, 
\begin{align*}
& \ell_{\theta}(\theta;u_1,u_2,\delta_1,\delta_2) \\
= & \frac{\delta_1\delta_2}{1+\theta} - \log\left(u_1^{\delta_1}u_2^{\delta_2}\right) +\frac{\psi(u_1,u_2;\theta)}{\theta^2} + \left(\delta_1+\delta_2 + \frac{1}{\theta}\right) \frac{u_1^{-\theta}\log u_1 +u_2^{-\theta}\log u_2 }{\psi(u_1,u_2;\theta)},
\end{align*}
and
\begin{align*}
& \ell_{\theta\theta}(\theta;u_1,u_2,\delta_1,\delta_2)  = -\frac{\delta_1\delta_2}{(1+\theta)^2} -\frac{2 \psi(u_1,u_2;\theta)}{\theta^3} - \frac{2(u_1^{-\theta}\log u_1 +u_2^{-\theta}\log u_2)}{\theta^2\psi(u_1,u_2;\theta)}\\
& +\left(\delta_1+\delta_2 + \frac{1}{\theta}\right) \left\{\frac{(u_1^{-\theta}\log u_1 +u_2^{-\theta}\log u_2)^2}{\psi(u_1,u_2;\theta)^2}-\frac{u_1^{-\theta}(\log u_1)^2 + u_2^{-\theta}(\log u_2)^2}{\psi(u_1,u_2;\theta)}\right\}.
\end{align*}

\subsection{Frank copula}

The Frank copula function is
$$
\Cbbm(u_1,u_2;\theta) = -\frac{1}{\theta}\log\left(1-\frac{(1-e^{-\theta u_1})(1-e^{-\theta u_2})}{1-e^{-\theta}}\right).
$$
The relationship between Kendall's $\tau$ and the copula parameter $\theta$ is
$$
\tau = 1 - \frac{4}{\theta}\left[1-D_1(\theta)\right],
$$
where $D_1(\theta) = \theta^{-1}\int_0^\infty \frac{x}{e^x-1}dx$.

Define $\zeta = \frac{(1-e^{-\theta u_1})(1-e^{-\theta u_2})}{1-e^{-\theta}}$, and the copula function can be expressed as $\Cbbm = -\frac{1}{\theta}\log(1-\zeta)$. We have
\begin{align}
\label{equ:f-c1} \cbbm_1(u_1,u_2;\theta) & = (1-\zeta)^{-1} \frac{e^{-\theta u_1}(1-e^{-\theta u_2})}{1-e^{-\theta}}\\ 
\label{equ:f-c2}\cbbm_2(u_1,u_2;\theta) & =   (1-\zeta)^{-1} \frac{e^{-\theta u_2}(1-e^{-\theta u_1})}{1-e^{-\theta}},\\
\label{equ:f-c}\cbbm(u_1,u_2;\theta) & =  (1-\zeta)^{-2}\frac{\theta e^{-\theta u_1}e^{-\theta u_2}}{1-e^{-\theta}}.
\end{align}
The log-likelihood function is 
$$
\ell = \delta_1\delta_2 \widetilde \cbbm + \delta_1 (1-\delta_2) \widetilde \cbbm_1 + (1-\delta_1)\widetilde \cbbm_2 + (1-\delta_1)(1-\delta_2)\widetilde \Cbbm,
$$
with
$$
\ell_{\theta} = \delta_1\delta_2 \widetilde \cbbm_{\theta} + \delta_1 (1-\delta_2) \widetilde \cbbm_{1,\theta} + (1-\delta_1)\widetilde \cbbm_{2,\theta} + (1-\delta_1)(1-\delta_2)\widetilde \Cbbm_\theta,
$$
and
$$
\ell_{\theta\theta} = \delta_1\delta_2 \widetilde \cbbm_{\theta\theta} + \delta_1 (1-\delta_2) \widetilde \cbbm_{1,\theta\theta} + (1-\delta_1)\widetilde \cbbm_{2,\theta\theta} + (1-\delta_1)(1-\delta_2)\widetilde \Cbbm_{\theta\theta}.
$$
In the following, we derive the first-order and second-order partial derivatives of $\widetilde \Cbbm$, $\widetilde \cbbm_1$, $\widetilde \cbbm_2$, and $\widetilde \cbbm$ w.r.t. $\theta$. 

\paragraph{Derivation of $\widetilde \Cbbm_{\theta}$ and $\widetilde \Cbbm_{\theta\theta}$.} The first-order and second-order derivatives of $\widetilde \Cbbm$ w.r.t. $\theta$ are given as $\widetilde \Cbbm_{\theta} = \Cbbm_{\theta}/\Cbbm$ and $\widetilde \Cbbm_{\theta\theta} = \Cbbm_{\theta\theta}/\Cbbm - \Cbbm_{\theta}^2/\Cbbm^2$, where 
\begin{align*}
\Cbbm_{\theta} & = (1/\theta^2)\log(1-\zeta) + (1/\theta)\zeta_{\theta}(1-\zeta)^{-1}, \\
\Cbbm_{\theta\theta} & = -(2/\theta^3)\log(1-\zeta) - (2/\theta^2)\zeta_{\theta}(1-\zeta)^{-1} + (1/\theta)\zeta_{\theta\theta}(1-\zeta)^{-1} + (1/\theta)\zeta_{\theta}^2(1-\zeta)^{-2},
\end{align*}
where $\zeta_{\theta}$ and $\zeta_{\theta\theta}$ are the first-order and second-order derivatives of $\zeta$ w.r.t. $\theta$. To derive the expressions of $\zeta_{\theta}$ and $\zeta_{\theta\theta}$, we consider taking the derivatives of $\widetilde \zeta = \log \zeta$, which is given as
$$
\widetilde \zeta = \log (1-e^{-\theta u_1}) + \log (1-e^{-\theta u_2}) - \log(1-e^{-\theta}).
$$
The first-order derivative of $\widetilde \zeta$ is
\begin{align*}
\widetilde \zeta_{\theta} & = \frac{e^{-\theta u_1}u_1}{1-e^{-\theta u_1}} + \frac{e^{-\theta u_2}u_2}{1-e^{-\theta u_2}} - \frac{e^{-\theta}}{1-e^{-\theta}} \\
& = u_1 \left(1-e^{-\theta u_1}\right)^{-1} + u_2 \left(1-e^{-\theta u_2}\right)^{-1} - \left(1-e^{-\theta }\right)^{-1} + 1 -u_1 - u_2.
\end{align*}
On the other hand, $\widetilde \zeta_{\theta} = \partial \log \zeta / \partial \theta = \zeta_{\theta}/\zeta$, which gives $\zeta_{\theta} = \zeta \widetilde \zeta_{\theta}$. In addition,
$$
\widetilde \zeta_{\theta\theta} = - u_1^2 \left(1-e^{-\theta u_1}\right)^{-2} e^{-\theta u_1} - u_2^2 \left(1-e^{-\theta u_2}\right)^{-2} e^{-\theta u_2} + \left(1-e^{-\theta }\right)^{-2} e^{-\theta}.
$$
On the other hand, $\widetilde \zeta_{\theta\theta} = \partial^2 \log \zeta/\partial \theta^2 = \zeta_{\theta\theta}/\zeta - \zeta_{\theta}^2/\zeta^2$, which gives $\zeta_{\theta\theta} = \zeta\left(\widetilde \zeta_{\theta\theta} + \zeta_{\theta}^2/\zeta^2\right) = \zeta\left(\widetilde \zeta_{\theta\theta} + \widetilde \zeta_{\theta}^2\right)$.

\paragraph{Derivation of $\widetilde \cbbm_{1,\theta}$ and $\widetilde \cbbm_{1,\theta\theta}$.} By Equation (\ref{equ:f-c1}), $\widetilde \cbbm_{1}= \log \cbbm_{1}$ is given as
\begin{align*}
\widetilde \cbbm_{1} & = - \log(1-\zeta) - \theta u_1 + \log(1-e^{-\theta u_2}) - \log(1-e^{-\theta}).
\end{align*}
Its first-order and second-order derivatives are 
\begin{align*}
\widetilde \cbbm_{1,\theta} & = \zeta_{\theta}(1-\zeta)^{-1} - u_1 + u_2(1-e^{-\theta u_2})^{-1} -u_2 - (1-e^{-\theta})^{-1} + 1,\\
\widetilde \cbbm_{1,\theta\theta} & = \zeta_{\theta\theta}(1-\zeta)^{-1} + \zeta_{\theta}^2(1-\zeta)^{-2} - (1-e^{-\theta u_2})^{-2}e^{-\theta u_2}u_2^2 + (1-e^{-\theta})^{-2}e^{-\theta}.
\end{align*}

\paragraph{Derivation of $\widetilde \cbbm_{2,\theta}$ and $\widetilde \cbbm_{2,\theta\theta}$.} We use the similar arguments for $\widetilde \cbbm_{1,\theta}$ and $\widetilde \cbbm_{1,\theta\theta}$. By Equation (\ref{equ:f-c2}), $\widetilde \cbbm_{2}= \log \cbbm_{2}$ is given as
\begin{align*}
\widetilde \cbbm_{2} & = - \log(1-\zeta) - \theta u_2 + \log(1-e^{-\theta u_1}) - \log(1-e^{-\theta}).
\end{align*}
Its first-order and second-order derivatives are 
\begin{align*}
\widetilde \cbbm_{2,\theta} & = \zeta_{\theta}(1-\zeta)^{-1} - u_2 + u_1(1-e^{-\theta u_1})^{-1} -u_1 - (1-e^{-\theta})^{-1} + 1,\\
\widetilde \cbbm_{2,\theta\theta} & = \zeta_{\theta\theta}(1-\zeta)^{-1} + \zeta_{\theta}^2(1-\zeta)^{-2} - (1-e^{-\theta u_1})^{-2}e^{-\theta u_1}u_1^2 + (1-e^{-\theta})^{-2}e^{-\theta}.
\end{align*}

\paragraph{Derivation of $\widetilde \cbbm_{\theta}$ and $\widetilde \cbbm_{\theta\theta}$.} By Equation (\ref{equ:f-c}), we have
$$
\widetilde \cbbm = \log \cbbm = -2\log(1-\zeta) + \log \theta - \theta u_1 - \theta u_2 - \log(1-e^{-\theta}).
$$
Its first-order and second-order derivatives are 
\begin{align*}
\widetilde \cbbm_{\theta} & = 2\zeta_{\theta}(1-\zeta)^{-1} + 1/\theta - u_1 - u_2 - (1-e^{-\theta})^{-1} + 1,\\
\widetilde \cbbm_{\theta\theta} & = 2\zeta_{\theta\theta}(1-\zeta)^{-1} + 2\zeta_{\theta}^2(1-\zeta)^{-2} - 1/\theta^2 + (1-e^{-\theta})^{-2}e^{-\theta}.
\end{align*}

\def \ubar {\bar u}
\subsection{Joe copula} 
The Joe copula function is given as
$$
\Cbbm(u_1,u_2;\theta) = 1 - \left(\ubar_1^\theta + \ubar_2^\theta - \ubar_1^\theta \ubar_2^\theta\right)^{1/\theta},
$$
with $\ubar_1 = 1-u_1$ and $\ubar_2 = 1 - u_2$. The relationship between Kendall's $\tau$ and the copula parameter $\theta$ is
$$
\tau = 1 + \frac{4}{\theta^2}\int_0^1 x log(x) (1-x)^{2(1-\theta)/\theta}dx.
$$

Let $\Gamma = \ubar_1^\theta + \ubar_2^\theta - \ubar_1^\theta \ubar_2^\theta$ with the first-order and second-order partial derivatives w.r.t. $\theta$ 
\begin{align}
\label{equ:Gamma-theta}\Gamma_{\theta} & = \ubar_1^\theta\log\ubar_1 + \ubar_2^\theta\log\ubar_2 - \ubar_1^\theta\ubar_2^\theta(\log \ubar_1 + \log\ubar_2),\\
\label{equ:Gamma-thetatheta}\Gamma_{\theta\theta} & = \ubar_1^\theta(\log\ubar_1)^2 + \ubar_2^\theta(\log\ubar_2)^2 -\ubar_1^\theta\ubar_2^\theta(\log \ubar_1 + \log\ubar_2)^2.
\end{align}
From the above copula function, we have
\begin{align}
\label{equ:c1}\cbbm_1(u_1,u_2;\theta) & = \Gamma^{1/\theta-1}(1-\ubar_2^\theta)\ubar_1^{\theta-1},\\ 
\label{equ:c2}\cbbm_2(u_1,u_2;\theta) & = \Gamma^{1/\theta-1}(1-\ubar_1^\theta)\ubar_2^{\theta-1},\\
\label{equ:c} \cbbm(u_1,u_2;\theta) &  = \theta\Gamma^{1/\theta-1}\ubar_1^{\theta-1}\ubar_2^{\theta-1} -(1-\theta)\Gamma^{1/\theta-2}\ubar_1^{\theta-1}\ubar_2^{\theta-1}(1-\ubar_1^\theta)(1-\ubar_2^\theta).
\end{align}
The log-likelihood function is 
$$
\ell = \delta_1\delta_2 \widetilde \cbbm + \delta_1 (1-\delta_2) \widetilde \cbbm_1 + (1-\delta_1)\widetilde \cbbm_2 + (1-\delta_1)(1-\delta_2)\widetilde \Cbbm,
$$
with
$$
\ell_{\theta} = \delta_1\delta_2 \widetilde \cbbm_{\theta} + \delta_1 (1-\delta_2) \widetilde \cbbm_{1,\theta} + (1-\delta_1)\widetilde \cbbm_{2,\theta} + (1-\delta_1)(1-\delta_2)\widetilde \Cbbm_\theta,
$$
and
$$
\ell_{\theta\theta} = \delta_1\delta_2 \widetilde \cbbm_{\theta\theta} + \delta_1 (1-\delta_2) \widetilde \cbbm_{1,\theta\theta} + (1-\delta_1)\widetilde \cbbm_{2,\theta\theta} + (1-\delta_1)(1-\delta_2)\widetilde \Cbbm_{\theta\theta}.
$$
In the following, we derive the first-order and second-order partial derivatives of $\widetilde \Cbbm$, $\widetilde \cbbm_1$, $\widetilde \cbbm_2$, and $\widetilde \cbbm$ w.r.t. $\theta$. 

\paragraph{Derivatives of $\widetilde \Cbbm_{\theta}$ and $\widetilde \Cbbm_{\theta\theta}$.}
Define $\Gamma_1 = \Gamma^{1/\theta}$, and $\Cbbm = 1 - \Gamma_1$. We obtain the first-order partial derivative of $\widetilde \Cbbm$: $\widetilde \Cbbm_{\theta}=\frac{\Cbbm_{\theta}}{\Cbbm} = \frac{-\Gamma_{1,\theta}}{1-\Gamma_1}$, where $\Gamma_{1,\theta}$ is the first-order derivative of $\Gamma_1$ w.r.t. $\theta$. To obtain $\Gamma_{1,\theta}$, we consider taking the first-order derivative of $\widetilde \Gamma_ 1= \log\Gamma_1 = (1/\theta) \log \Gamma$ with its first-order derivative $
\widetilde \Gamma_{1,\theta} = - (1/\theta^2)\log \Gamma  + (1/\theta) \Gamma_{\theta}/\Gamma$, where $\Gamma_{\theta}$ is given in Equation (\ref{equ:Gamma-theta}). On the other hand, $\widetilde \Gamma_{1,\theta} = \frac{\partial}{\partial \theta}\log\Gamma_1  = \Gamma_{1,\theta}/\Gamma_1$. 
Thus, the first-order derivative of $\Gamma_1$ is 
$$
\Gamma_{1,\theta} = - (1/\theta^2)\Gamma^{1/\theta}\log \Gamma  + (1/\theta) \Gamma_{\theta}\Gamma^{1/\theta-1}.
$$ 
The second-order partial derivative of $\widetilde \Cbbm$ w.r.t. $\theta$ is $\widetilde \Cbbm_{\theta\theta} = \frac{\Cbbm_{\theta\theta}}{\Cbbm} - \frac{\Cbbm_{\theta}^2}{\Cbbm^2} = \frac{-\Gamma_{1,\theta\theta}}{1-\Gamma_1} - \frac{\Gamma_{1,\theta}^2}{(1-\Gamma_1)^2}$. To obtain $\Gamma_{1,\theta\theta}$, we have
$$
\Gamma_{1,\theta\theta} = \frac{\partial \Gamma_{1,\theta}}{\partial \theta} = \frac{2\Gamma^{1/\theta}\log \Gamma}{\theta^3} - \frac{\Gamma_{1,\theta}\log \Gamma + 2 \Gamma^{1/\theta-1} \Gamma_{\theta}}{\theta^2} + \frac{\Gamma_{\theta\theta}\Gamma^{1/\theta-1}+\Gamma_{\theta}\Gamma_{1,\theta}/\Gamma - \Gamma_{\theta}^2\Gamma^{1/\theta-2}}{\theta},
$$
where $\Gamma_{\theta\theta}$ is given in Equation (\ref{equ:Gamma-thetatheta}). 

\paragraph{Derivatives of $\widetilde \cbbm_{1,\theta}$ and $\widetilde \cbbm_{1,\theta\theta}$.}
By Equation (\ref{equ:c1}), $\widetilde \cbbm_1 = \log \cbbm_1$ is given as
$$
\widetilde \cbbm_1  = (1/\theta-1)\log \Gamma + \log(1-\ubar_2^\theta) + (\theta-1) \log \ubar_1.
$$
The first-order derivative of $\widetilde \cbbm_1$ is
\begin{align*}
\widetilde \cbbm_{1,\theta} &= -(1/\theta^2)\log \Gamma + (1/\theta-1)\Gamma_{\theta}/\Gamma - \frac{\ubar_2^\theta \log\ubar_2}{1-\ubar_2^\theta} + \log \ubar_1.
\end{align*}
The second-order derivative of $\widetilde \cbbm_1$ is
\begin{align*}
\widetilde \cbbm_{1,\theta\theta} &= (2/\theta^3)\log \Gamma - (2/\theta^2)\Gamma_{\theta}/\Gamma  + (1/\theta-1)\Gamma_{\theta\theta}/\Gamma - (1/\theta-1)\Gamma_\theta^2/\Gamma^2- \frac{\ubar_2^\theta(\log \ubar_2)^2}{(1-\ubar_2^\theta)^2}. 
\end{align*}

\paragraph{Derivatives of $\widetilde \cbbm_{2,\theta}$ and $\widetilde \cbbm_{2,\theta\theta}$.} We follow the similar arguments for $\widetilde \cbbm_1$. By Equation (\ref{equ:c2}), $\widetilde \cbbm_2 = \log \cbbm_2$ is given as
$$
\widetilde \cbbm_2  = (1/\theta-1)\log \Gamma + \log(1-\ubar_1^\theta) + (\theta-1) \log \ubar_2.
$$
The first-order derivative of $\widetilde \cbbm_2$ is
\begin{align*}
\widetilde \cbbm_{2,\theta} &= -(1/\theta^2)\log \Gamma + (1/\theta-1)\Gamma_{\theta}/\Gamma - \frac{\ubar_1^\theta \log\ubar_1}{1-\ubar_1^\theta} + \log \ubar_2.
\end{align*}
The second-order derivative of $\widetilde \cbbm_1$ is
\begin{align*}
\widetilde \cbbm_{2,\theta\theta} &= (2/\theta^3)\log \Gamma - (2/\theta^2)\Gamma_{\theta}/\Gamma  + (1/\theta-1)\Gamma_{\theta\theta}/\Gamma - (1/\theta-1)\Gamma_\theta^2/\Gamma^2- \frac{\ubar_1^\theta(\log \ubar_1)^2}{(1-\ubar_1^\theta)^2}. 
\end{align*}

\paragraph{Derivation of $\widetilde \cbbm_{\theta}$ and $\widetilde \cbbm_{\theta\theta}$.} The first-order and second-order derivatives of $\widetilde \cbbm$ w.r.t. $\theta$ are
$$
\widetilde \cbbm_{\theta} = \frac{\cbbm_{\theta}}{\cbbm}\quad \text{and}\quad \widetilde \cbbm_{\theta\theta} = \frac{\cbbm_{\theta\theta}}{\cbbm} - \frac{\cbbm_{\theta}^2}{\cbbm^2}.
$$ 
Let $\Gamma_2 = \Gamma^{1/\theta-1}$ and $\Gamma_3=\Gamma^{1/\theta-2}$. 
By Equation (\ref{equ:c}), 
$$
\cbbm = \theta \Gamma_2 \eta_1 - (1-\theta)\Gamma_3 \eta_2
$$
where $\eta_1=\ubar_1^{\theta-1}\ubar_2^{\theta-1}$ and $\eta_2 = \ubar_1^{\theta-1}\ubar_2^{\theta-1} - \ubar_1^{2\theta-1}\ubar_2^{\theta-1} - \ubar_1^{\theta-1}\ubar_2^{2\theta-1} + \ubar_1^{2\theta-1}\ubar_2^{2\theta-1}$. The first-order and second-order derivatives of $\cbbm$ are
\begin{align*}
\cbbm_{\theta} = & \Gamma_2\eta_1 + \theta \Gamma_{2,\theta}\eta_1 + \theta \Gamma_2\eta_{1,\theta} + \Gamma_3\eta_2 + (\theta-1)\Gamma_{3,\theta}\eta_2 + (\theta-1)\Gamma_3 \eta_{2,\theta},\\
\cbbm_{\theta\theta} = & \Gamma_{2,\theta}\eta_1 + \Gamma_2\eta_{1,\theta} + \Gamma_{2,\theta}\eta_1 + \theta \Gamma_{2,\theta\theta}\eta_1 + \theta \Gamma_{2,\theta} \eta_{1,\theta} + \Gamma_2\eta_{1,\theta} + \theta \Gamma_{2,\theta}\eta_{1,\theta} + \theta \Gamma_2\eta_{1,\theta\theta} \\
& + \Gamma_{3,\theta}\eta_2 + \Gamma_3\eta_{2,\theta} + \Gamma_{3,\theta}\eta_2  + (\theta-1)\Gamma_{3,\theta\theta}\eta_2  + (\theta-1)\Gamma_{3,\theta}\eta_{2,\theta} \\
& + \Gamma_3 \eta_{2,\theta} + (\theta-1)\Gamma_{3,\theta} \eta_{2,\theta} + (\theta-1)\Gamma_3 \eta_{2,\theta\theta},
\end{align*}
where $\Gamma_{2,\theta}$ and $\Gamma_{3,\theta}$ are the first-order derivatives of $\Gamma_2$ and $\Gamma_3$, $\Gamma_{2,\theta\theta}$ and $\Gamma_{3,\theta\theta}$ are the second-order derivatives of $\Gamma_2$ and $\Gamma_3$, $\eta_{1,\theta}$ and $\eta_{2,\theta}$ are the first-order derivatives of $\eta_1$ and $\eta_2$, and $\eta_{1,\theta\theta}$ and $\eta_{2,\theta\theta}$ are the second-order derivatives of $\eta_1$ and $\eta_2$.

In the following, we will derive $\Gamma_{2,\theta}$, $\Gamma_{3,\theta}$, $\Gamma_{2,\theta\theta}$, $\Gamma_{3,\theta\theta}$, $\eta_{1,\theta}$, $\eta_{2,\theta}$, $\eta_{1,\theta\theta}$, and $\eta_{2,\theta\theta}$. The first-order derivatives of $\Gamma_2$ and $\Gamma_3$ are
\begin{align*}
\Gamma_{2,\theta} & = -(1/\theta^2)\Gamma_2\log \Gamma + (1/\theta-1)\Gamma_{\theta}\Gamma_3,\\
\Gamma_{3,\theta} & = -(1/\theta^2)\Gamma_3\log \Gamma + (1/\theta-2)\Gamma_{\theta}\Gamma_4,
\end{align*}
where $\Gamma_4 = \Gamma^{1/\theta-3}$ with its first-order derivative 
$$
\Gamma_{4,\theta} = -(1/\theta^2)\Gamma^{1/\theta-3}\log \Gamma + (1/\theta-3)\Gamma_{\theta}\Gamma^{1/\theta-4}.
$$
The second-order derivatives of $\Gamma_2$ and $\Gamma_3$ are
\begin{align*}
\Gamma_{2,\theta\theta} & = (2/\theta^3)\Gamma_{2}\log \Gamma - (1/\theta^2) \Gamma_{2,\theta}\log \Gamma - (2/\theta^2)\Gamma_3\Gamma_{\theta} + (1/\theta-1)\Gamma_{\theta\theta}\Gamma_3 + (1/\theta-1)\Gamma_{\theta}\Gamma_{3,\theta},\\
\Gamma_{3,\theta\theta} & = (2/\theta^3)\Gamma_{3}\log \Gamma - (1/\theta^2) \Gamma_{3,\theta}\log \Gamma - (2/\theta^2)\Gamma_4\Gamma_{\theta} + (1/\theta-2)\Gamma_{\theta\theta}\Gamma_4 + (1/\theta-2)\Gamma_{\theta}\Gamma_{4,\theta}.
\end{align*}
In addition, the first-order and second-order derivatives of $\eta_1$ are
\begin{align*}
\eta_{1,\theta} & = \ubar_1^{\theta-1}\ubar_2^{\theta-1}(\log \ubar_1 + \log \ubar_2),\\
\eta_{1,\theta\theta} & = \ubar_1^{\theta-1}\ubar_2^{\theta-1}(\log \ubar_1 + \log \ubar_2)^2.
\end{align*}
In addition, the first-order and second-order derivatives of $\eta_1$ are
\begin{align*}
\eta_{2,\theta} & = \ubar_1^{\theta-1}\ubar_2^{\theta-1}(\log \ubar_1 + \log \ubar_2) - \ubar_1^{2\theta-1}\ubar_2^{\theta-1}(2\log \ubar_1 + \log \ubar_2)\\
& \quad - \ubar_1^{\theta-1}\ubar_2^{2\theta-1}(\log \ubar_1 + 2 \log \ubar_2) + \ubar_1^{2\theta-1}\ubar_2^{2\theta-1}(\log \ubar_1 + \log \ubar_2)\\
\eta_{2,\theta\theta} & = \ubar_1^{\theta-1}\ubar_2^{\theta-1}(\log \ubar_1 + \log \ubar_2)^2 - \ubar_1^{2\theta-1}\ubar_2^{\theta-1}(2\log \ubar_1 + \log \ubar_2)^2\\
& \quad - \ubar_1^{\theta-1}\ubar_2^{2\theta-1}(\log \ubar_1 + 2 \log \ubar_2)^2 + \ubar_1^{2\theta-1}\ubar_2^{2\theta-1}(\log \ubar_1 + \log \ubar_2)^2.
\end{align*}

\subsection{Gaussian copula} 
The Gaussian copula function is
$$
\Cbbm(u_1,u_2;\theta) = \Phi_2\left(\Phi^{-1}(u_1), \Phi^{-1}(u_2); \theta\right),
$$
where $\Phi_2(z_1,z_2;\theta)$ is the joint CDF of the bivariate standard normal distribution with correlation $\theta$, and $\Phi(z)$ is the CDF of the standard normal distribution. The relationship between Kendall's $\tau$ and the copula parameter is
$$
\tau = \frac{2}{\pi}\arcsin(\theta).
$$

Based on the above expression of $\Cbbm$, we have
\begin{align*}
\cbbm_1(u_1,u_2;\theta) & = \frac{\int_0^{\Phi^{-1}(u_2)}\phi_{2}\left(\Phi^{-1}(u_1),z_2;\theta\right)dz_2}{\phi\left(\Phi^{-1}(u_1)\right)}\\ 
\cbbm_2(u_1,u_2;\theta) & = \frac{\int_0^{\Phi^{-1}(u_1)}\phi_2\left(z_1,\Phi^{-1}(u_2);\theta\right)dz_1}{\phi\left(\Phi^{-1}(u_2)\right)},\\
\cbbm(u_1,u_2;\theta) & = \frac{\phi_2\left(\Phi^{-1}(u_1),\Phi^{-1}(u_2);\theta\right)}{\phi\left(\Phi^{-1}(u_1)\right)\phi\left(\Phi^{-1}(u_2)\right)},
\end{align*}
where $\phi_2(z_1,z_2;\theta) = \frac{1}{2\pi\sqrt{1-\theta^2}}\exp\left[-\frac{z_1^2+z_2^2-2\theta z_1 z_2}{2(1-\theta^2)}\right]$ is the joint PDF of the bivariate standard normal distribution with correlation $\theta$, and $\phi(z) = \frac{1}{\sqrt{2\pi}}\exp\left(-z^2/2\right)$ is the PDF of the standard normal distribution.

The log-likelihood function is 
$$
\ell = \delta_1\delta_2 \widetilde \cbbm + \delta_1 (1-\delta_2) \widetilde \cbbm_1 + (1-\delta_1)\widetilde \cbbm_2 + (1-\delta_1)(1-\delta_2)\widetilde \Cbbm,
$$
with
$$
\ell_{\theta} = \delta_1\delta_2 \widetilde \cbbm_{\theta} + \delta_1 (1-\delta_2) \widetilde \cbbm_{1,\theta} + (1-\delta_1)\widetilde \cbbm_{2,\theta} + (1-\delta_1)(1-\delta_2)\widetilde \Cbbm_\theta,
$$
and
$$
\ell_{\theta\theta} = \delta_1\delta_2 \widetilde \cbbm_{\theta\theta} + \delta_1 (1-\delta_2) \widetilde \cbbm_{1,\theta\theta} + (1-\delta_1)\widetilde \cbbm_{2,\theta\theta} + (1-\delta_1)(1-\delta_2)\widetilde \Cbbm_{\theta\theta}.
$$
In the following, we derive the first-order and second-order partial derivatives of $\widetilde \Cbbm$, $\widetilde \cbbm_1$, $\widetilde \cbbm_2$, and $\widetilde \cbbm$ w.r.t. $\theta$. 

\paragraph{Derivation of $\widetilde \Cbbm_{\theta}$ and $\widetilde \Cbbm_{\theta\theta}$.} We have $\widetilde \Cbbm_{\theta} = \Cbbm_{\theta}/\Cbbm$ and $\Cbbm_{\theta\theta} = \Cbbm_{\theta\theta}/\Cbbm - \Cbbm_{\theta}^2/\Cbbm^2$. We derive the expressions of $\Cbbm_{\theta}$ and $\Cbbm_{\theta\theta}$. The function $\Cbbm$ can be expressed as
$$
\Cbbm(u_1,u_2;\theta) = \int_0^{\Phi^{-1}(u_1)}\int_0^{\Phi^{-1}(u_2)}\phi_2(z_1,z_2;\theta) dz_1 dz_2.
$$
By the interchangeability between the derivative and integral, we can obtain
\begin{align*}
\Cbbm_{\theta}(u_1,u_2;\theta) & = \int_0^{\Phi^{-1}(u_1)}\int_0^{\Phi^{-1}(u_2)}\phi_{2,\theta}(z_1,z_2;\theta) dz_1 dz_2,\\
\Cbbm_{\theta\theta}(u_1,u_2;\theta) & = \int_0^{\Phi^{-1}(u_1)}\int_0^{\Phi^{-1}(u_2)}\phi_{2,\theta\theta}(z_1,z_2;\theta) dz_1 dz_2,
\end{align*} 
where $\phi_{2,\theta}$ and $\phi_{2,\theta\theta}$ are the first-order and second-order derivatives of $\phi_2$ w.r.t. $\theta$. To obtain their expression, we consider taking derivatives of $\widetilde \phi_2 = \log \phi_2$ given as
$$
\widetilde \phi_2 = \log \phi_2 = - \log(2\pi) - (1/2) \log(1-\theta^2) - \frac{z_1^2+z_2^2-2\theta z_1z_2}{2(1-\theta^2)}. 
$$
The first-order derivative of $\widetilde \phi_2$ w.r.t. $\theta$
\begin{align*}
\widetilde \phi_{2,\theta} & = \frac{\theta(1-\theta^2)-\theta\left(z_1^2+z_2^2\right)+(1+\theta^2)z_1z_2}{(1-\theta^2)^2},
\end{align*}
and the second-order derivative is
\begin{align*}
\widetilde \phi_{2,\theta\theta} & = \frac{1+\theta^2}{(1-\theta^2)^2}+\frac{(6\theta+2\theta^3)z_1z_2-(1+3\theta^2)(z_1^2+z_2^2)}{(1-\theta^2)^3}.
\end{align*}
On the other hand, $\widetilde \phi_{2,\theta} = \partial \log \phi_2/\partial \theta = \phi_{2,\theta}/\pi_2$. Thus,
\begin{equation}\label{equ:phi2-theta}
\phi_{2,\theta} = \phi_2 \widetilde \phi_{2,\theta}. 
\end{equation}
In addition, $\widetilde \phi_{2,\theta\theta} = \partial^2 \log \phi_2/\partial \theta^2 = \phi_{2,\theta\theta}/\phi_2 - \phi_{2,\theta}^2/\phi_2^2$. Thus, 
\begin{equation}\label{equ:phi2-thetatheta}
\phi_{2,\theta\theta} = \phi_2[\widetilde \phi_{2,\theta\theta}+\phi_{2,\theta}^2/\phi_2^2] = \phi_2[\widetilde \phi_{2,\theta\theta}+\widetilde \phi_{2,\theta}^2].
\end{equation}

\paragraph{Derivation of $\widetilde \cbbm_{1,\theta}$ and $\widetilde \cbbm_{1,\theta\theta}$.} We have $\widetilde \cbbm_{1,\theta} = \cbbm_{1,\theta}/\cbbm_1$ and $\cbbm_{1,\theta\theta} = \cbbm_{1,\theta\theta}/\cbbm_1 - \cbbm_{1,\theta}^2/\cbbm_1^2$ with
\begin{align*}
\cbbm_{1,\theta}(u_1,u_2;\theta) & = \frac{\int_0^{\Phi^{-1}(u_2)}\phi_{2,\theta}(\Phi^{-1}(u_1),z_2;\theta) dz_2}{\phi\left(\Phi^{-1}(u_1)\right)}\\\cbbm_{1,\theta\theta}(u_1,u_2;\theta) & = \frac{\int_0^{\Phi^{-1}(u_2)}\phi_{2,\theta\theta}(\Phi^{-1}(u_1),z_2;\theta) dz_2}{\phi\left(\Phi^{-1}(u_1)\right)},
\end{align*}
where $\phi_{2,\theta}$ is given in Equation (\ref{equ:phi2-theta}) and $\phi_{2,\theta\theta}$ is given in Equation (\ref{equ:phi2-thetatheta}).

\paragraph{Derivation of $\widetilde \cbbm_{2,\theta}$ and $\widetilde \cbbm_{2,\theta\theta}$.} We have $\widetilde \cbbm_{2,\theta} = \cbbm_{2,\theta}/\cbbm_2$ and $\cbbm_{2,\theta\theta} = \cbbm_{2,\theta\theta}/\cbbm_2 - \cbbm_{2,\theta}^2/\cbbm_2^2$ with
\begin{align*}
\cbbm_{2,\theta}(u_1,u_2;\theta) & = \frac{\int_0^{\Phi^{-1}(u_1)}\phi_{2,\theta}(z_1,\Phi^{-1}(u_2);\theta) dz_1}{\phi\left(\Phi^{-1}(u_2)\right)},\\
\cbbm_{2,\theta\theta}(u_1,u_2;\theta) & = \frac{\int_0^{\Phi^{-1}(u_1)}\phi_{2,\theta\theta}(z_1,\Phi^{-1}(u_2);\theta) dz_1}{\phi\left(\Phi^{-1}(u_2)\right)},
\end{align*}
where $\phi_{2,\theta}$ is given in Equation (\ref{equ:phi2-theta}) and $\phi_{2,\theta\theta}$ is given in Equation (\ref{equ:phi2-thetatheta}).

\paragraph{Derivation of $\widetilde \cbbm_{\theta}$ and $\widetilde \cbbm_{\theta\theta}$.} We have $\widetilde \cbbm_{\theta} = \cbbm_{\theta}/\cbbm$ and $\cbbm_{\theta\theta} = \cbbm_{\theta\theta}/\cbbm - \cbbm_{\theta}^2/\cbbm^2$ with
\begin{align*}
\cbbm_{\theta}(u_1,u_2;\theta) & = \frac{\phi_{2,\theta}(\Phi^{-1}(u_1),\Phi^{-1}(u_2);\theta)}{\phi\left(\Phi^{-1}(u_1)\right)\phi\left(\Phi^{-1}(u_2)\right)},\\
\cbbm_{\theta\theta}(u_1,u_2;\theta) & = \frac{\phi_{2,\theta\theta}(\Phi^{-1}(u_1),\Phi^{-1}(u_2);\theta)}{\phi\left(\Phi^{-1}(u_1)\right)\phi\left(\Phi^{-1}(u_2)\right)},
\end{align*}
where $\phi_{2,\theta}$ is given in Equation (\ref{equ:phi2-theta}) and $\phi_{2,\theta\theta}$ is given in Equation (\ref{equ:phi2-thetatheta}).

%
%
%

